\documentclass[twocolumn,runningheads]{svjour2}
\smartqed  
\usepackage[active]{srcltx}
\usepackage{graphicx}
\usepackage{mathptmx}   
\usepackage{amsmath}
\usepackage{amsfonts}
\usepackage{amssymb}
\usepackage{latexsym}
\usepackage[authoryear]{natbib}
\usepackage{journals}

\journalname{Astrophysics and Space Science (CoRoT/ESTA Volume)}
\newcommand{\msol}{\mbox{${M}_{\odot}$}}

\newcommand{\corot}{{\small CoRoT}}

\newcommand{\esta}{{\small ESTA}}

\newcommand{\task}{{\small TASK}}

\newcommand{\CESAM}{{\small\bf CESAM}}
\newcommand{\CLES}{{\small\bf CL\'ES}}

\newcommand{\cesam}{{\small CESAM}}
\newcommand{\cles}{{\small CL\'ES}}

\newcommand{\PMS}{{\small PMS}}
\newcommand{\MS}{{\small MS}}
\newcommand{\ZAMS}{{\small ZAMS}}
\newcommand{\TAMS}{{\small TAMS}}
\newcommand{\SGB}{{\small SGB}}

\newcommand{\adipls}{{\small ADIPLS}}

\newcommand{\losc}{{\small LOSC}}
\newcommand{\thisapss}{Astrophys. Space Sci. (CoRoT/ESTA Volume)}

\begin{document}

\title{Thorough analysis of  input physics in CESAM and CL\'ES codes
}
\subtitle{}

\titlerunning{Grids:\cles+\losc}        

\author{  Josefina Montalb\'an        \and
          Yveline Lebreton            \and
	  Andrea Miglio               \and
	  Richard Scuflaire           \and
	  Pierre~~Morel                \and
	  Arlette Noels
}


\institute{J.~Montalb\'an \and A.~Miglio \and R.~Scuflaire \and A.~Noels
            \at 
            Institut d'Astrophysique et Geophysique, Universit\'e de Li\`ege,
            all\'ee du 6 Ao\^ut 17, B-4000 Li\`ege, Belgium\\
           \email{J.Montalban, A.Miglio, R.Scuflaire@ulg.ac.be}
	    \and
	   Y. Lebreton \at Observatoire de Paris, GEPI, CNRS UMR 8111, 5 Place Janssen, 92195 Meudon, France \\
              \email{Yveline.Lebreton@obspm.fr}
	    \and
	   P. Morel \at D\'epartement Cassiop\'ee,CNRS UMR 6202, Observatoire de la C\^ote d'Azur,  Nice, France\\
             \email{Pierre.Morel@obs-nice.fr}  
}

\date{Received: date / Accepted: date}

\maketitle
\begin{abstract}
This contribution is not about the quality of the agreement between stellar models computed by \cesam\ and \cles\ codes, but more interesting, on  what \esta-Task~1 run has taught us about these codes and about  the input physics they use. We also quantify the effects of different implementations of the same  physics on the seismic properties of the stellar models, that in fact is the main aim of \esta\ experiments.

\end{abstract}


 \section{Introduction}\label{sec:intro}

The goal of \esta-Task~1 experiment is  to check the evolution codes and, if necessary,  to improve them.  The results of Task~1 comparisons were presented in \citet{monteiro06} and \citet{yl2-apss} for a set of stellar models representative of potential \corot\ targets. The models calculated for \task~1 were based on rather simple input physics. Moreover,  a great  effort  was  done to reduce at maximum the differences between computations by fixing the values of fundamental constants and the physics to be used \citep[see][]{yl1-apss}. In spite of that, some differences among stellar models computed by different codes persist.

In  \cesam\ and \cles\ computations we paid  attention to adopt, not only the same fundamental constants and metal mixture \citep[][thereafter GN93]{GN93}, but also the same isotopic ratios and atomic mass values.   Nevertheless, even if the same metal mixture, opacity tables and equation of state were adopted, there is still some freedom on their implementation. In  this paper we  analyze these different implementations and   estimate the consequent effects on the stellar structure and on the seismic properties of the  theoretical models.  In section~\ref{sec:eos} we study the equation of state and  in section~\ref{sec:opa} the differences in the opacity tables.  The nuclear reaction rates are discussed in Sect.~\ref{sec:nuc} and the effect of different surface boundary conditions in Sect.~\ref{sec:atm}. Finally, in Sect.~\ref{sec:num} we analyze the differences due to different numerical techniques.

\section{Equation of State}
\label{sec:eos}
As fixed in  \esta\, we used the OPAL2001 \citep{2002ApJ...576.1064R} equation of state which is  provided in a tabular form.  In  \cesam\  the  quantities: density, $\rho$, internal energy, $E$, the compressibilities $\chi_T=(\partial \ln P/\partial \ln T)_{\rho}$ and $\chi_{\rm \rho}=(\partial \ln P/\partial \ln \rho)_{\rm T}$, the adiabatic indices $\Gamma_1$,  $\Gamma_2/(\Gamma_2-1)$,  $\Gamma_3-1$, and the specific heat at constant volume $C_V$, are obtained  from the variables $P$, $T$, $X$  and $Z$ (respectively pressure, temperature, hydrogen and heavy element mass fraction) using the interpolation package  provided on the OPAL web site, and the specific heat at constant pressure ($C_p$) is derived from $(\Gamma_3-1)$.  On the other hand, \cles\  interpolates only $C_V$, $P$, $\chi_\rho$  and $\chi_T$ in the OPAL EOS tables by a method ensuring the continuity of first derivatives  at cell boundaries  in the four-dimensional space defined by the variables $\rho$, $T$, $X$ and $Z$.  The other thermodynamic quantities   $\Gamma_1$, $(\Gamma_3 -1)$ and $C_p$ are derived from the values of $C_V$, $P$, $\chi_\rho$ and   $\chi_T$ by means of the thermodynamic relations.

\begin{figure}
\centering
\includegraphics[scale=0.35]{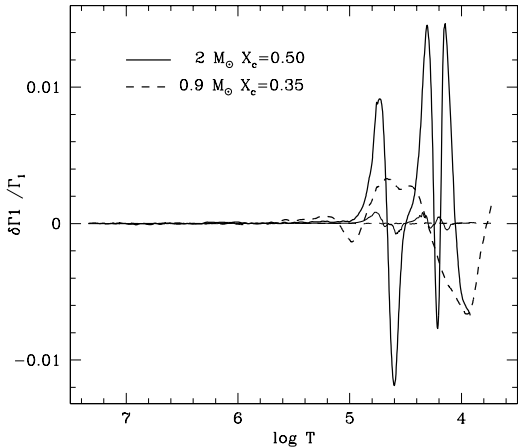}
\caption{Relative differences in the $\Gamma_1$ values provided for a given ($\rho$,$T$) structure by \cesam\ and \cles\ EOS routines. The thick lines correspond to $\Gamma_1$ values derived from $C_V$ tabulated values while the thin ones are derived from the tabulated $\Gamma_1$.  Solid lines corresponds to a 2~\msol\ model with a central  hydrogen mass fraction  $X_{\rm c}=0.50$, and dashed ones to a 0.9~\msol\  star in the middle of the main-sequence.}
\label{fig:eoscesam-cles}
\end{figure} 

As a first step we want to disentangle the differences in the thermodynamic quantities from  their effects on  the stellar structure. We estimate therefore the intrinsic differences between the equation of state used in \cesam\ and in \cles. To this purpose we computed the differences between the thermodynamic quantities from the corresponding EoS routines, for a stellar structure defined by $\rho$, $T$, $X$ and $Z$ values. In Fig.~\ref{fig:eoscesam-cles} (thick lines) we show the result of $\Gamma_1$  comparison for two different  stellar models, a 2~\msol\ model with a mass fraction of hydrogen in the center $X_{\rm c}=0.50$ (solid line)  and a 0.9~\msol\  model with $X_{\rm c}=0.35$ (dashed line). By comparing also the other thermodynamic quantities we found that the largest discrepancies between \cesam\ and \cles\ EoS occurs for $\log T < 5$  (corresponding to the partial He and H ionization regions), and they are, at maximum, of the order of 2\% for  $\Gamma_1$, and  of 5\% for $\nabla_{\rm ad}$ and $C_p$. By using the OPAL interpolation routine in \cles, we verified that  the different interpolation schemes used in \cesam\ and \cles\  can only account for an uncertainty of 0.05\% in $P$, 0.2\% in $\Gamma_1$, and 0.5\% in $\nabla_{\rm ad}$\ and $C_p$.  These remaining differences are probably explained by the fact that \cesam\ uses as variables ($P,T$) and uses  subroutine {\it rhoofp}  of OPAL-package to transform ($P,T$) into  ($\rho,T$), while \cles\ uses directly ($\rho,T$).  Nevertheless,  those discrepancies  are  an order of magnitude smaller than the differences between \cesam\ and \cles\ EoS.

\begin{figure}
\centering
\includegraphics[scale=0.35]{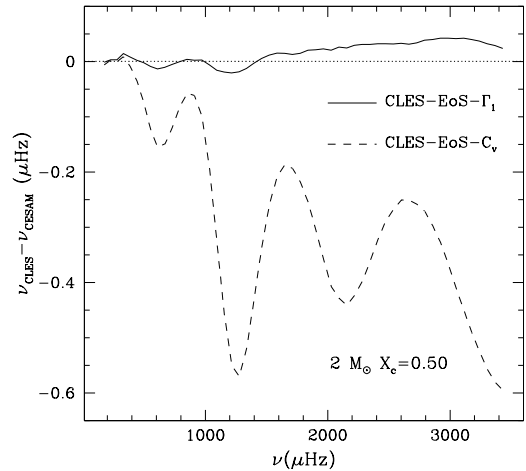}
\caption{Frequency differences between \cesam\ and \cles\ 2~\msol\ models. The two different curves correspond to CLES models computed by using the two different EoS tables (see text).}
\label{fig:eoscesam-cles1.5}
\end{figure} 

\begin{figure}
\centering
\includegraphics[scale=0.35]{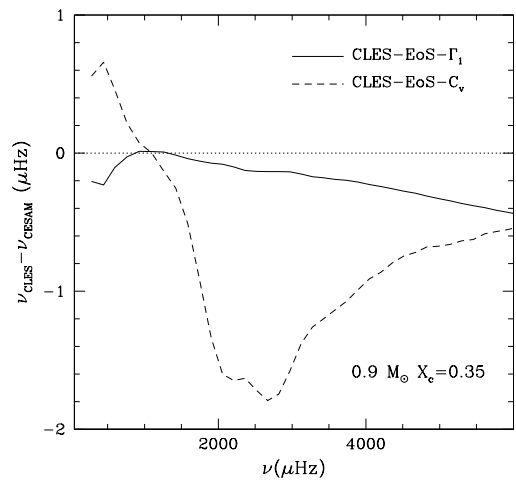}
\caption{As Fig.~\ref{fig:eoscesam-cles1.5} for 0.9~\msol\ model.}
\label{fig:eoscesam-cles1.1}
\end{figure} 

As it was already noted by \citet{boothroydsackmann03},  some inconsistencies existed between thermodynamic quantities tabulated in OPAL EOS: {\it ``for the OPAL EOS \citep{rogersetal96},  we found that there were significant inconsistencies when we compared their tabulated values of  $\Gamma_1$, $\Gamma_2/(\Gamma_2-1)$, and $(\Gamma_3 -1)$ to the values calculated from their  tabulated values of $P$, $C_V$, $\chi_{\rho}$, and $\chi_T$. ... Preliminary tests indicate that this OPAL2001 EOS has larger but smoother inconsistencies in its tabulated thermodynamic quantities...''}. As a consequence of these inconsistencies, the choice of the basic thermodynamic quantities  is not irrelevant, and it was shown by Roxburgh (2005, priv. communication), that the choice done in \cles\ was the worst one. A direct comparison with the  values of $C_V$ computed from the  derivative of the internal energy as tabulated in OPAL EOS, showed that the OPAL tabulated $C_V$ was affected by a large inaccuracy.

The OPAL team acknowledged afterwards  the $C_V$-issue and recommended  not to use it. The EoS tables used in \cles\ have then been changed by replacing the  tabulated $C_V$ value by that obtained from the tabulated values of $P$, $\chi_{\rho}$, $\chi_T$, and  $\Gamma_1$. The remaining discrepancies  ($\sim0.2\%$) between the \cesam\ $\Gamma_1$  values and those from the new \cles--EoS table (hereafter called \cles-{\small EoS-$\Gamma_1$} to tell it apart from the original one \cles-{\small EoS-$C_V$}) are due to the different interpolation routine.  As shown  in Fig.~\ref{fig:eoscesam-cles} (thin lines) the discrepancies are  much smaller for a solar like than for a 2~\msol\ model  and they appear mainly in the ionization regions.

Concerning  the quantities that in \cles\  are obtained from thermodynamic relations and  in \cesam\  from interpolation in  OPAL tables, the differences come in part  from the interpolation routine and in part from the remaining, even if much smaller, inconsistencies between the tabulated  values of  $\Gamma_1$, $\Gamma_2/(\Gamma_2-1)=\nabla_{\rm ad}^{-1}$, and $(\Gamma_3 -1)$. For instance, the values of $C_V$ derived from $\Gamma_1$ may differ by 0.5\% from the corresponding value obtained from  $(\Gamma_3 -1)$, and that occurs  always in the H and He ionization regions. The problem is that even if the thermodynamic relations to derive the adiabatic indices seem more physical, there is some numerical incoherence. In fact, the derivatives of interpolated (very often polynomial) quantities  do not fit in those of the interpolated functions (whose behavior is far from  polynomial one). 

All the \cles\ models involved  in Task~1 and Task~3 comparisons \citep{yl2-apss,mm2-apss} were  recomputed with {\small CL\'ES-EoS}-$\Gamma_1$, but the models used for comparisons presented in \cite{monteiro06} were not. In fact, most of the frequency differences found in that paper came from {\small CL\'ES-EoS-$C_V$}. The effect of EoS differences on the seismic properties are illustrated in Fig.~\ref{fig:eoscesam-cles1.5} for a 2~\msol\ model and in Fig.~\ref{fig:eoscesam-cles1.1}  for the solar like model. In those figures we plot the frequency differences of $\ell=0$ modes for \cesam\ models and two types of \cles\ ones:  those computed with {\small EoS-$C_V$} (dashed lines) and those computed with {\small EoS-$\Gamma_1$} (solid line). The period of the oscillatory signature shown by   $\Delta\nu$ ($\nu_{\rm\small CLES}-\nu_{\rm \small CESAM}$) in  Fig.~\ref{fig:eoscesam-cles1.5} is related to the acoustic depth  where models differ. A Fourier transform of $\Delta\nu$ shows clearly  that the oscillation is linked to the $\Gamma_1$ differences. Moreover, the comparison of Fig.~\ref{fig:eoscesam-cles1.1} with the Fig.6  in \citet{monteiro06} confirms that also for the Case1.1 model \citep[see e.g.][]{yl2-apss}, the maximum difference of almost 2~$\mu$Hz   between \cesam\ and \cles\ models  found by \citet{monteiro06} was due to the inconsistency between the tabulated $C_V$ and adiabatic indices.

\section{ Opacities}
\label{sec:opa}
\esta\ specifications require the use of  OPAL96 opacity tables \citep{ir96} complemented at low temperatures by the \citet{af94} (thereafter AF94) tables. \cesam\  uses OPAL tables provided  by C. Iglesias, prior to their availability on the web site, and interpolates in the opacity tables by means of a four-point Lagrangian interpolation. The OPAL opacity tables used by \cles\  were picked up later on the OPAL web site and smoothed according to the prescription found there ({\small xztrin21.f} routine), we will call them  thereafter {\sc OPAL96-S}. Furthermore, the interpolation method in \cles\ opacity routine is the same as that  used in EoS table interpolation.  In both codes the metal mixture adopted in  the opacity tables is the GN93 one.

\begin{figure}
\centering
\includegraphics[scale=0.35]{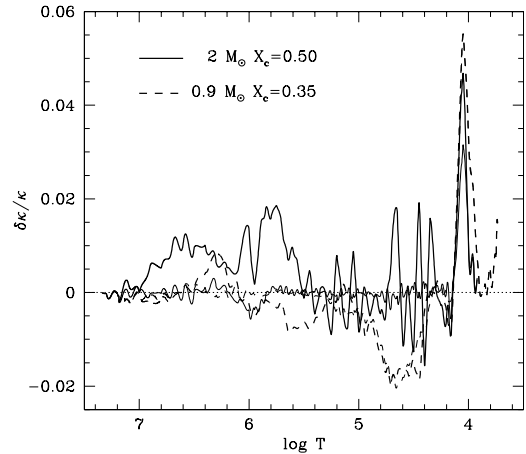}
\caption{Relative differences in the opacity  values provided for a given ($\rho$,$T$) structure by \cesam\ and \cles\ opacity  routines. The thick lines correspond to the $\kappa$ values derived from the smoothed OPAL tables,  while the thin ones were obtained by including in the \cles\ opacity routine the OPAL tables without smoothing. Solid lines corresponds to a 2~\msol\ model with $X_{\rm c}=0.50$, and dashed ones to a 0.9~\msol\ star in the middle of the main-sequence.}
\label{fig:difopa-t}
\end{figure}

\begin{figure}
\centering
\includegraphics[scale=0.35]{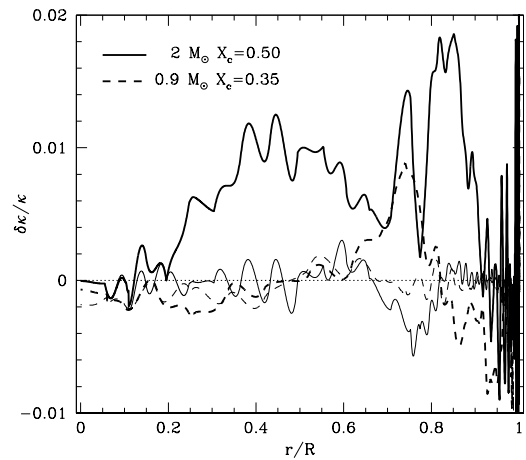}
\caption{The same differences as in  Fig.~\ref{fig:difopa-t} but plotted as a function of the relative stellar radius.}
\label{fig:difopa-r}
\end{figure}

To disentangle the differences in the opacity computations from the differences in the stellar structure, we  proceed as in EoS table analysis, that is, we estimate the intrinsic differences in the opacity ($\kappa$) by comparing the $\kappa$ values provided by \cesam\ and by \cles\ routines for the same  stellar structure. The results of these comparisons are shown in Figs.~\ref{fig:difopa-t} and ~\ref{fig:difopa-r}, where we plot for two different stellar models the opacity relative differences ($\kappa_{\rm \small CESAM}-\kappa_{\rm \small CLES})/\kappa$ as a function of the local temperature and of the relative radius. From comparisons of different models it results that the opacity discrepancies depend on the mass of the stellar model and, for a given mass, on the evolutionary state as well. Moreover, a peaked  feature at $\log T \simeq 4$ which can reach values of the order of 5\%,  appears in all the comparisons. This is a consequence of the differences between OPAL and AF94 opacities in the domain [9000~K--12000~K] and of the different method used in \cles\ and \cesam\ to assemble AF94 and OPAL tables. \cles\ uses the procedure described in \citet{rs1-apss} that ensures a smooth passage between both tables, while \cesam\ searches for the point of minimum  discrepancy  between OPAL and AF94. In the interior regions the differences between \cesam\ and \cles\ opacities do not present the  oscillatory behavior that we would expect if these differences resulted from the interpolation schemes. On the contrary, the \cesam\ opacities are systematically larger (by 1-2\%) than the \cles\ ones in the region $\log T \in [5.5,7]$ of 2~\msol\ model.

\begin{figure}[b]
\centering
\includegraphics[scale=0.35]{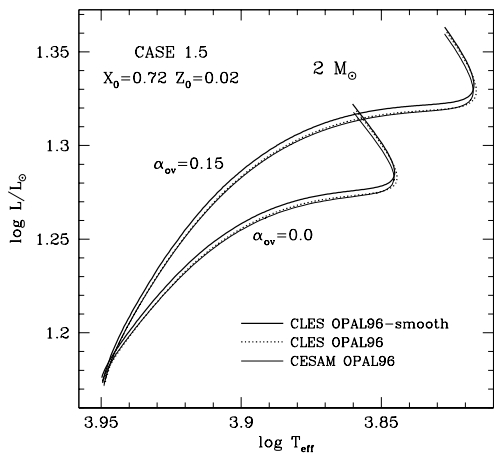}
\caption{Evolutionary tracks for stellar parameters corresponding to the case C1.5 in Task~1, and also without overshooting. Solid thick lines: \cles\ models with the default opacity tables (OPAL96-S); solid thin lines: \cesam\ models ; dotted lines: \cles\ models where the opacity tables have been recomputed without using the opal smoothing filter (OPAL96).}
\label{fig:hr}
\end{figure}

Even if the metal mixture to be used in opacity computations is fixed (GN93), there may be some uncertainties in its definition. For instance,  OPAL uses atomic masses that do not correspond to the values given by the isotopic ratios in \citet{ag89}. In particular there is a difference of 0.5\% for Neon, and 10\% for Argon. Moreover, OPAL opacity tables are computed for 19 elements, while the  GN93 mixture contains 23 elements. There are two options: either to ignore the mass fraction of F, Sc, V, and Co, or to allot the abundances of these elements among the close neighbors. We have analyzed the effects of these uncertainties on the opacity values, but they turned out to be  of the same order of the accuracy in OPAL data (0.1--0.2\%). Hence, they  cannot account for the discrepancy between \cesam\ and \cles\ opacities.

The other important difference between \cesam\ and  \cles\  opacities is  on whether they use the OPAL smoothing routine or not. In fact, the OPAL opacity tables are affected by somewhat random numerical errors of a few percent. To overcome undesirable effects the  OPAL web site  suggests  to pass the original tabular data through a smoothing filter before  interpolating for  $Z$, $X$, $\log T$, and $R$ (with $R=\rho/T^3_6$).  A direct comparison between the original  and smoothed opacity values have shown a difference larger than 2\% for $\log R=-4$ and $ -3.5$  and $\log T \in [5.5,7]$. These differences decrease for larger and smaller values of $\log R$.

We have computed new opacity tables for the \cles\ opacity routine without passing through the smoothing filter (\cles-{\small OPAL96} instead of \cles-{\small OPAL96-S}). The comparison between \cesam\ and new \cles\ opacity computations are also shown in Figs.~\ref{fig:difopa-t},~\ref{fig:difopa-r} (thin lines). We note that when both codes use similar OPAL96 tables, the discrepancies in the internal regions almost  disappear. The remaining differences are due to the interpolation schemes and to the small differences in GN93 definition. The feature at $\log T \sim 4$ is still there  since the method used in \cles\ to assemble AF94 and OPAL96 tables is the same as in \cles-{\small OPAL96-S}.

At variance with the EOS tables, where an error was detected and acknowledged by the OPAL team, we do not have any argument to prefer the smoothed to the original opacity tables, and we think that the  differences between both groups of results must be considered  an estimate of the precision of current stellar modeling.  Therefore, the Li\'ege group decided to provide for Task~1 and Task~3 comparisons \citep{yl2-apss} the modes computed with the standard tables  in \cles, that is {\sc OPAL96-S}. A part of the differences between \cesam\ and \cles\ models that were reported in \citet{yl2-apss} should be hence  due to the opacity tables we used. In order to estimate these effects we have re-computed with \cles\ and  OPAL96 tables (without smoothing) the models for all the cases in \task~1 \citep[see Table~1.][]{yl2-apss}. In the next three sections we present the effect of the opacity uncertainty on: the global parameters, the stellar structure, and on the seismic properties.

\subsection{Effects on global stellar parameters}
\label{sec:globalpar}
In general, the change of opacity tables  decreases  the discrepancies between the stellar global parameters provided by \cles\ and \cesam. In table~1 we collect the differences (in percent) in radius, luminosity,  central density, and central temperature between pairs of models for the Task~1 targets. Columns labeled A give the differences $X_{\rm CLES-OPAL96-S}-X_{\rm CLES-OPAL96}$, and  columns B and C the differences with \cesam, that is, $X_{\rm CLES-OPAL96-S}-X_{\rm CESAM}$ and  $X_{\rm CLES-OPAL96}-X_{\rm CESAM}$ respectively.

We note that for the most evolved models (C1.3 and C1.5), the effect on the radius of changing the \cles\ opacity tables is small, and  that the agreement with \cesam\ gets even worse than with the original tables. There is however a significant decrease of the luminosity discrepancy. For the cases C1.4 and C1.6 (\PMS\ and \ZAMS\ respectively) the change from OPAL96-S to OPAL96 is particularly effective, leading to a decrease of $\Delta R$ and $\Delta L$ by a factor 4 and 6 respectively for C1.4, and dropping the discrepancy to values lower than $0.003$\% for C1.6.

\begin{figure}
\centering
\includegraphics[scale=0.35]{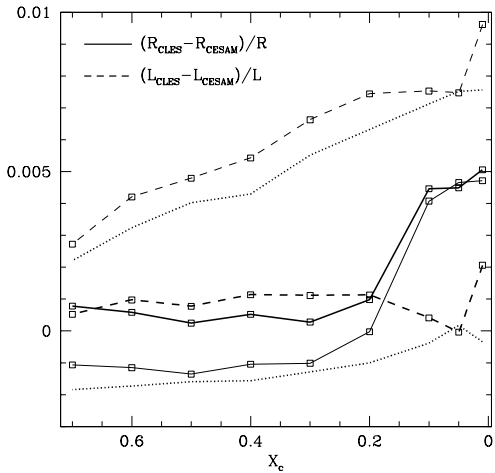}
\caption{Effect of opacity uncertainties on the stellar radius and luminosity along the  main-sequence evolution of a 2~\msol\ model. Thin lines correspond to differences between \cesam\ and \cles-OPAL96-S global parameters, and thick ones to differences between \cesam\ and \cles-OPAL96 ones. The dotted lines refer to differences in radius (lower curve) and in luminosity (upper curve) between \cles\ models computed with OPAL96-S and OPAL96 opacity tables.}
\label{fig:deltaHR}
\end{figure}

We also studied the effect of opacity tables on the main-sequence evolution of a 2~\msol\ star (parameters corresponding to C1.5). As shown in Fig.~\ref{fig:hr} the HR location of the \cles-{\sc OPAL96-S} evolutionary track is significantly modified by adopting OPAL96 tables, and except for the second gravitational contraction, the new track coincides quite well with the \cesam\ one.  The discrepancies in radius and luminosity along the \MS\ as well as the effect of opacity tables on their values are shown in Fig.~\ref{fig:deltaHR}.
These discrepancies are significantly reduced by switching from OPAL96-S to OPAL96, nevertheless the difference in the stellar radius at the end of \MS\ phase ($X_{\rm c} < 0.2$) is unchanged. That is not due to the uncertainties in opacity as shown by the dotted lines that correspond to the differences between \cles\ models computed with the two different opacity tables.  Neither it is a consequence of the treatment of overshooting  since stellar models computed without overshooting for the same stellar parameters show similar discrepancies. The reason is in  the treatment of the borders of convective regions in \cles\  that leads to a sort of  ``numerical diffusion'' \citep{rs2-apss}. For the stellar evolution, this diffusion at the border of the convective core works as a slightly larger overshooting. A significant increase of the number of mesh points used for computing the models reduces the numerical diffusion and improves the agreement \cles-\cesam. 

\begin{table*}[ht!]
\caption{Global parameter differences, in percent, between Task~1-target models computed by \cles\ with  OPAL opacity tables in the two versions: smoothed (OPAL96-S) and not smoothed (OPAL96). Stellar radius ($\Delta R$), luminosity ($\Delta L$), central density ($\Delta \rho_c$)  and central temperature ($\Delta T_c$). Columns labeled A give the differences $X_{\rm CLES-OPAL96-S}-X_{\rm CLES-OPAL96}$, and  columns B and C the differences with \cesam, that is, $X_{\rm CLES-OPAL96-S}-X_{\rm CESAM}$ and  $X_{\rm CLES-OPAL96}-X_{\rm CESAM}$ respectively.}
\centering
\label{tab:task1}
\begin{tabular}[h]{cclcccccccccccccccc}
\hline\noalign{\smallskip}
{\bf Case} & \boldmath$M/M_\odot$ & {\bf Type} &
\multicolumn{3}{c}{\boldmath$\Delta R\,/R$} &
\multicolumn{3}{c}{\boldmath$\Delta L/\,L$} &
\multicolumn{3}{c}{\boldmath$\Delta \rho_{\rm c}/\,\rho_{\rm c}$} &
\multicolumn{3}{c}{\boldmath$\Delta T_{\rm c}/\,T_{\rm c}$}  \\[3pt] \cline{4-6} \cline{7-9} \cline{10-12} \cline{13-15}
\\[-3pt]
 & & & A & B& C& A & B& C&  A & B& C&  A & B& C\\
\tableheadseprule\noalign{\smallskip}
{\bf C1.1}  & 0.9 & \MS\ & -0.01 & 0.10 & -0.08 & -0.03 & -0.05 & -0.02 & 0.00 & -0.17 & -0.17 & -0.01 & -0.02 & -0.007 \\ 
{\bf C1.2}  & 1.2 & \ZAMS\ & -0.05 & -0.12 & -0.07 & 0.04 & -0.06 & -0.10 & -0.001 & -0.17 & -0.17 & 0.006 & -0.04 & -0.05 \\ 
{\bf C1.3}  & 1.2 & \SGB\ & 0.01 & 0.33 & 0.31 & 0.36 & 0.90 & 0.53 & -0.30 & -2.5 & -2.2 & -0.02 & 0.55 & 0.57 \\ 
{\bf C1.4}  & 2.0 & \PMS\ & -0.16 & -0.13 & 0.04 & 0.26 & 0.20 & 0.04 & -0.16 & 0.47 & -0.30 & 0.00 & 0.00 & 0.00 \\ 
{\bf C1.5}  & 2.0 & \TAMS\ & -0.03 & 0.43 & 0.46 &  0.76 & 0.82 & 0.07 & -0.20 & -0.30 & -0.10 &  0.06 & 0.05 & -0.007 \\ 
{\bf C1.6}  & 3.0 & \ZAMS\ & -0.14 &  -0.14 & 0.00 & 0.14 & -0.14 & 0.00 & -0.06 & -0.05 & 0.00 &   0.01 & -0.008 & 0.01 \\ 
{\bf C1.7}  & 5.0 & \MS\ &   0.05 &  0.18 & 0.13 & 0.003 & 0.25 & 0.25 & -0.005 & -0.05 & -0.05 &  0.000 & 0.02 & 0.02 \\ 
\noalign{\smallskip}\hline
\end{tabular}
\end{table*}

\subsection{Effects on the stellar structure}

\begin{figure*}[ht!]
\centering
\resizebox{\hsize}{!}{\includegraphics{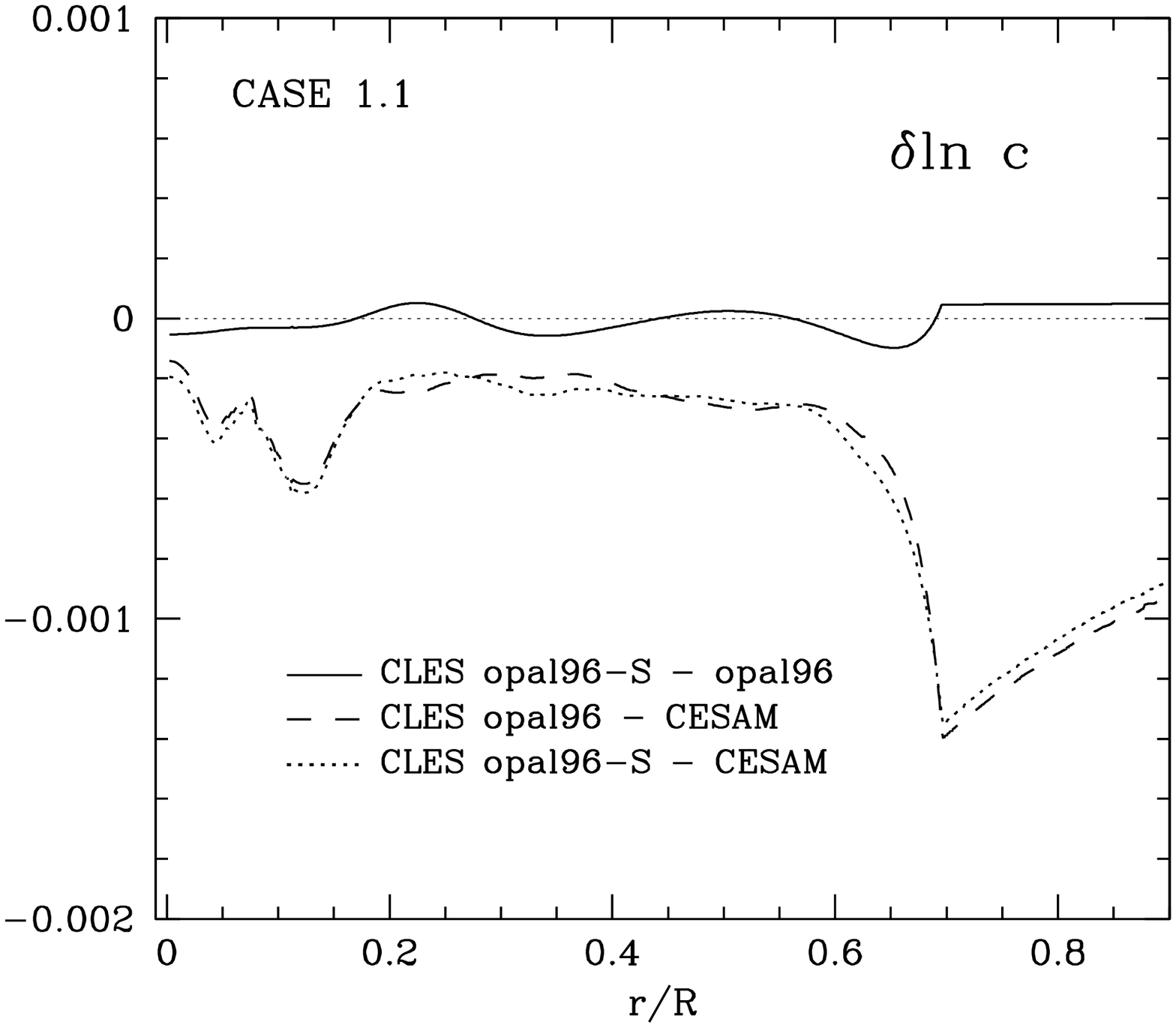}\includegraphics{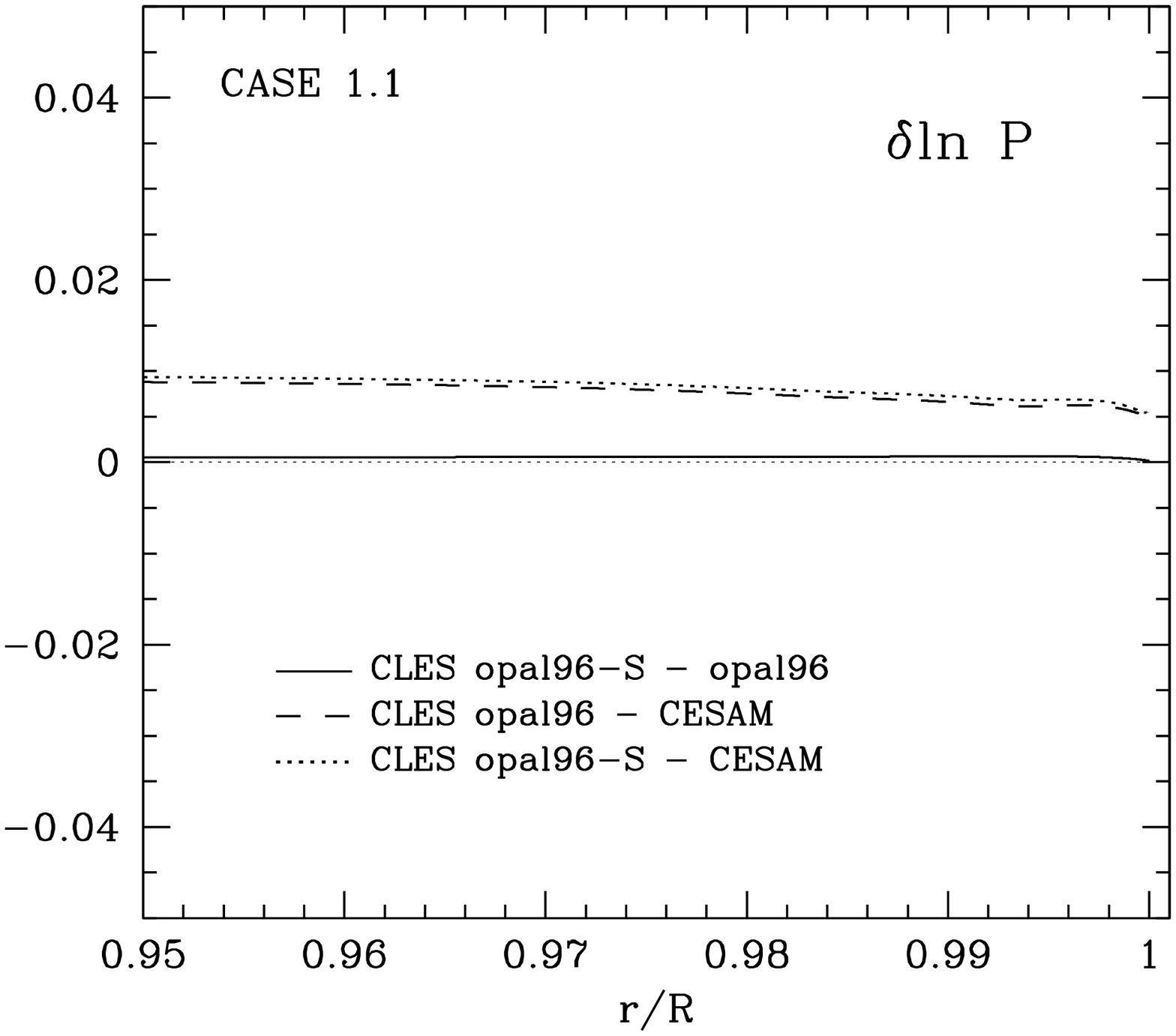}
\includegraphics{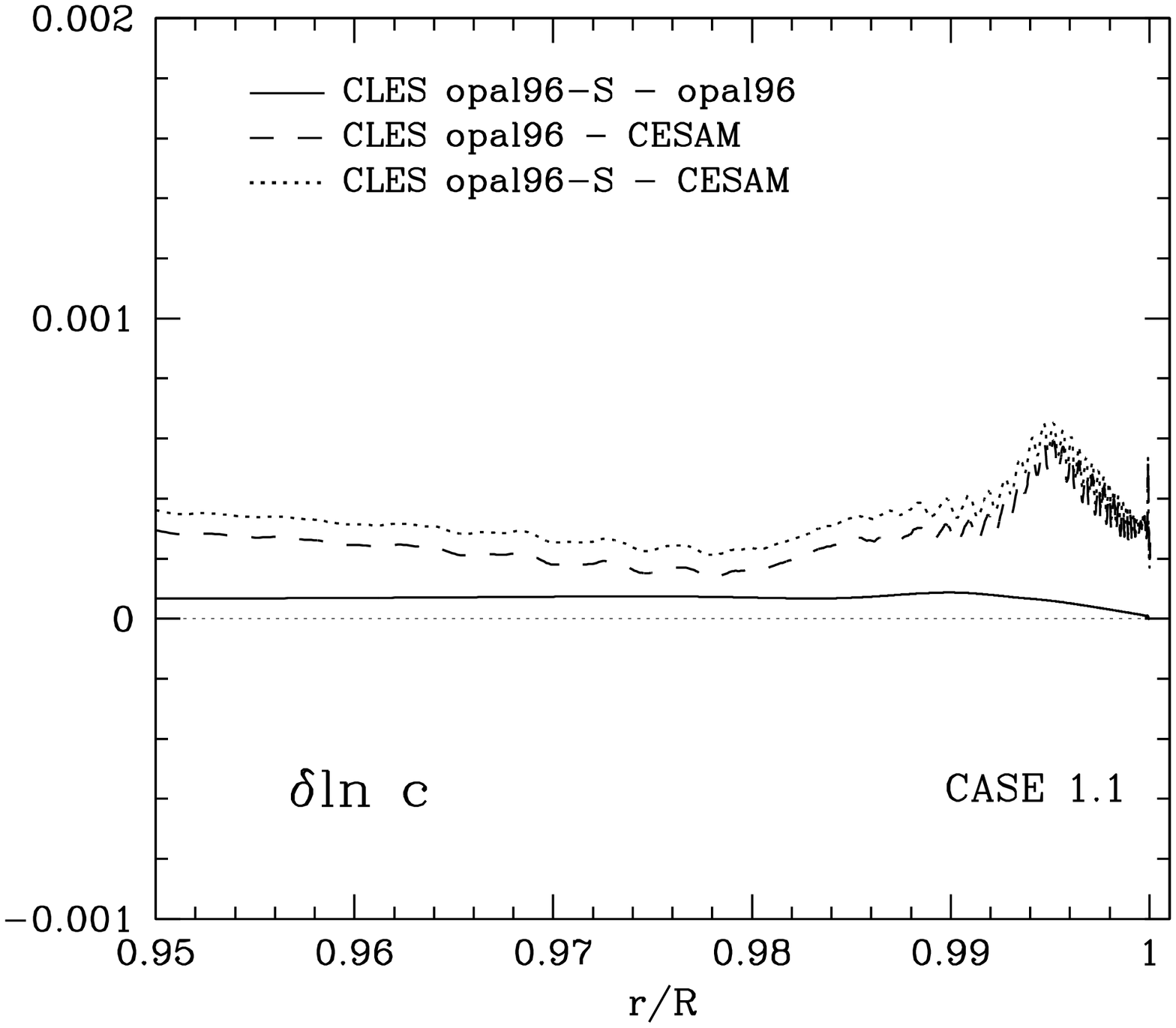} \includegraphics{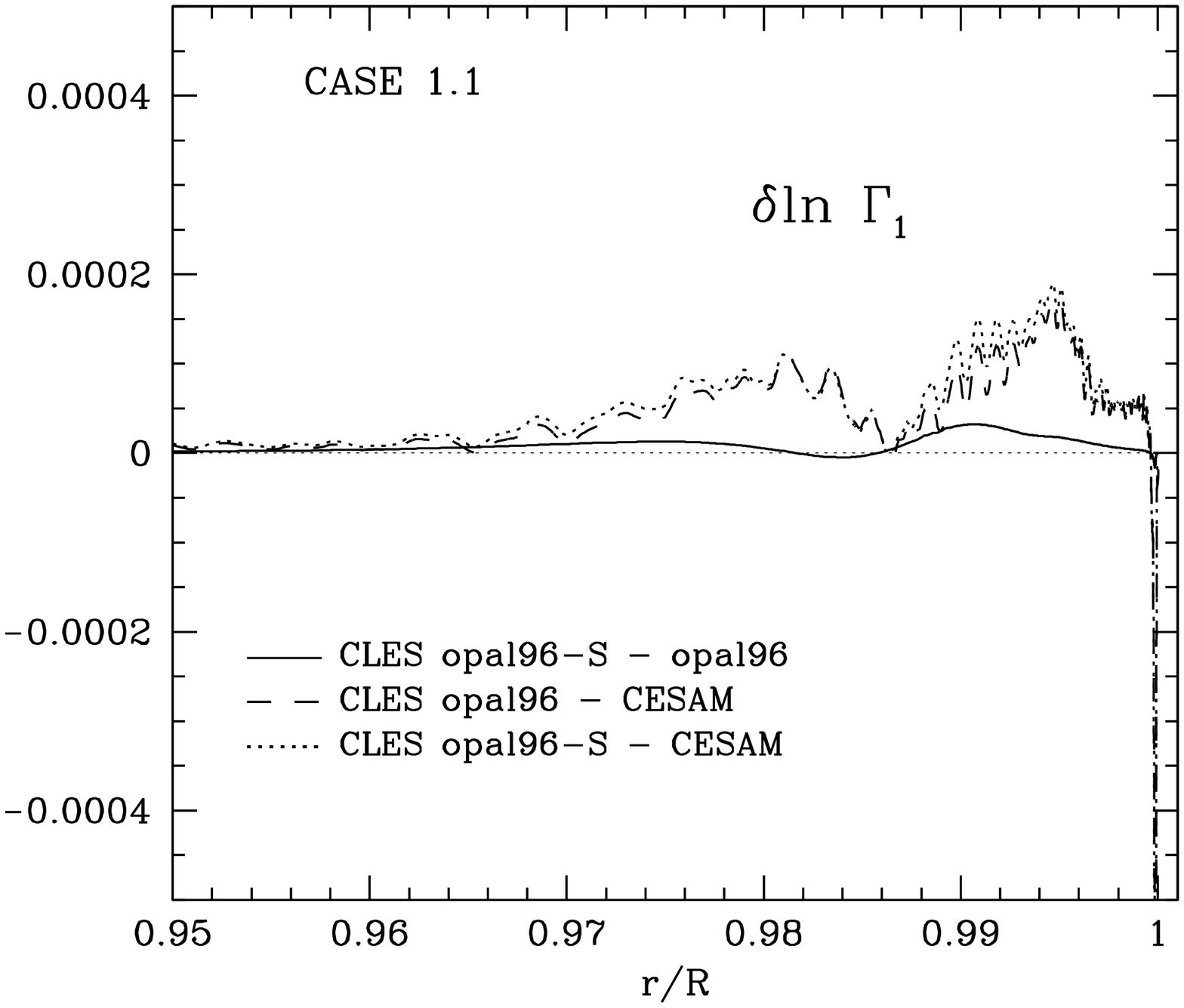}}
\resizebox{\hsize}{!}{\includegraphics{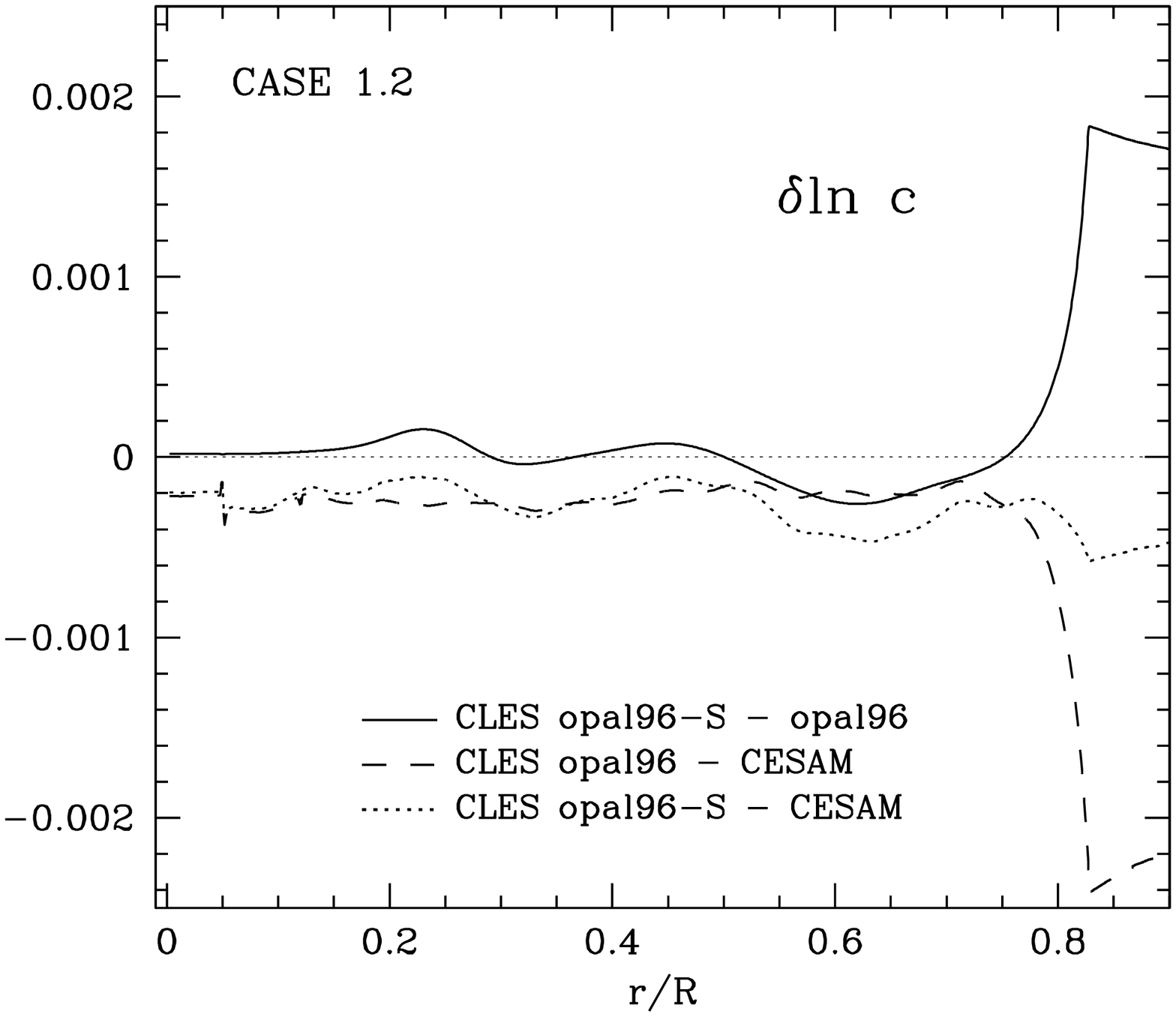}\includegraphics{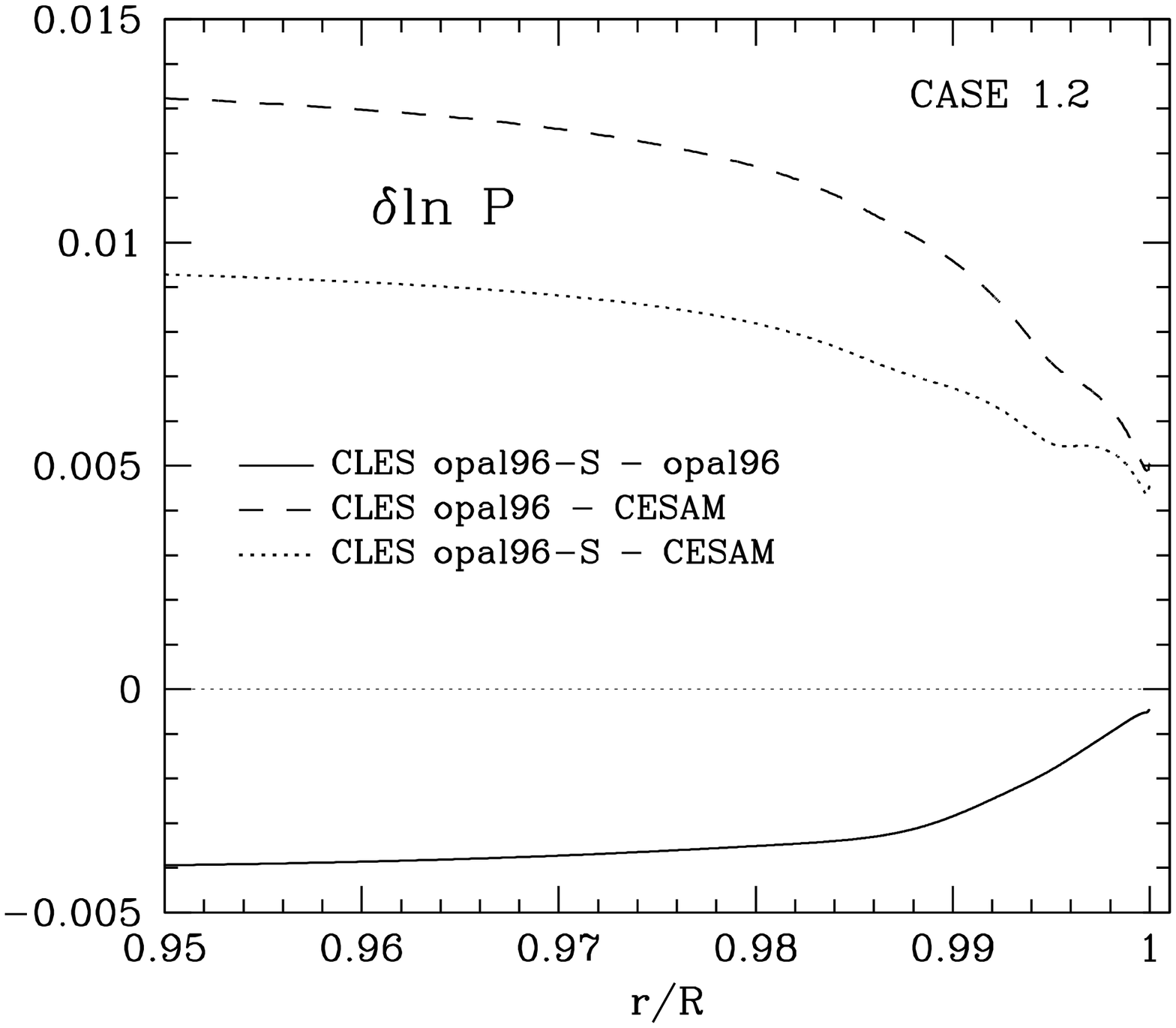}
\includegraphics{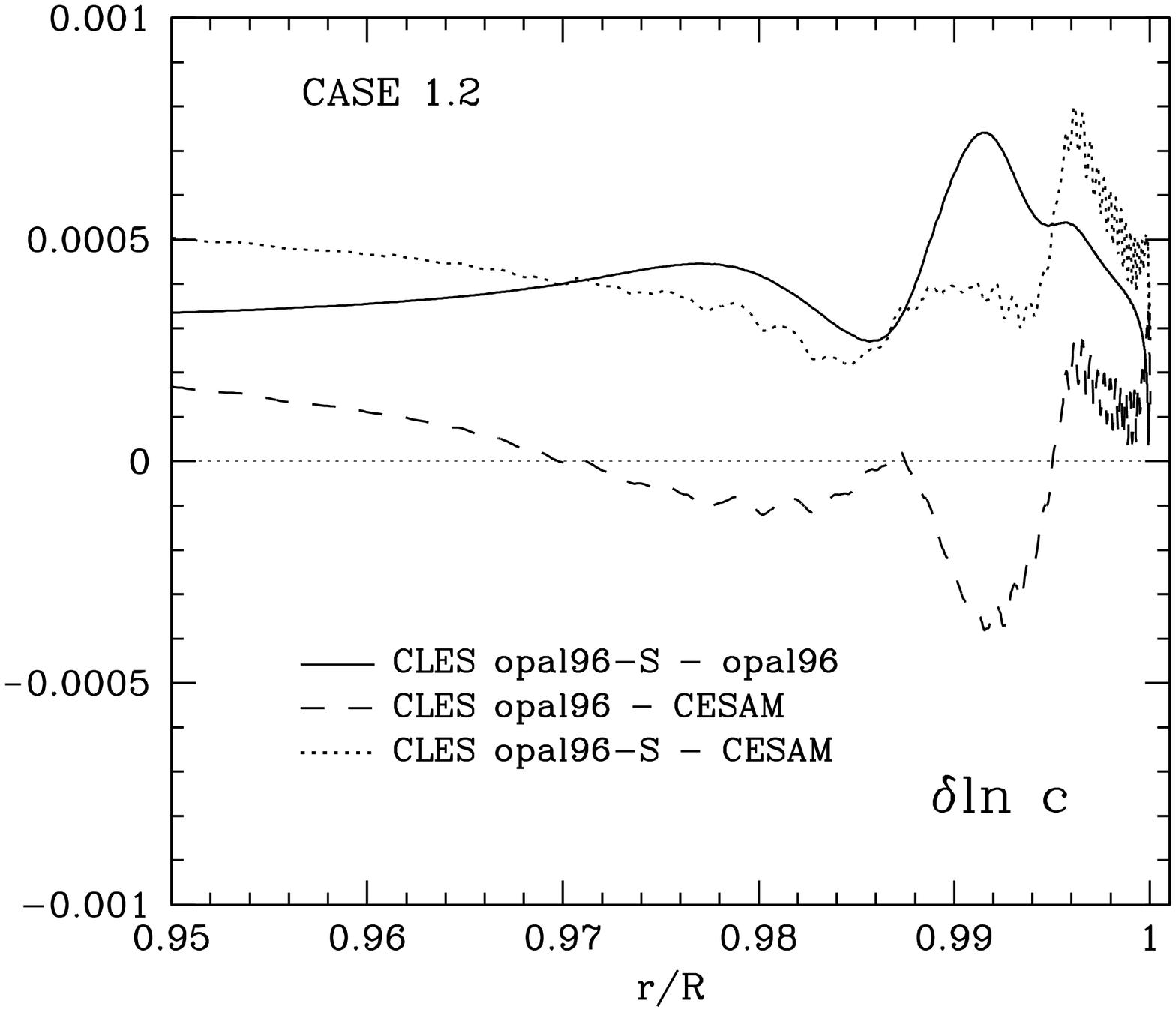} \includegraphics{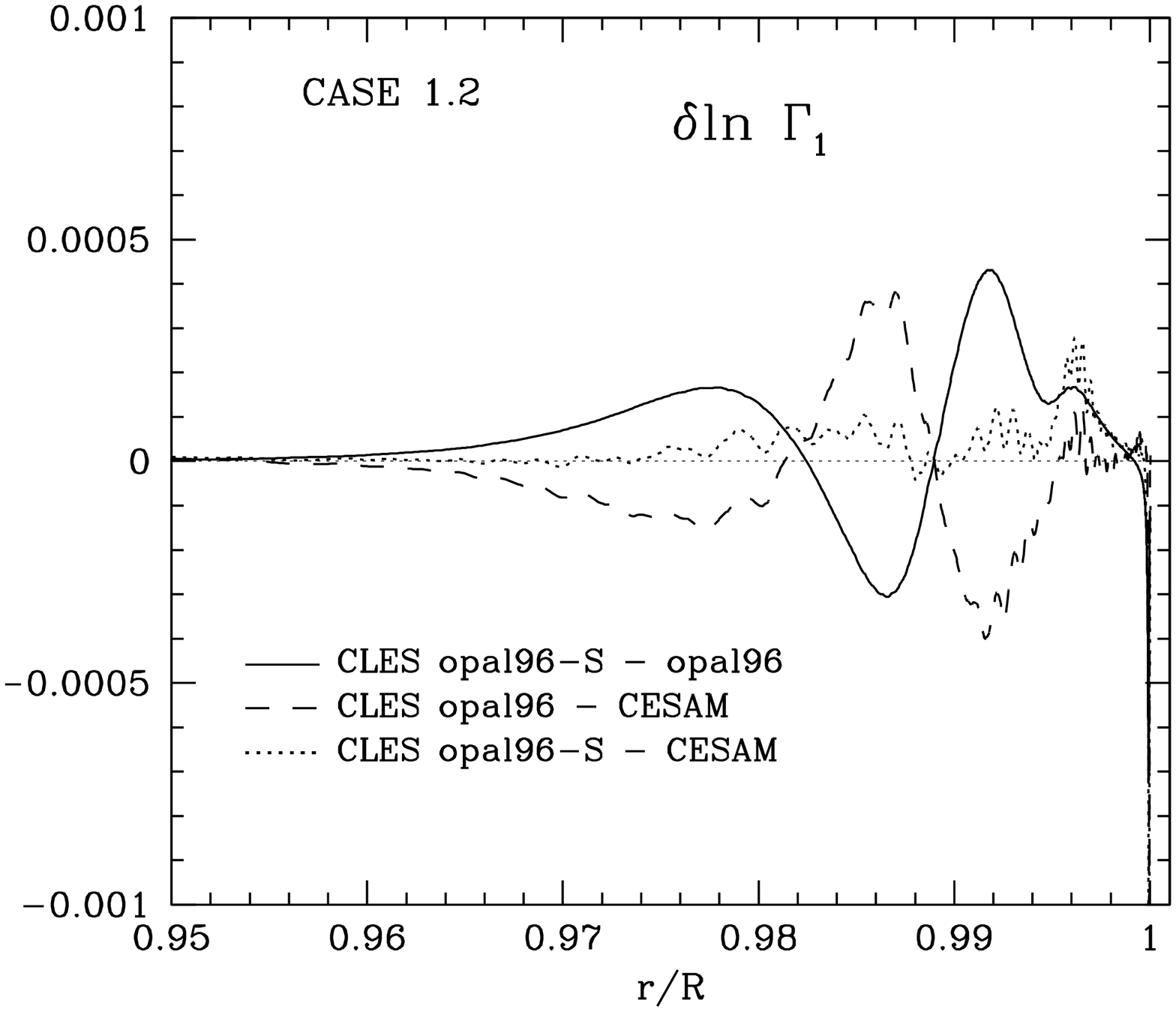}}
\resizebox{\hsize}{!}{\includegraphics{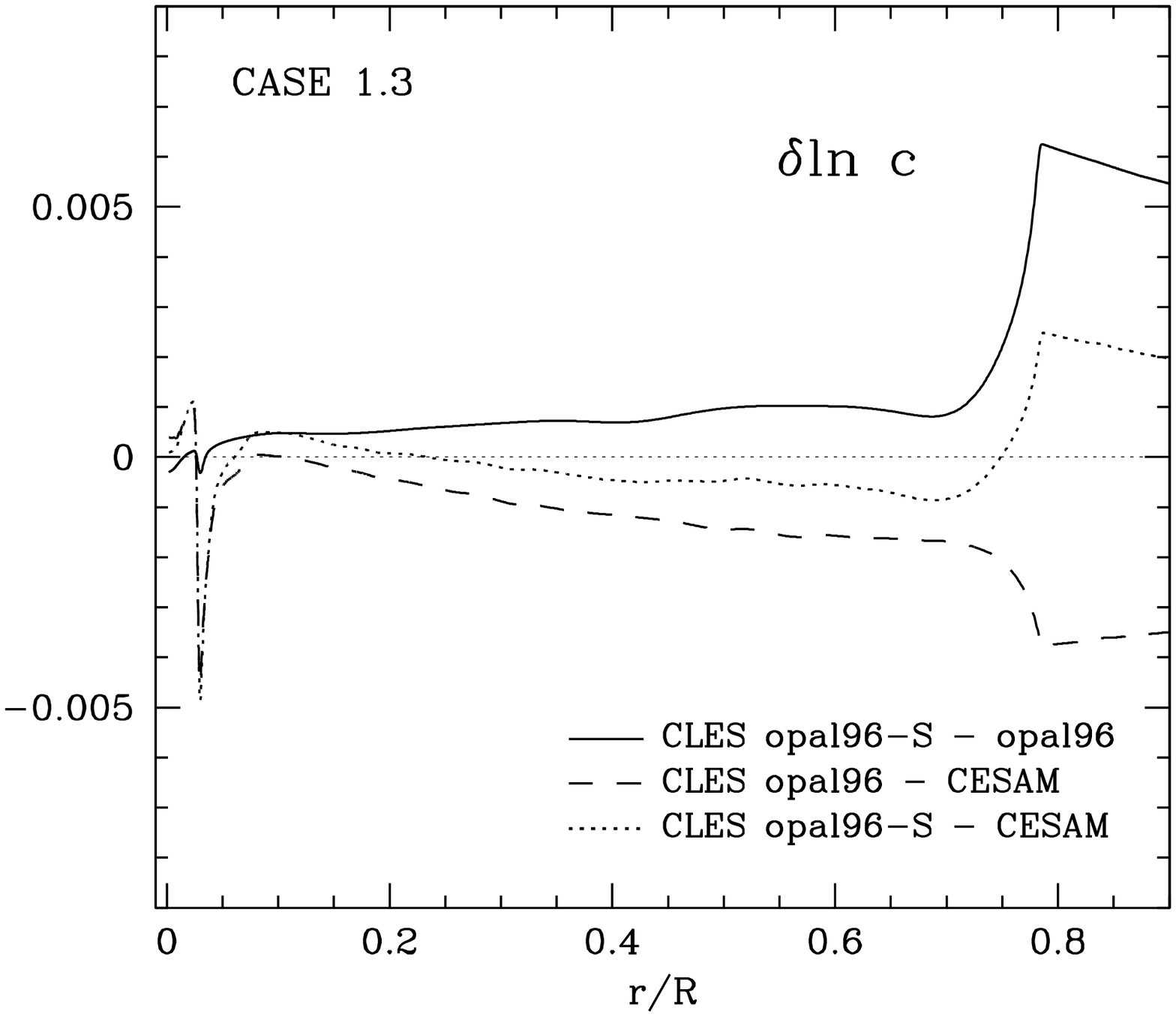}\includegraphics{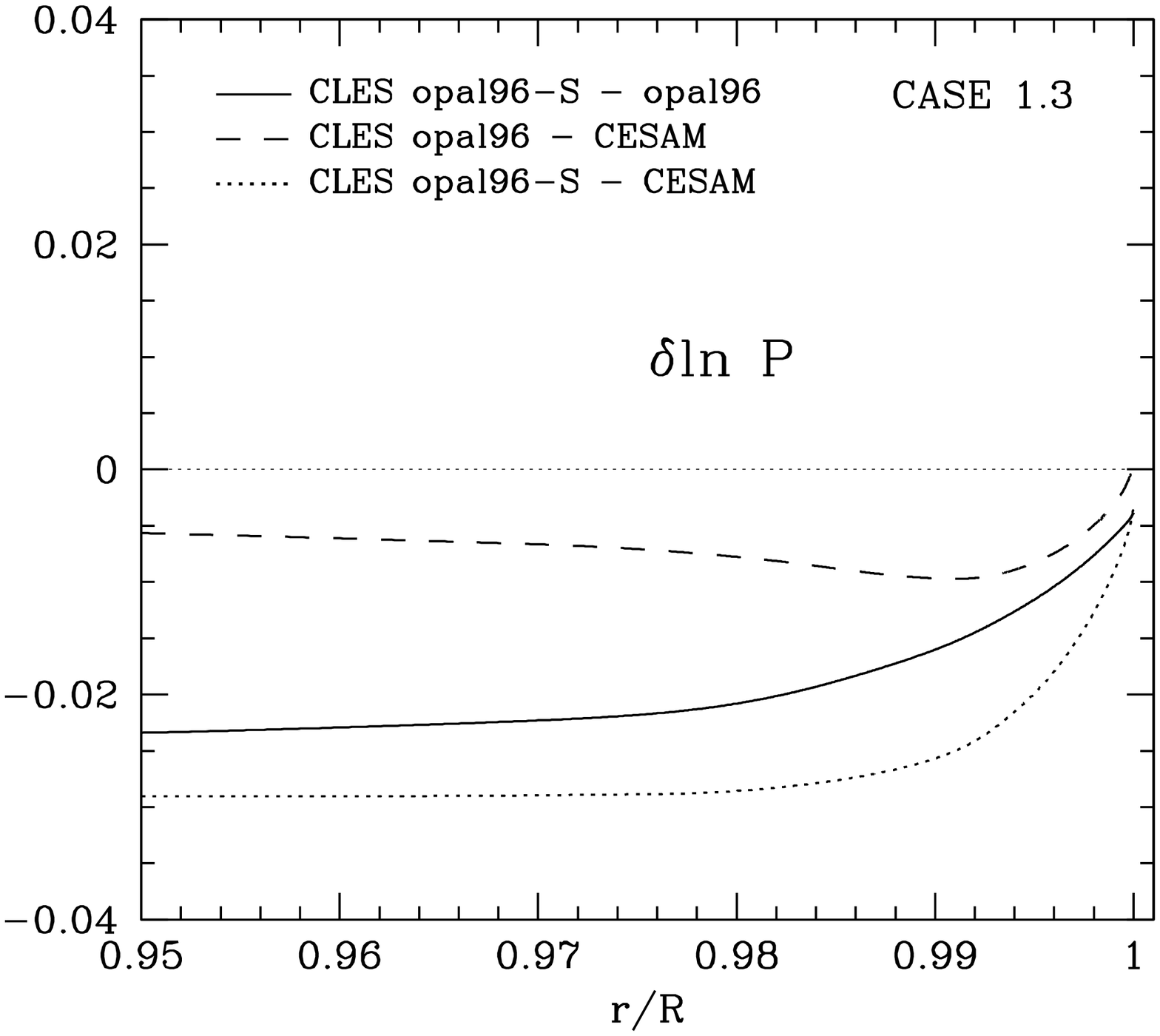}
\includegraphics{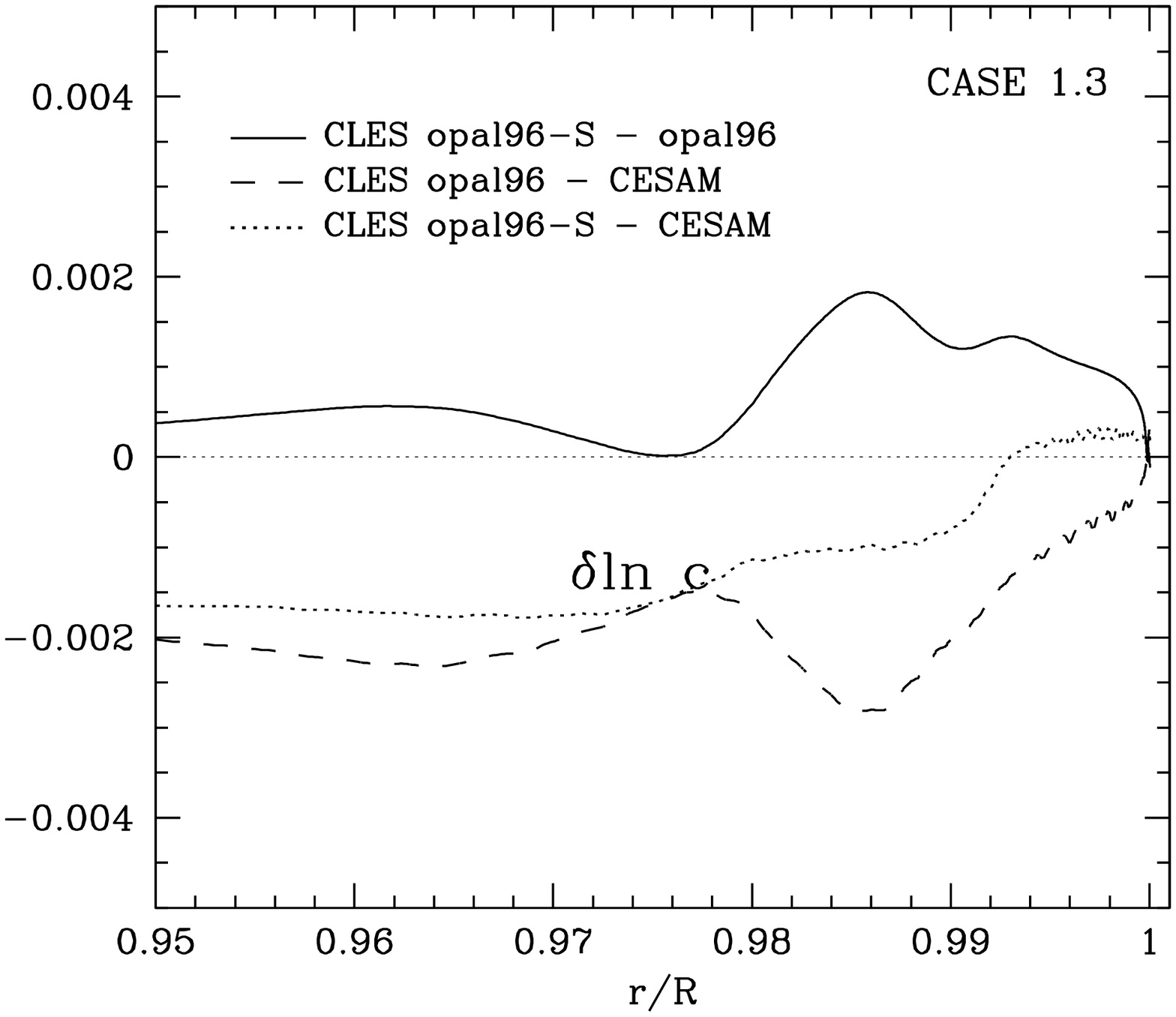} \includegraphics{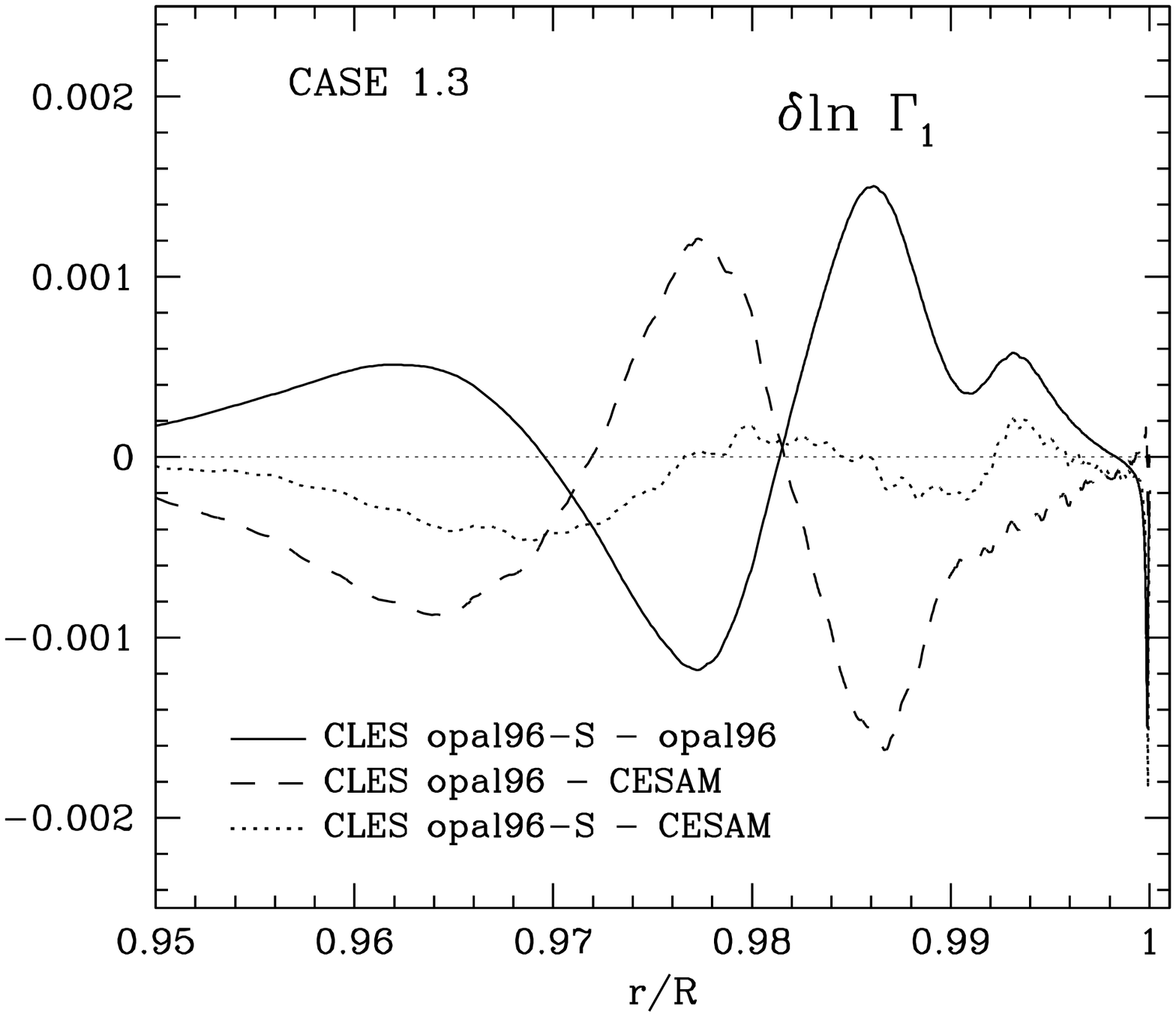}}
\vspace*{0.5cm}
\caption{Plots in terms of the relative radius of the differences at fixed relative mass (two left panels) for the internal regions, and at fixed radius (two right panels) for the outer layers, between models computed with different opacity routines for the Cases~1.1, 1.2 and 1.3.  Solid lines correspond to the difference between two types of \cles\ models: those obtained with the standard \cles\ version that uses the smoother OPAL opacity tables (opal96-S) and  those obtained by using an OPAL opacity table obtained without smoothing. Dotted lines: differences between the standard \cles\ models and those from \cesam. Dashed lines: differences between \cles\ models computed by using OPAL opacity tables without smoothing and \cesam\ models. {\it Left panel}: logarithmic sound speed differences. {\it Central left panel}: logarithmic pressure differences. {\it Central right panel}: logarithmic sound speed differences. {\it Right panel}: logarithmic adiabatic exponent differences.} 
\label{fig:dif-struc-opal}
\end{figure*}

\begin{figure*}[ht!]
\centering
\resizebox{\hsize}{!}{\includegraphics{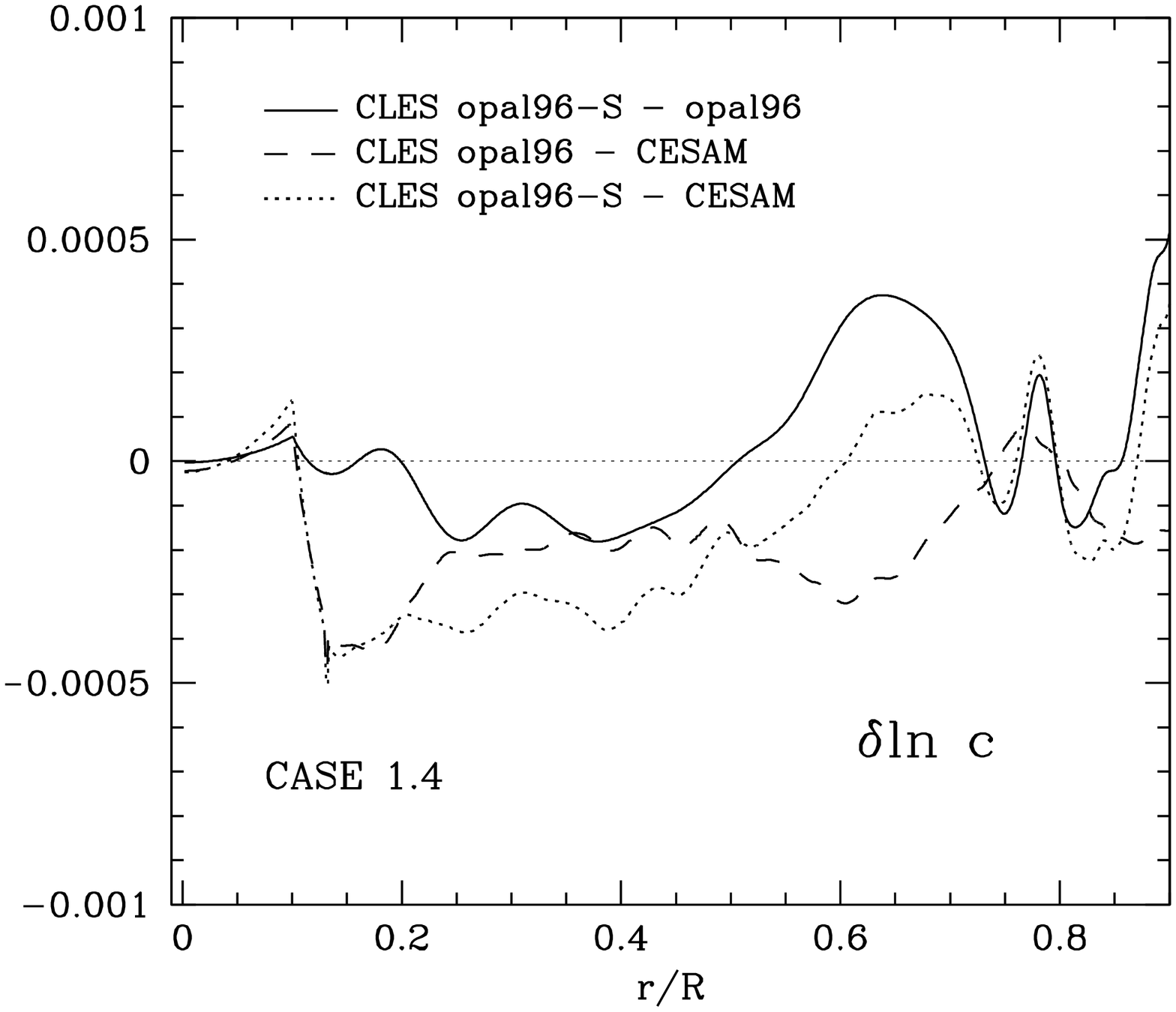}\includegraphics{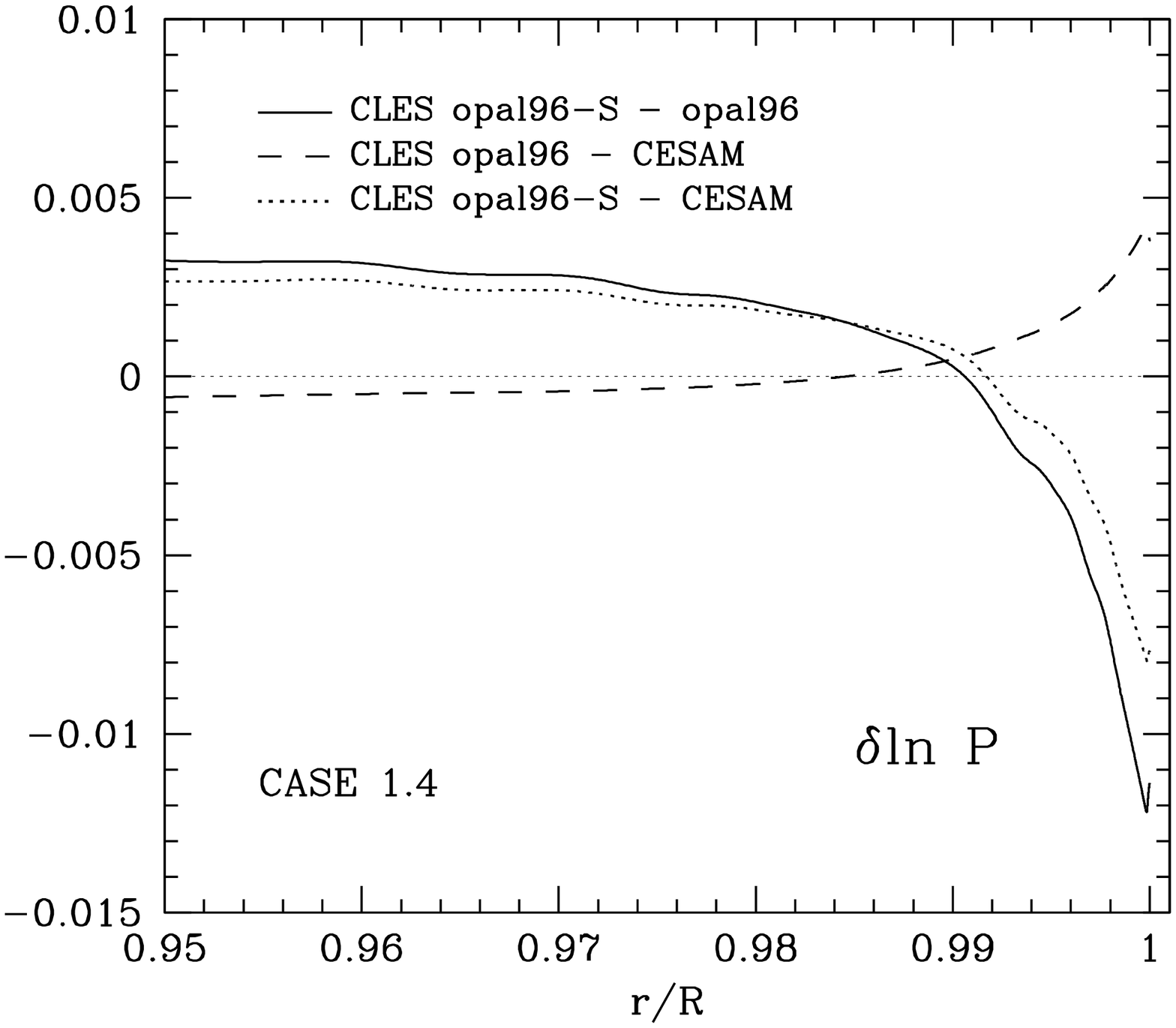}
\includegraphics{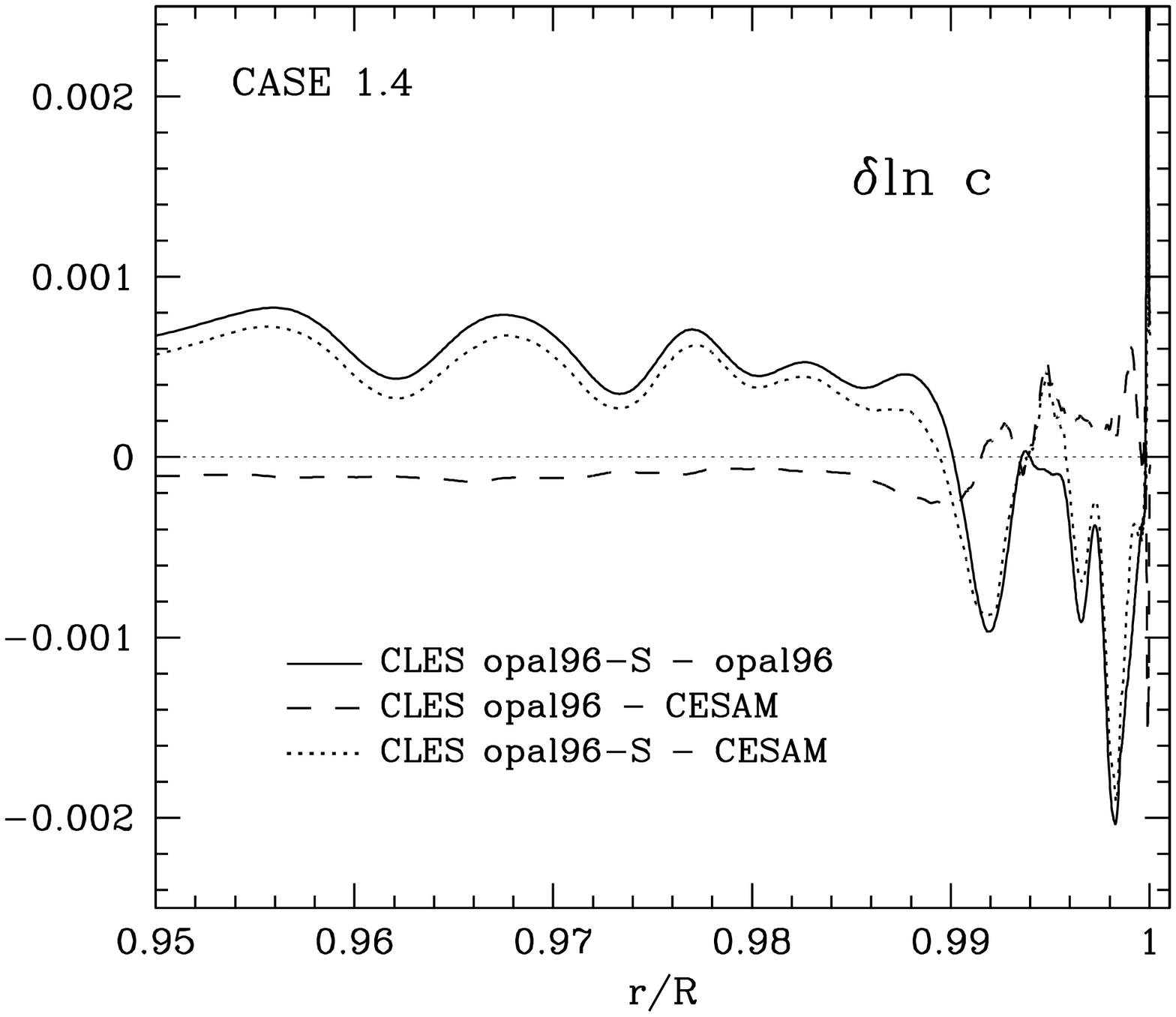} \includegraphics{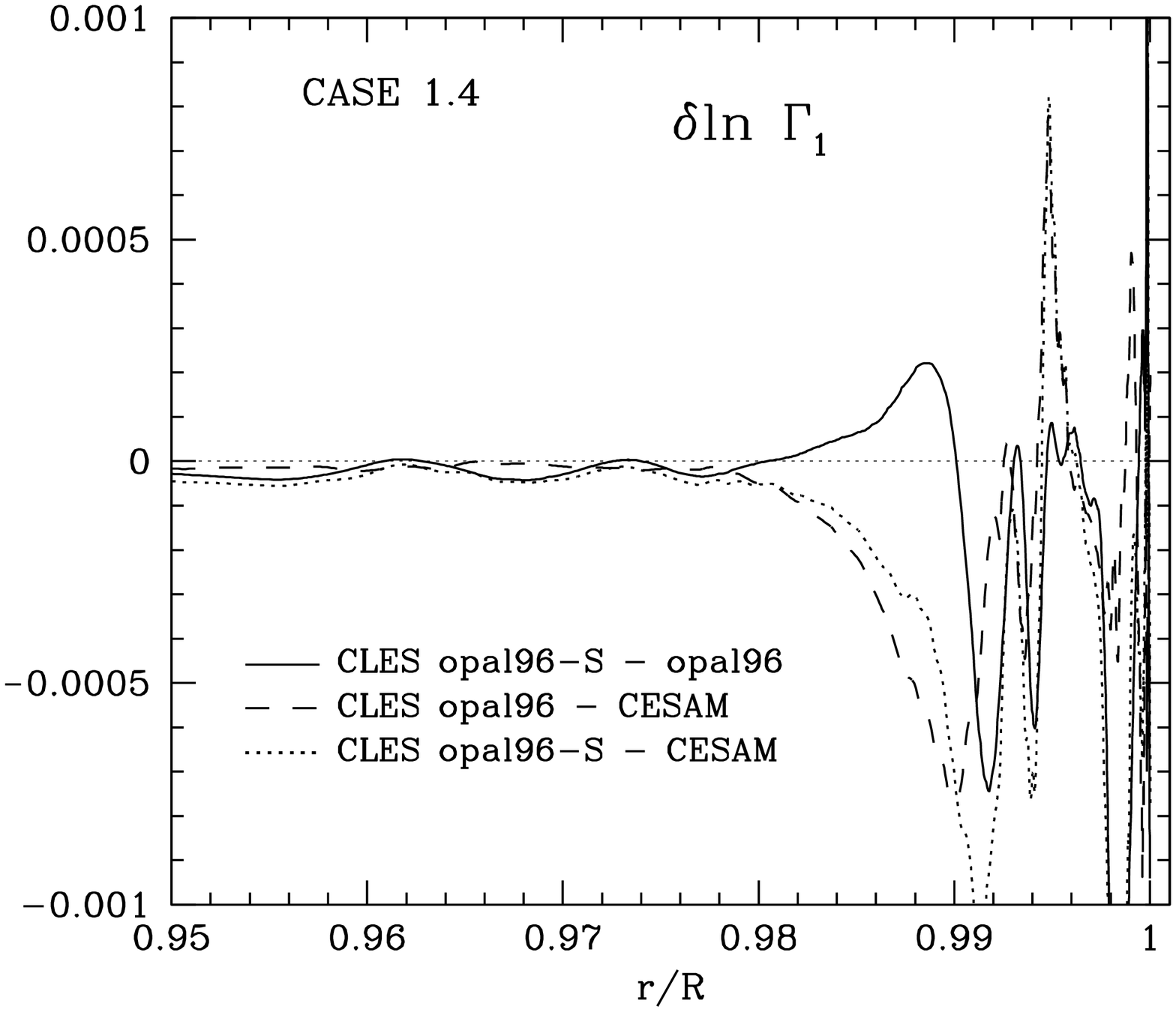}}
\resizebox{\hsize}{!}{\includegraphics{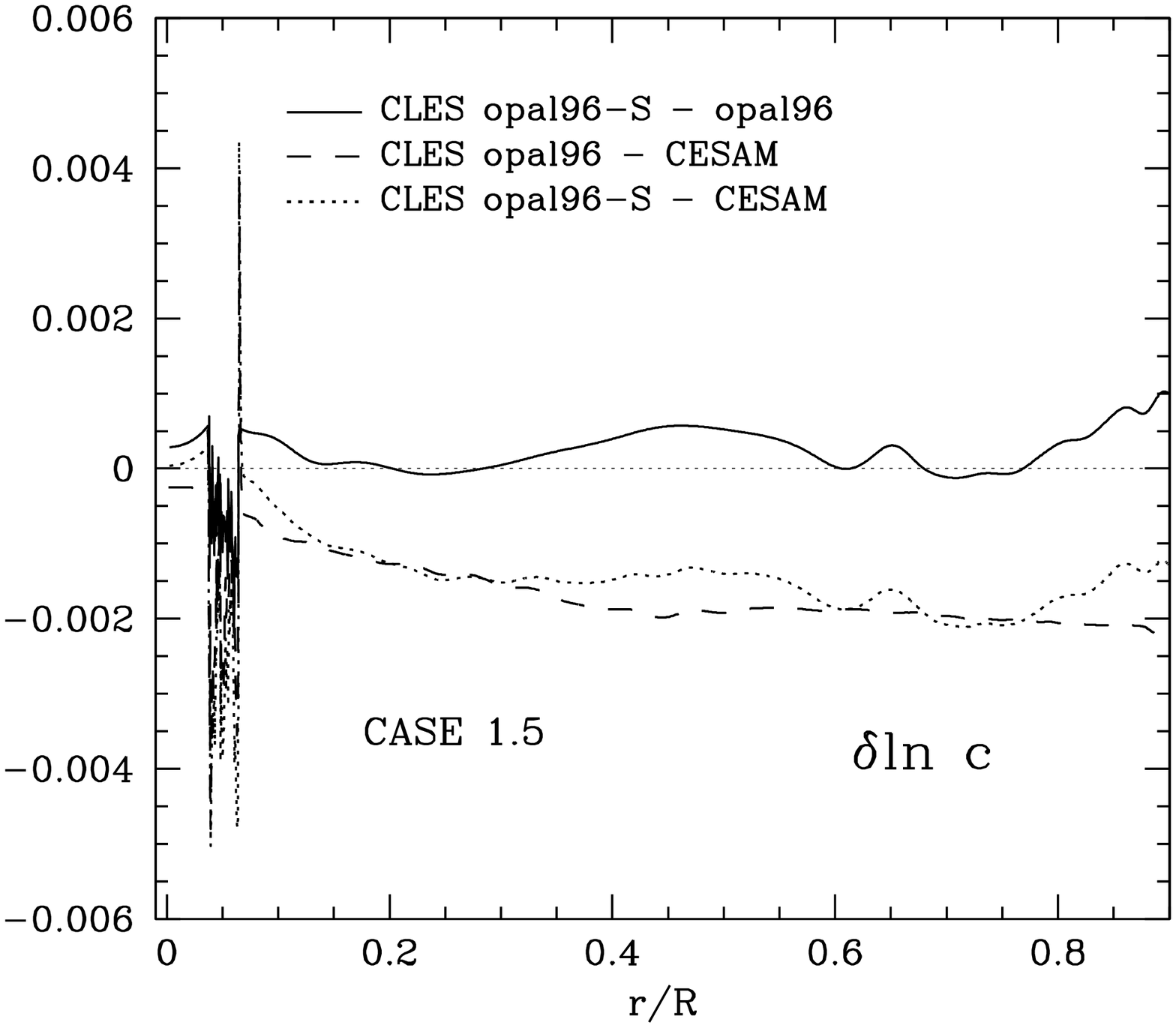}\includegraphics{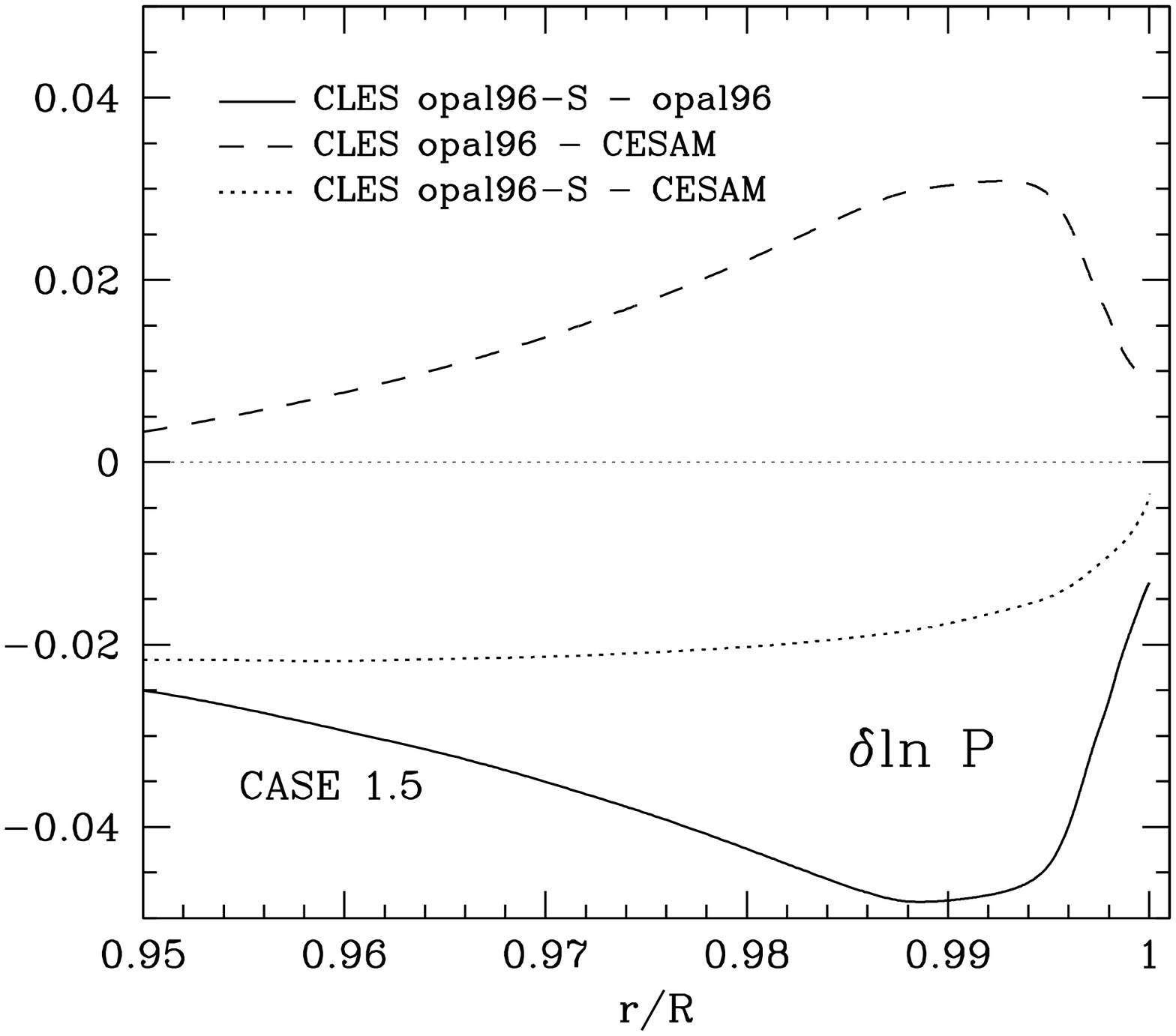}
\includegraphics{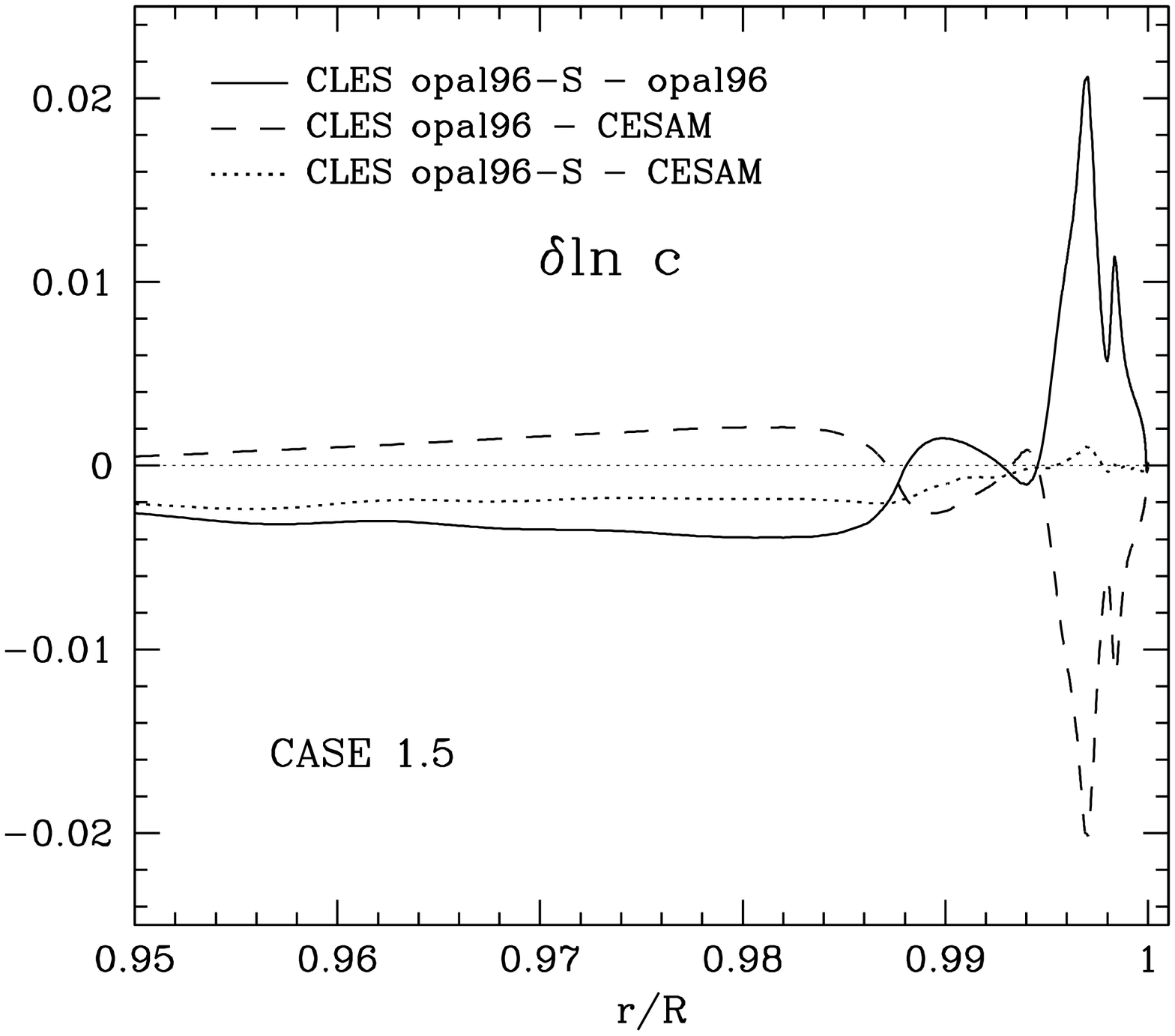} \includegraphics{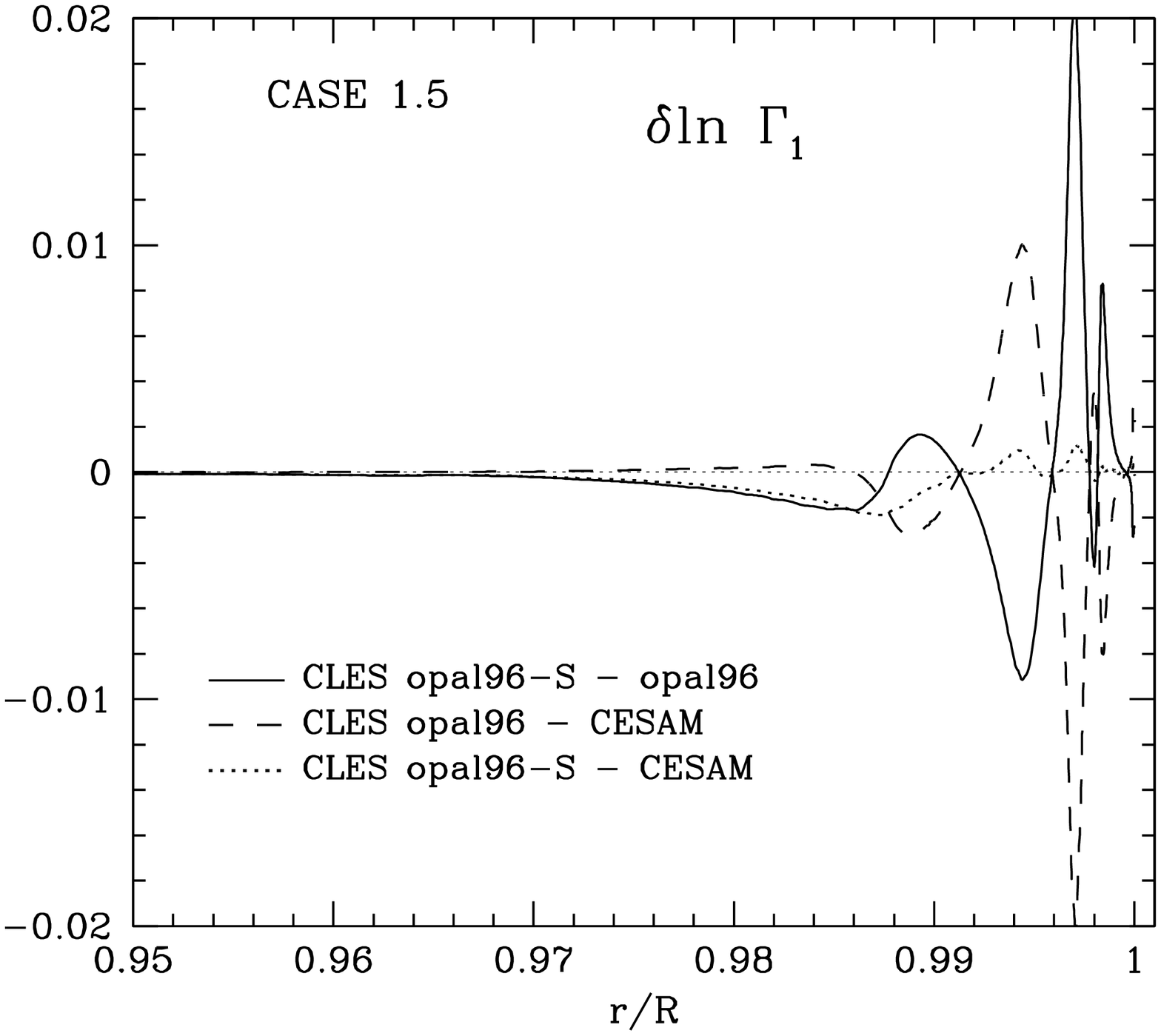}}
\resizebox{\hsize}{!}{\includegraphics{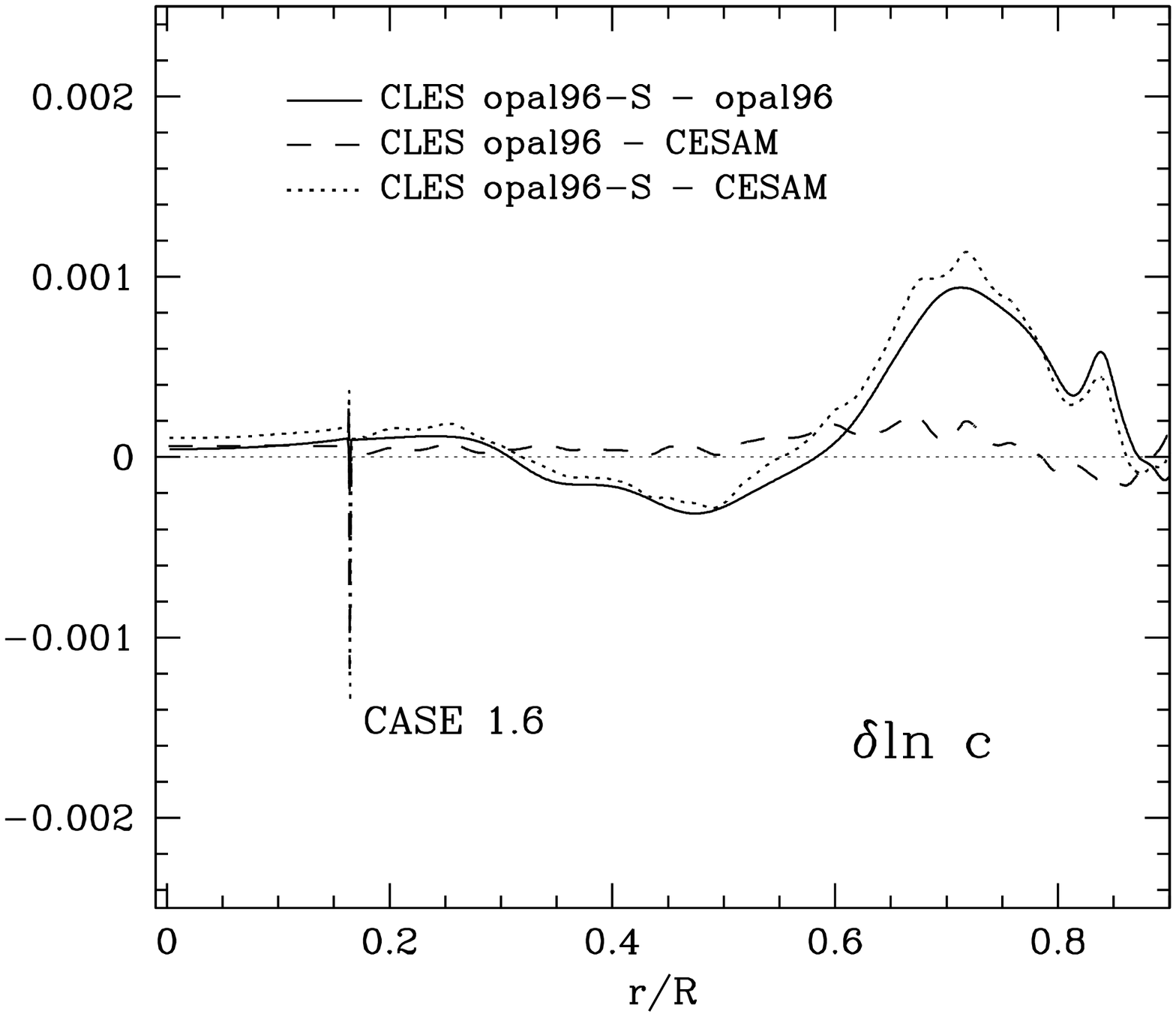}\includegraphics{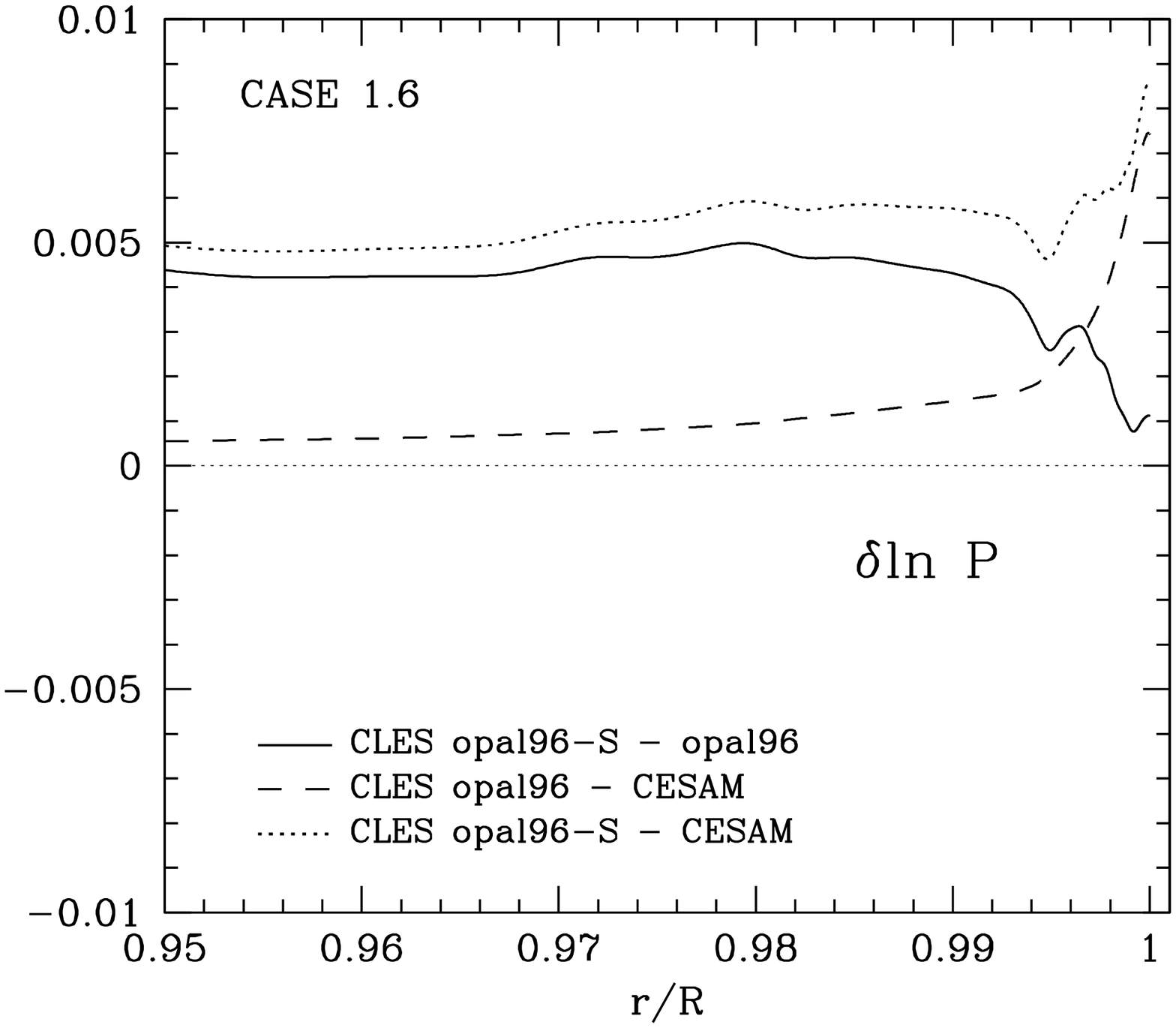}
\includegraphics{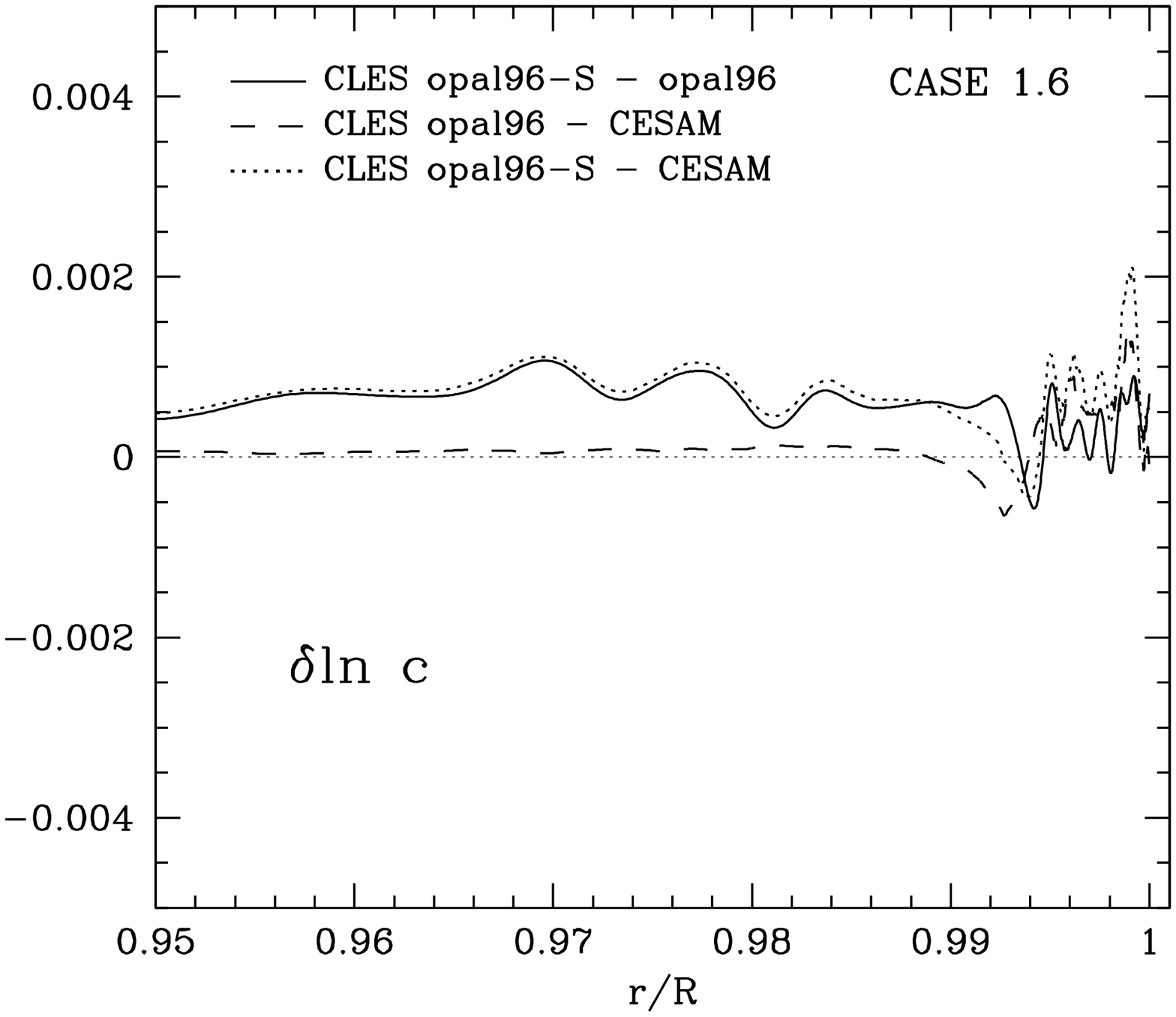} \includegraphics{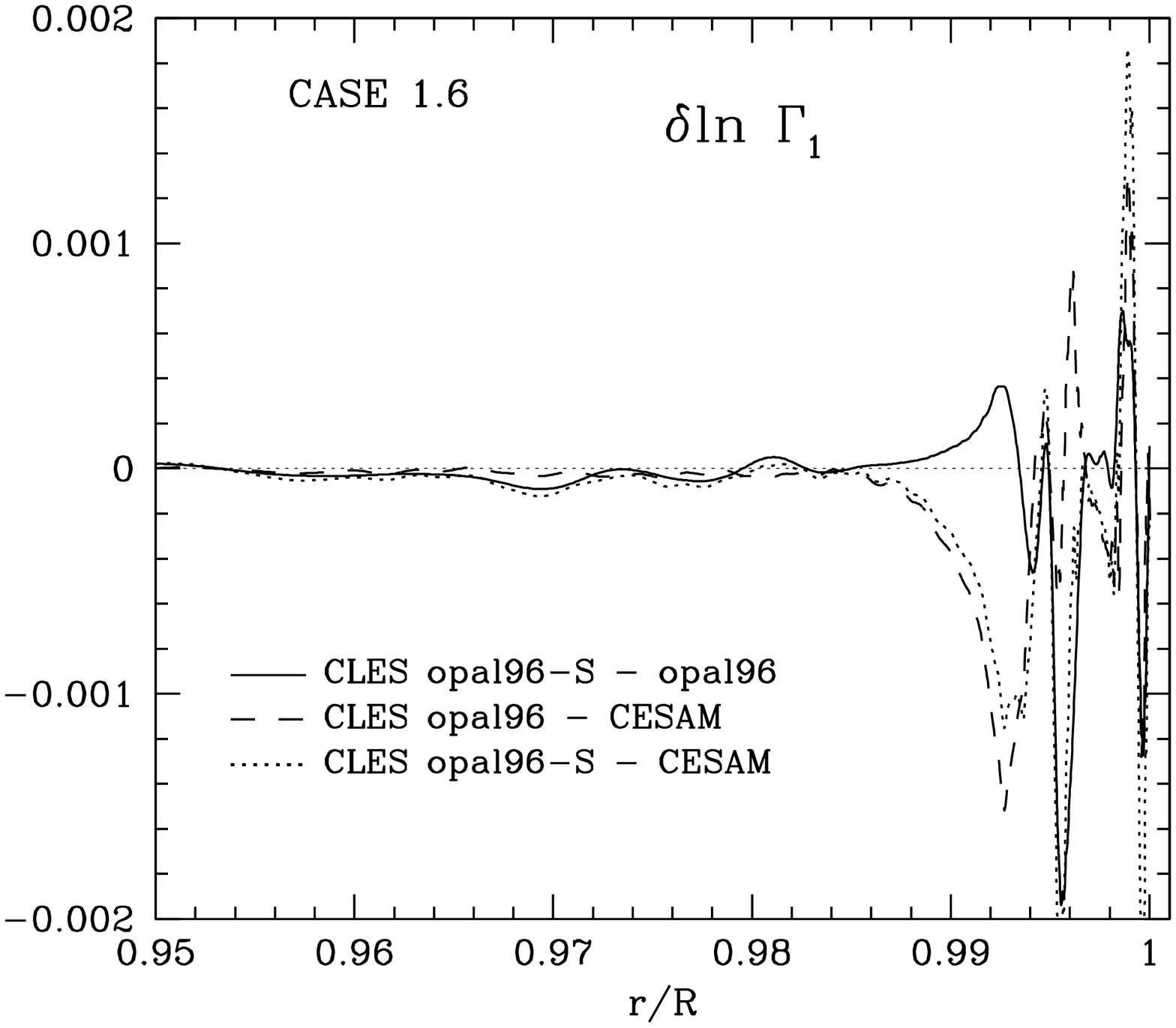}}
\resizebox{\hsize}{!}{\includegraphics{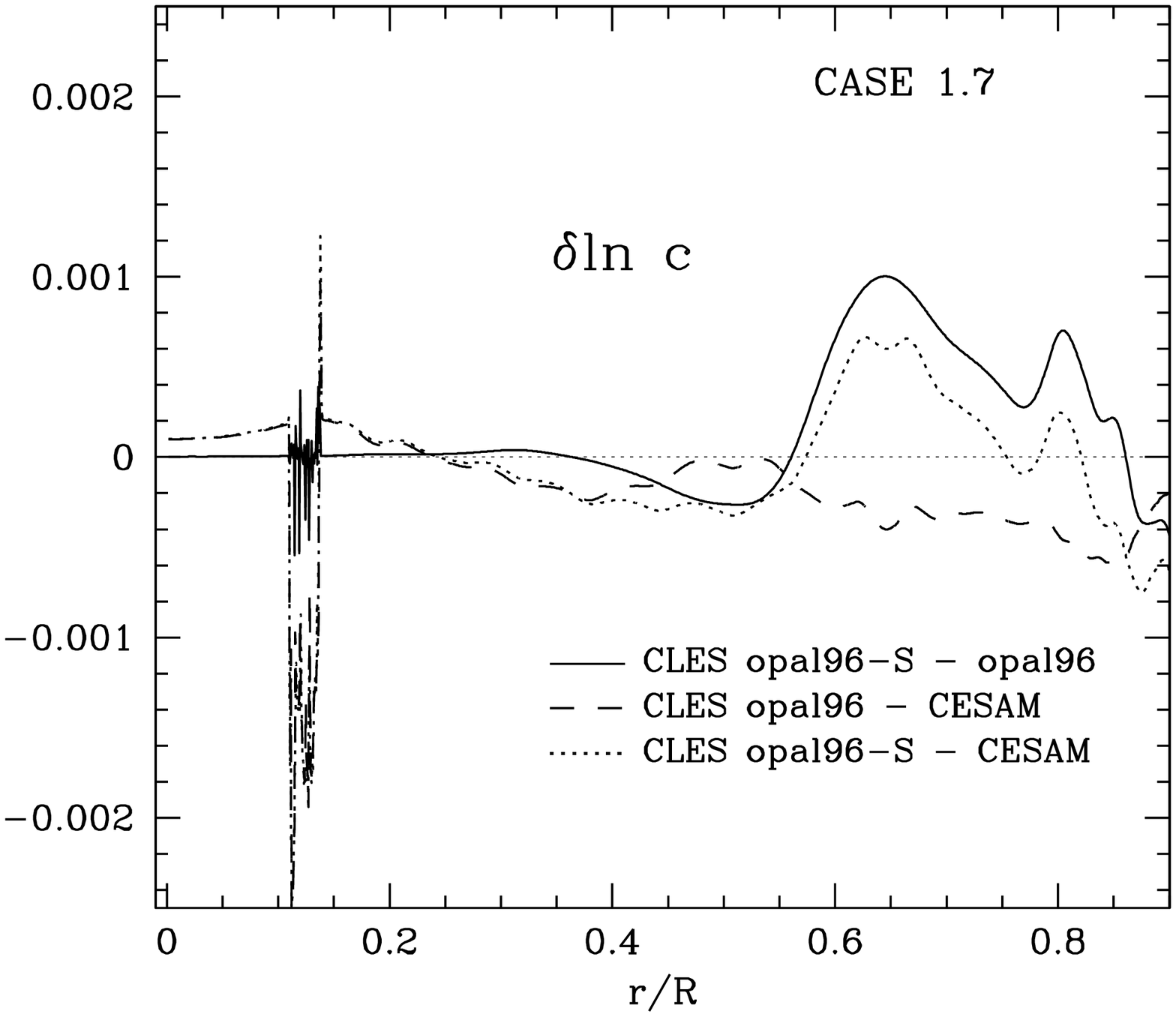}\includegraphics{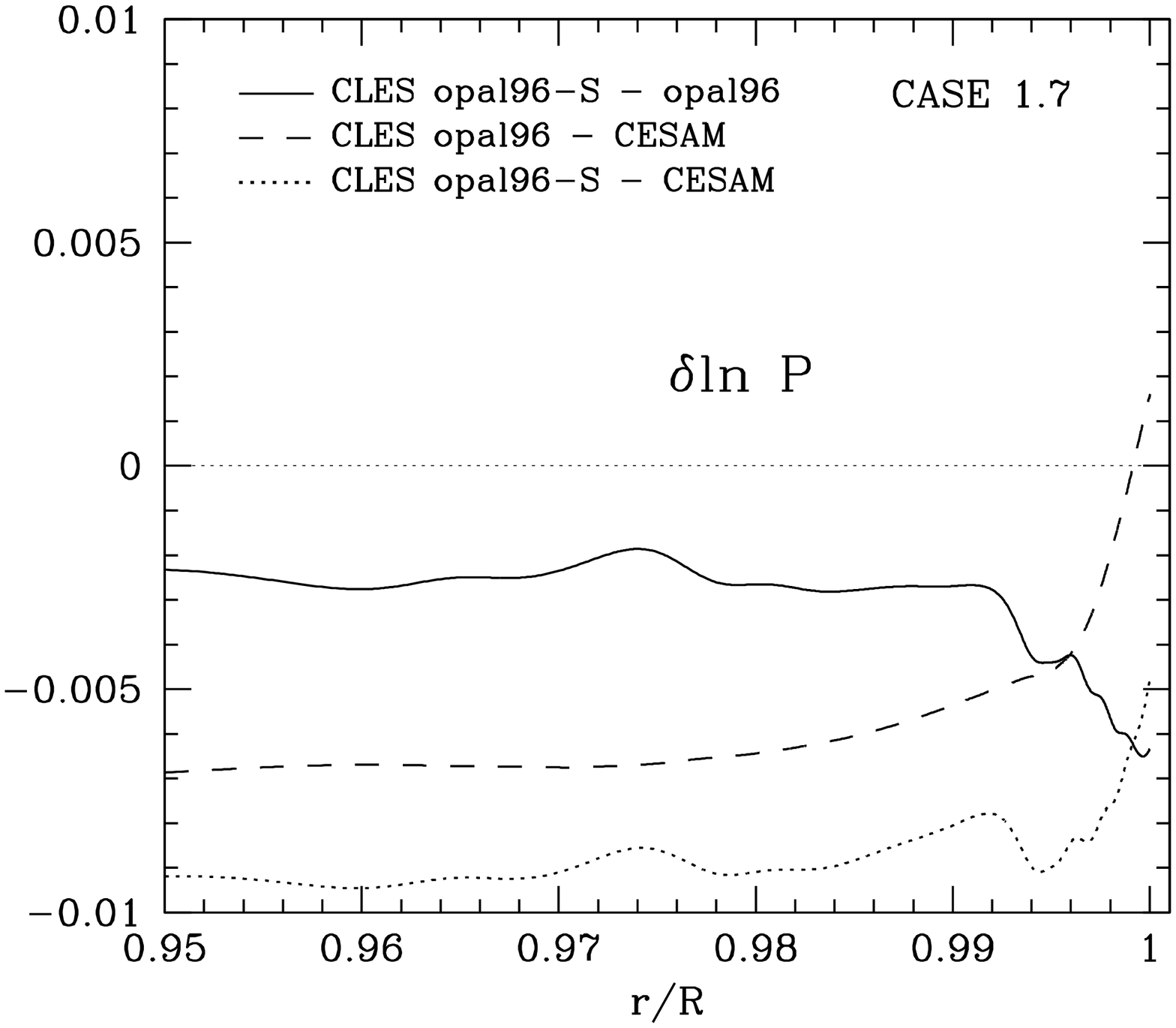}
\includegraphics{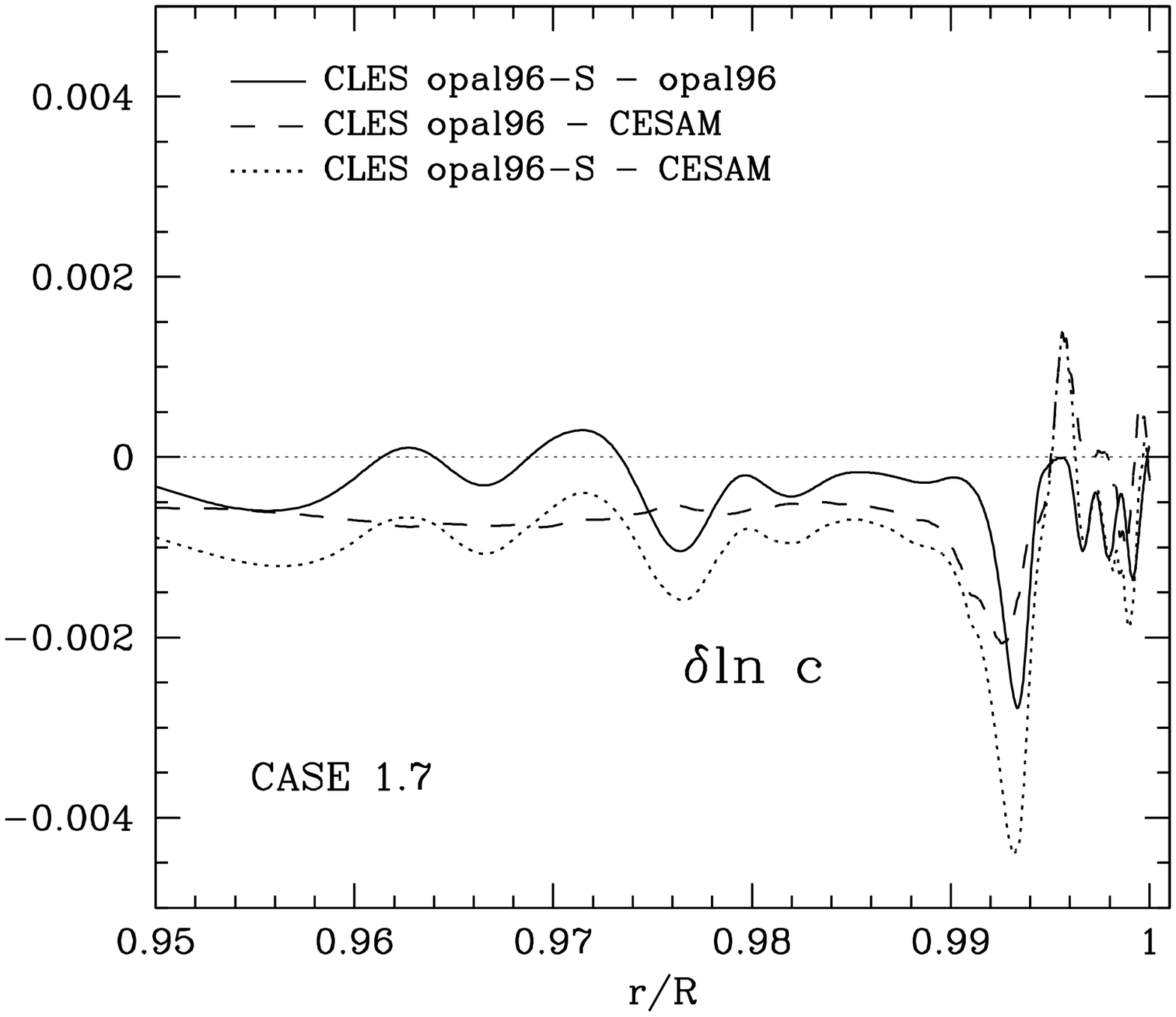} \includegraphics{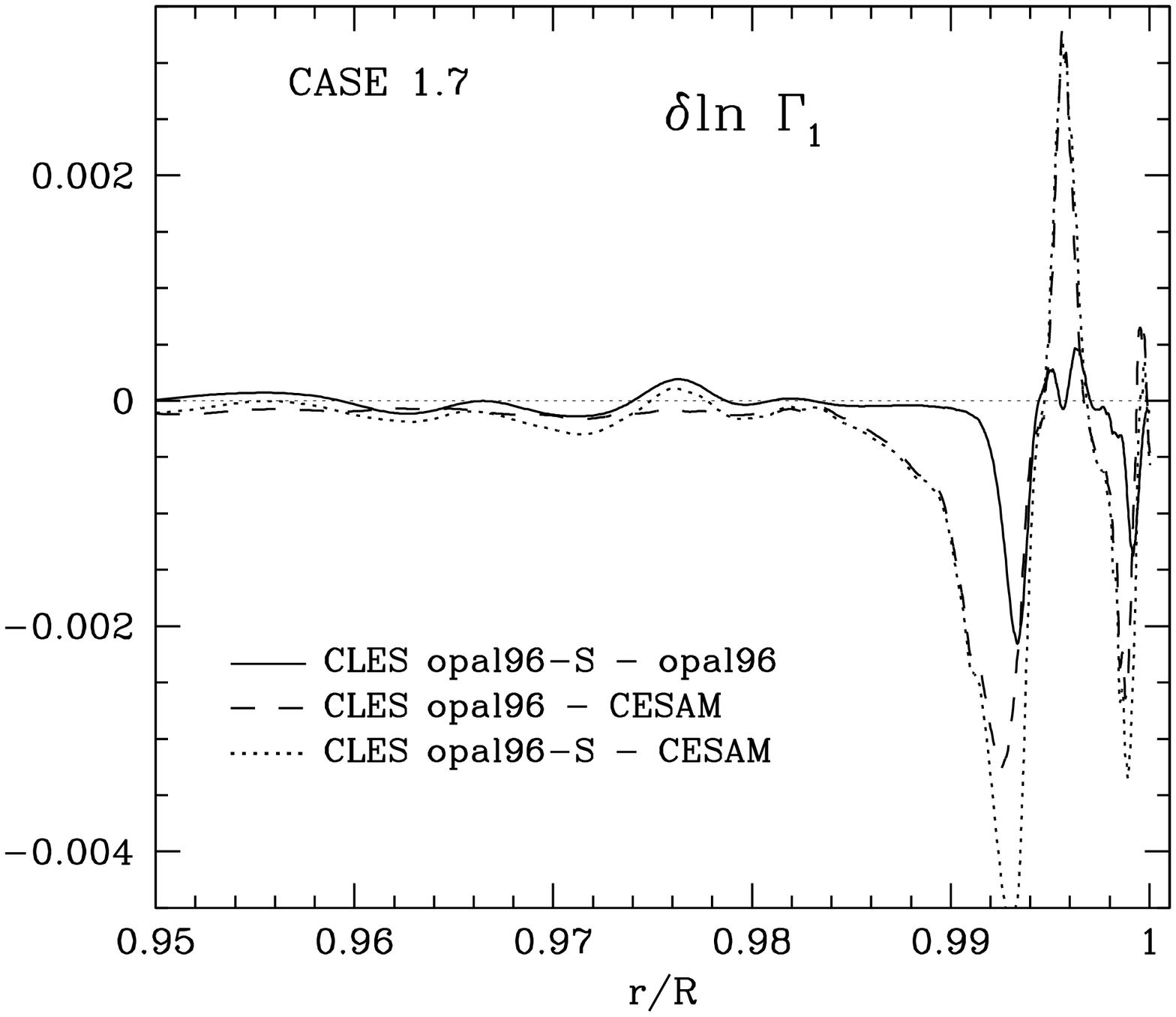}}
\vspace*{0.5cm}
\caption{As Fig.~\ref{fig:freqsopal}, for Cases~1.4, 1.5, 1.6 and 1.7.} 
\label{fig:dif-struc-opal2}
\end{figure*}

To analyze to which extent the differences reported in \cite{yl2-apss} come from the uncertainty in the opacity, we computed  for each \task~1 model the local differences in the physical variables at fixed relative mass and at fixed relative radius. To this purpose we use the so-called {\it diff-fgong.d} routine in the \adipls\ package\footnote{http://www.corot.pt/ntools}. In Figs.~\ref{fig:dif-struc-opal} and \ref{fig:dif-struc-opal2} we plot the logarithmic  differences of the sound speed, $c$, and pressure, $P$, for the stellar interior (two left panels) and of sound speed and the  adiabatic exponent $\Gamma_1$ in the external layers. In each panel there are three different curves that correspond to the comparisons labeled A, B, and C in previous section. So, if the differences shown in \citet{yl2-apss} come from the differences in the opacity tables, the solid and dotted lines should be close to each other. We call again the attention to the improvement got for the cases C1.4 and C1.6. 

\subsection{Effects on the frequencies}
Since the frequency of p-modes firstly depends  on the stellar radius, the improved  agreement between \cesam\ and \cles\ models that we obtained by changing from OPAL96-S to OPAL96 implies also a decrease in the frequency discrepancies. The values of $\Delta \nu$ ($\nu_{\rm CLES}-\nu_{\rm CESAM}$) change from 7.25 to 6.35~ $\mu$Hz (at 5300~$\mu$Hz) for the case C1.1; from 7 to 4.5~$\mu$Hz (at 4000~$\mu$Hz) for C1.2; from -5 to -4.5~$\mu$Hz at 1000~$\mu$Hz (and from -11 to -10 at 2300~$\mu$Hz) for C1.3; from 3.7 to -1~$\mu$Hz at 2200~$\mu$Hz for C1.4. The improvement is only of 0.3~$\mu$Hz for the case C1.5, and $\Delta\nu$ at 1200~$\mu$Hz is of the order of 7.5~$\mu$Hz. For C1.6 the initial $\Delta\nu\sim 3$~$\mu$Hz at 1500~$\mu$Hz drops to values lower than 0.05~$\mu$Hz, and for C1.7, $\Delta\nu$ change from 1.6~$\mu$Hz to 1.2~$\mu$Hz (at 600~$\mu$Hz).

\begin{figure*}[ht!]
\centering
\resizebox{\hsize}{!}{\includegraphics{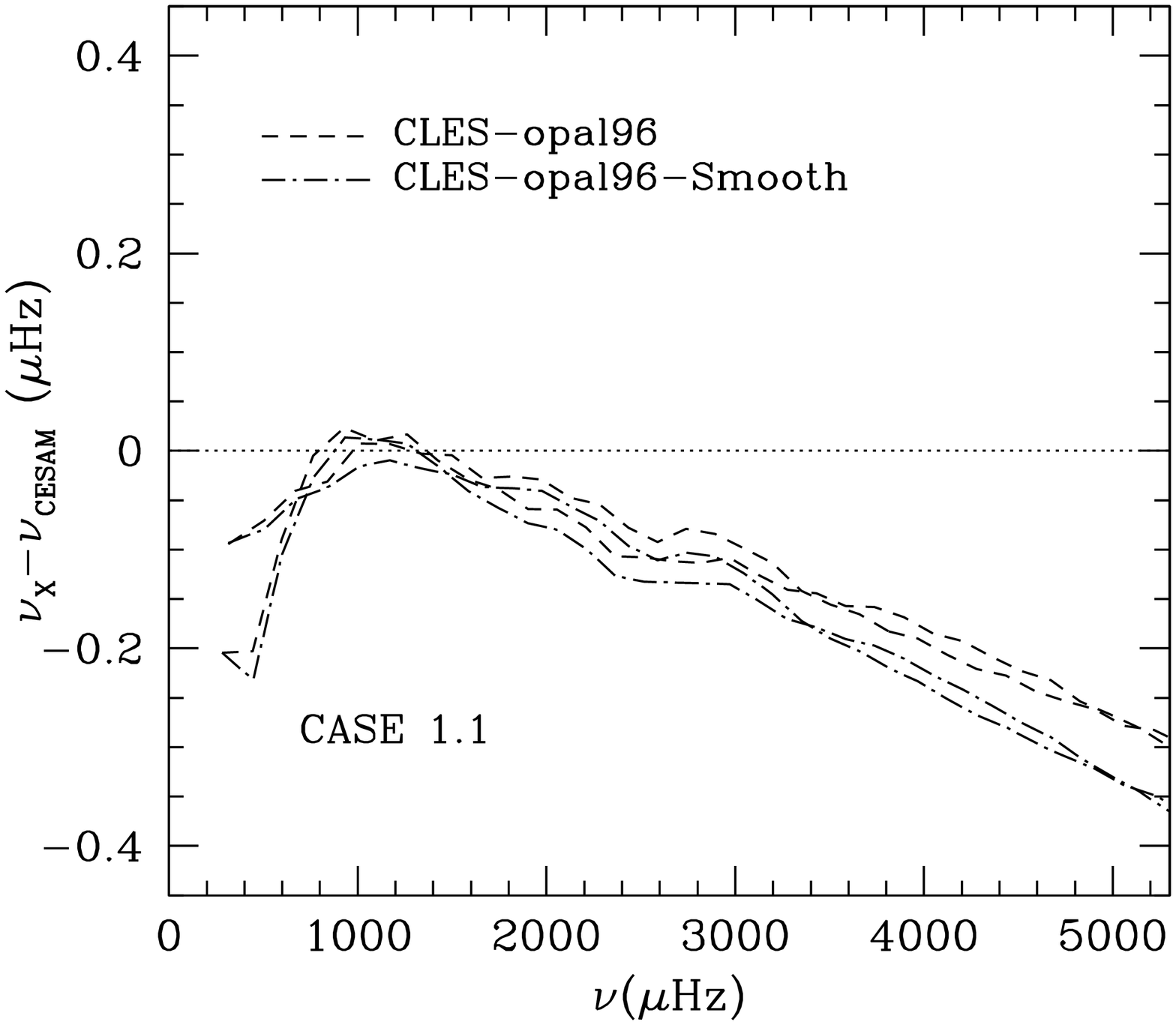}\includegraphics{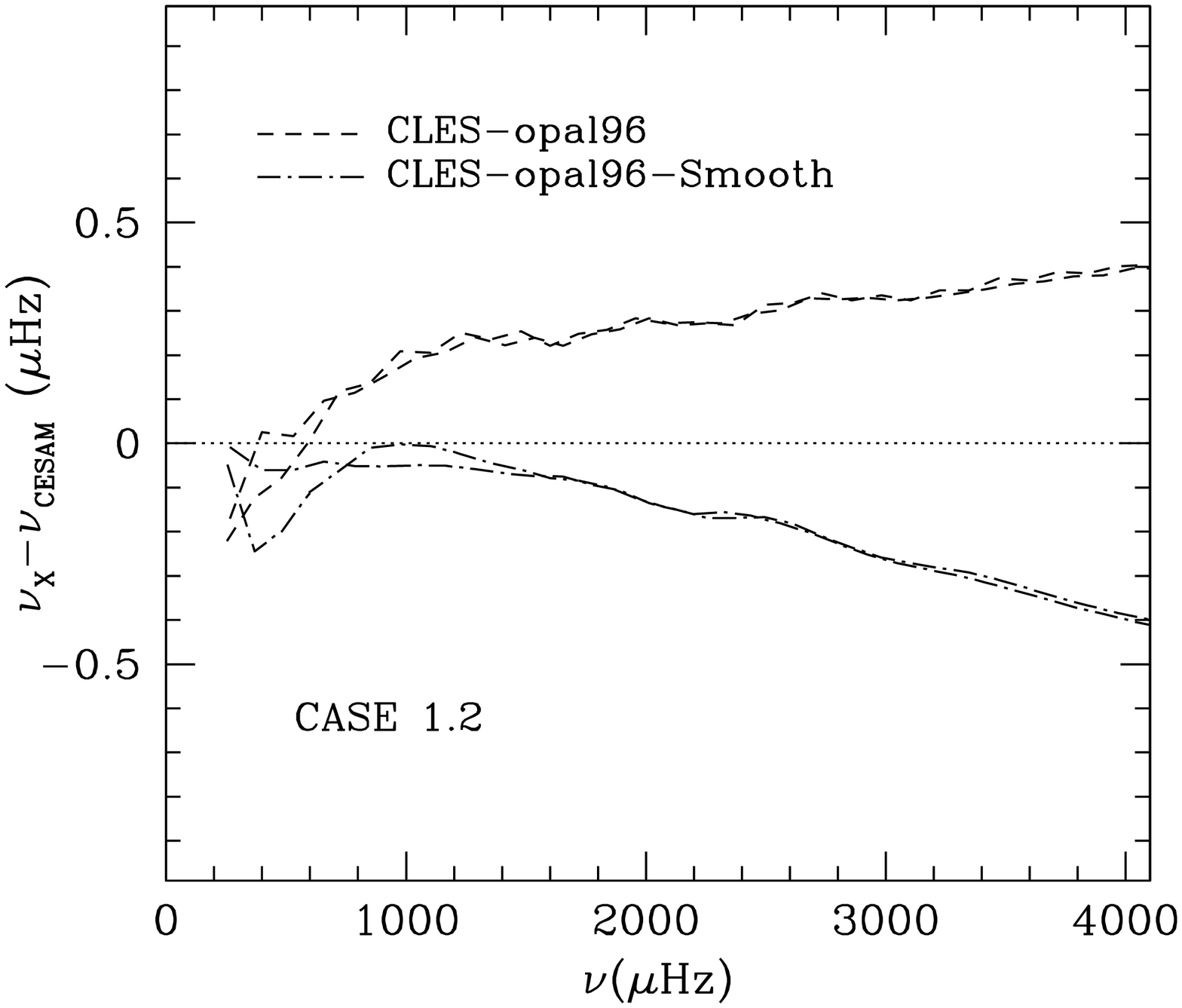}
\includegraphics{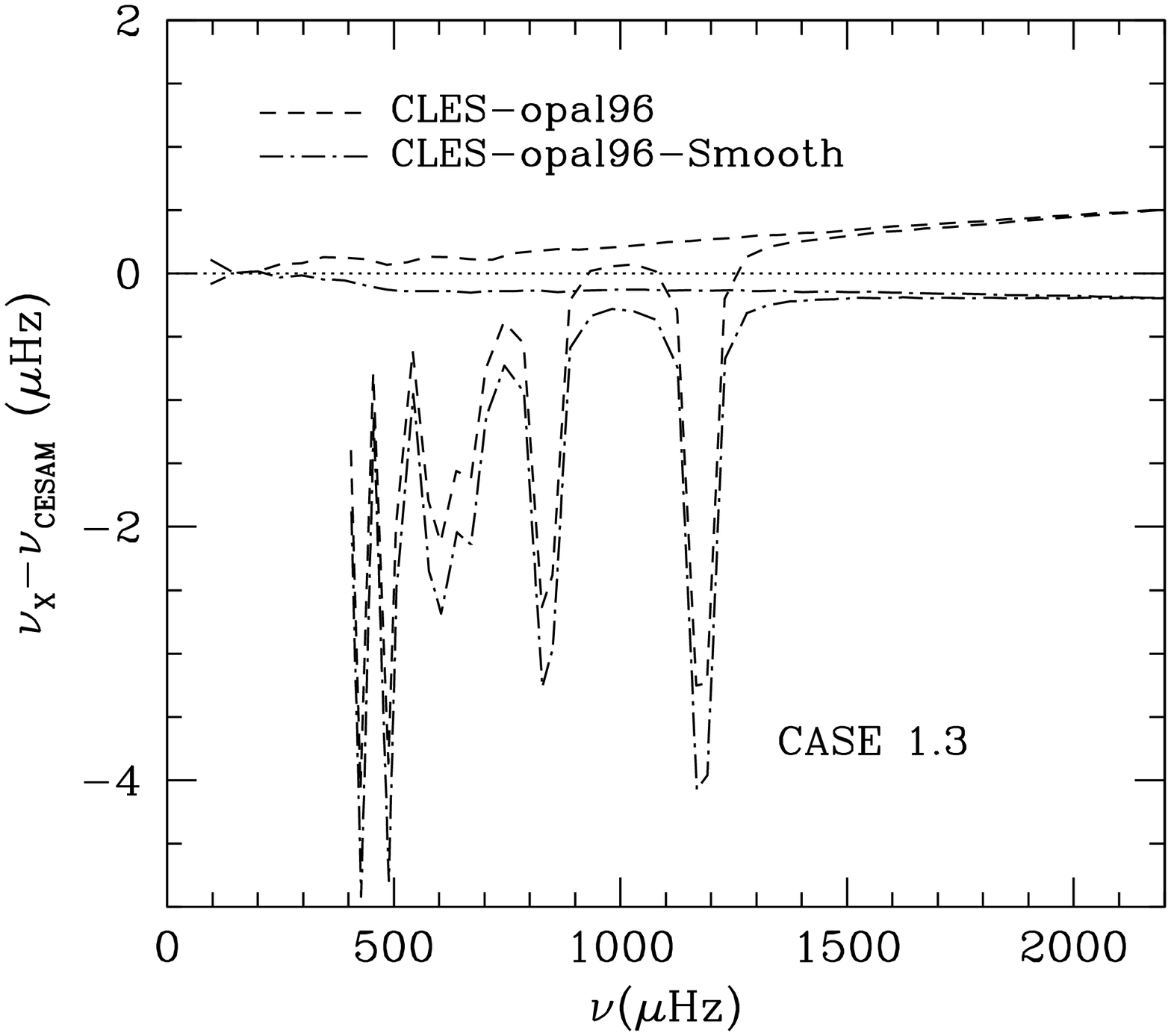} \includegraphics{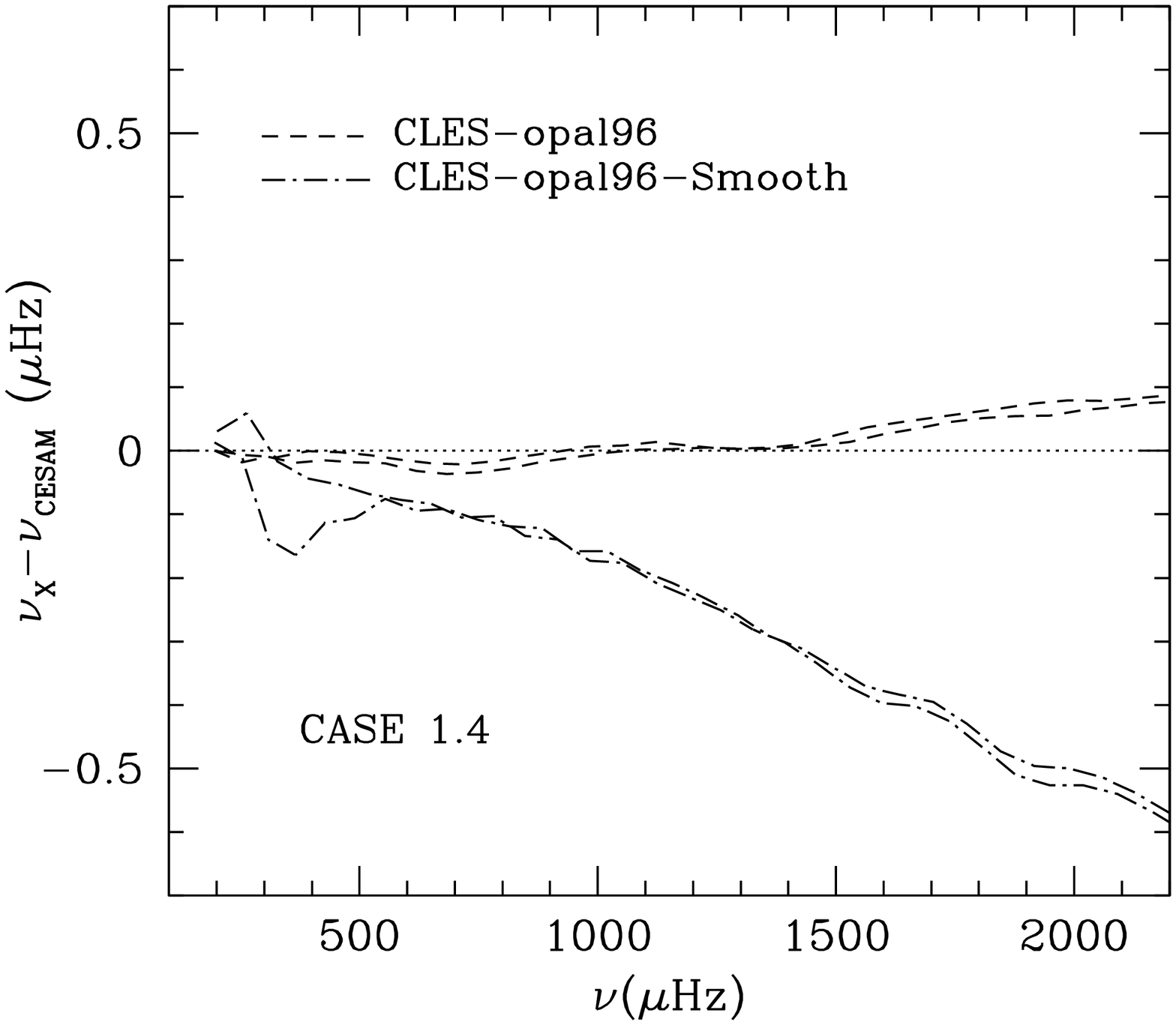}}
\resizebox{\hsize}{!}{\includegraphics{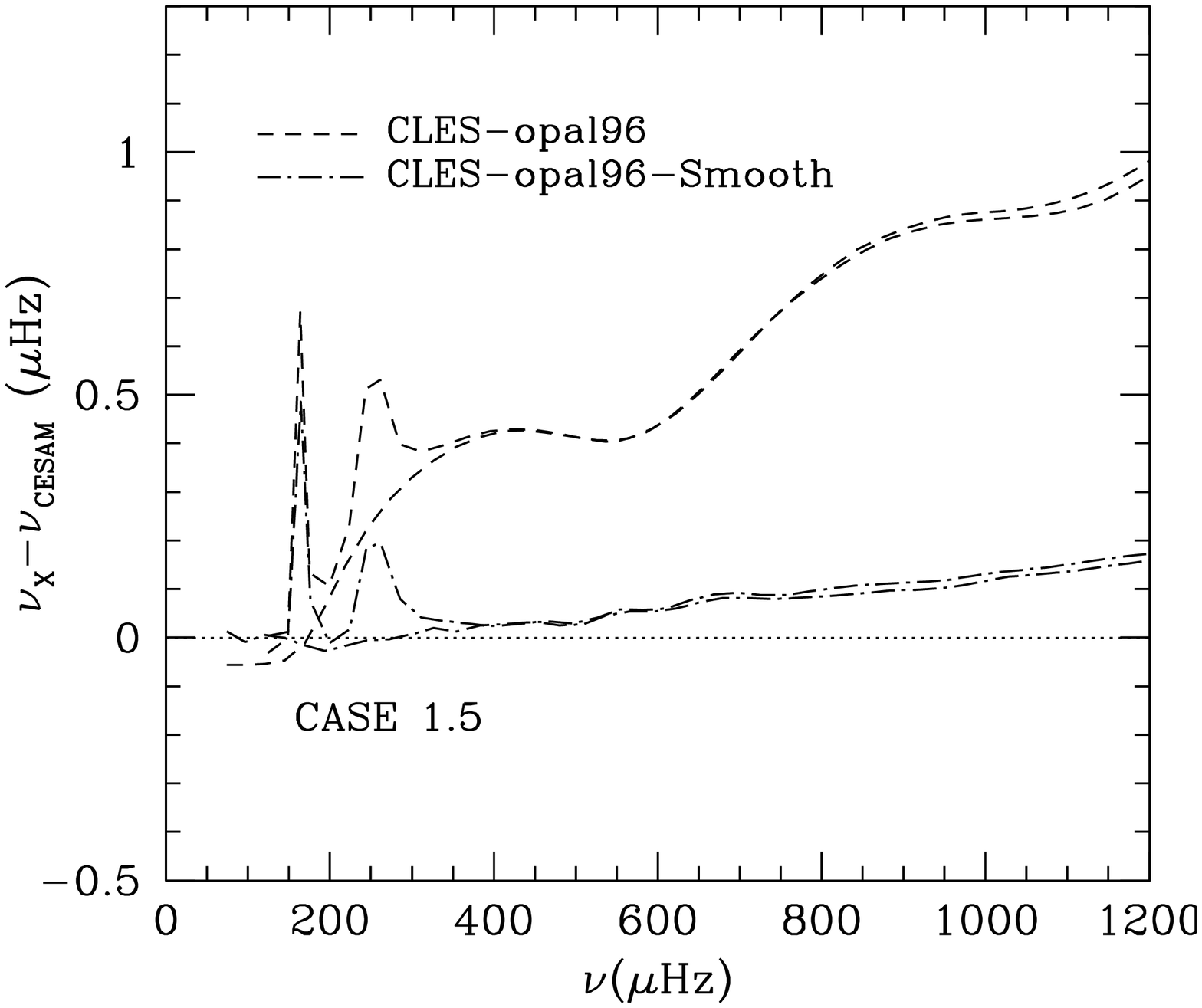}\includegraphics{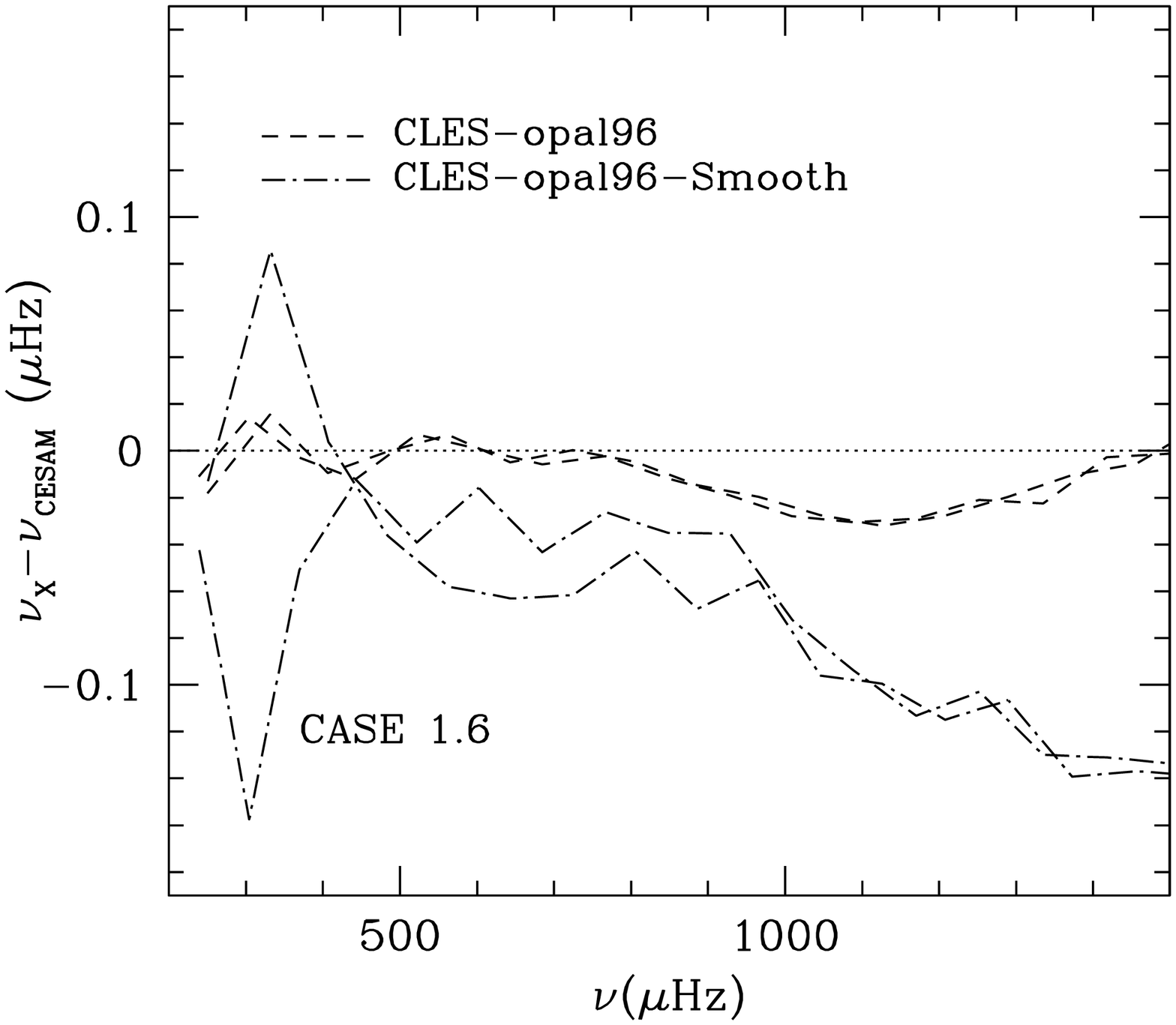}
\includegraphics{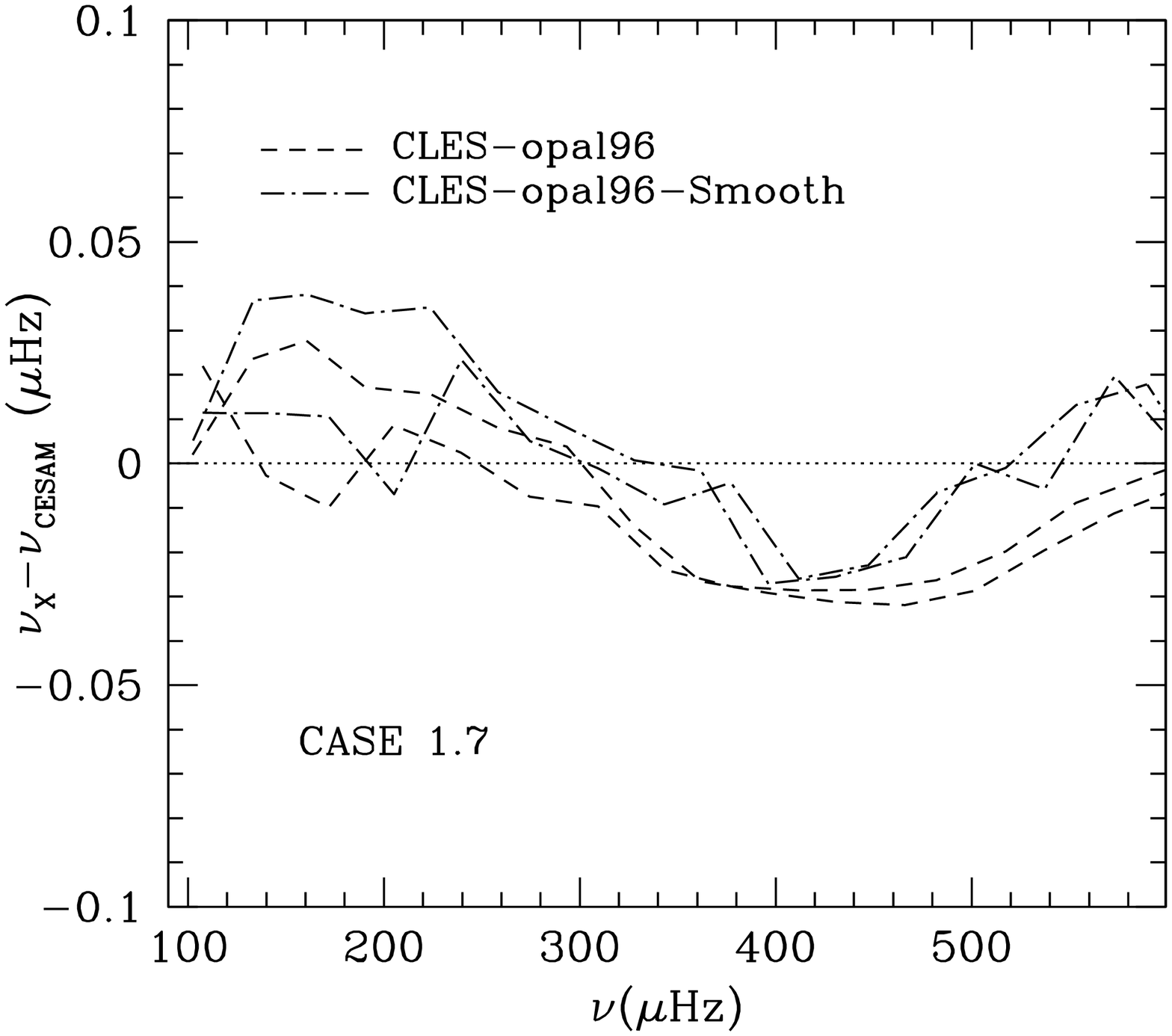}\includegraphics{./figs/nule.eps}}
\vspace*{0.5cm}
\caption{p-mode frequency differences between {\small TASK~1} models produced by \cles\ with the two kinds of opacity tables (dash-dotted lines: smoothed OPAL; and dashed lines: without smoothing) and \cesam. The latter is taken as reference, and the frequencies have been scaled  to remove the effect of different stellar radii. For each model we plot two curves corresponding to modes with degrees $\ell=0$ and $\ell=1$.} 
\label{fig:freqsopal}
\end{figure*}

By comparing frequencies that have been scaled to the same radius  we remove the effect of $\Delta R$ and make appear the differences due to discrepancies in the stellar structure. This was done in  Fig.~\ref{fig:freqsopal} where we plot the frequency differences for $\ell=0$ and 1 and for the cases considered in Task~1. There are two curves in each panel, one corresponding to the difference $\nu_{\rm CLES-OPAL96-S}-\nu_{\rm CESAM}$ \citep[that is, that appearing also in][]{yl2-apss}, and the second one corresponding to  $\nu_{\rm CLES-OPAL96}-\nu_{\rm CESAM}$.

The role of opacity on the oscillation frequencies is a intricate problem since the variations of $\kappa$ lead to changes of the temperature structure in the star, and therefore also of the value of $\Gamma_1$. As  can be seen in Fig.~\ref{fig:dif-struc-opal2} for the case 1.5, the differences in the outer layers might even increase for the model computed with similar opacity tables (OPAL96), and as consequence, the frequency differences scaled to the same radius (Fig.~\ref{fig:freqsopal}, lower-left panel) show a discrepancy even larger than with OPAL96-S. The oscillatory behavior is produced by the peak
in $\delta \ln c$ at $r/R\sim0.997$. A comparison between \cles-{\sc OPAL96} and \cles-{\sc OPAL96-S} clearly  shows the same oscillatory behavior, but the absolute frequency difference is only $\sim 0.3 \mu$Hz  (both models have similar radius). In the same way,  \cesam\ and \cles\  2~\msol\ models at $X_{\rm c}=0.50$ (whose radius differ by less than $3\times10^{-4}R_\ast$ once \cles\ adopts   OPAL96 tables) show  frequency differences of the order of 0.4$~\mu$Hz at 1200$~\mu$Hz. After normalizing to the same radius an oscillatory component (amplitude 0.01$~\mu$Hz) remains in the $\Delta \nu$ because of the differences in  $\Gamma_1$ in the outer layers.

\section{ Nuclear reaction rates} 
\label{sec:nuc}
 We used the basic pp and CNO reaction networks up to the $^{17}$O(p,$\alpha$)$^{14}$N reaction. In the present models the \cesam\ code takes $^7$Li, $^7$Be and $^2$H at equilibrium while \cles\ follows  entirely the combustion of $^7$Li and $^2$H. The nuclear reaction rates are computed using the analytical formulae  provided by the NACRE compilation \citep{1999NuPhA.656....3A}. \cesam\ uses a pre-computed table\footnote {By using pre-computed  tables the numerical coherence of derivatives required in the \cesam\ numerical scheme are guaranteed.}   while \cles\ uses directly the analytical expressions. Comparing the nuclear reaction rates for a 2~\msol\ stellar structure and a given chemical composition,  we found that the relative differences are of the order of  $3\times10^{-4}$--$2\times10^{-3}$ (except for nuclear reactions involving $^7$Li) if screening factors  are included, and of the order or $10^{-8}$ if not. In both codes weak screening is assumed under  \citet{1954AuJPh...7..373S}'s formulation. The screening factor is written   $f=\exp(A z_1 z_2 \sqrt{\frac{\rho \xi}{{T}^3}})$ where $z_1$ and $z_2$ are the charges of the interacting nuclei.   \cesam\ uses the expression 4-221 of \citet{1968psen.book.....C} where $A=1.88\ 10^8$, $\xi=\sum\limits_{i} z_i(1+z_i)x_i$, and  $x_i$ is the abundance per mole of element $i$. The standard version of  \cles\ code takes  $A=1.879\ 10^8$ and $\xi=\sum\limits_{i=1}^{4} z_i(1+z_i)x_i+ Z(1+Z)x(Z)$  where $x(Z)$ is the abundance of an ``average'' element containing all the elements different from hydrogen and helium, and $Z$ is the average charge of this element. This approximate estimation of $\xi$ has been changed in \cles\ by assuming full ionization and taking  the contribution from  each mixture element  into account. With this new prescription the differences between \cesam\ and \cles\ nuclear reaction rates are still of the same order, but with \cesam\ values larger than \cles\ ones at variance with what  was found with the standard \cles\ formulation.

All the Task~1 and Task~3 \cles-models were computed with the updated version.

\section {Atmosphere}

\label{sec:atm}
\begin{figure}[hb!]
\centering
\includegraphics[scale=0.35]{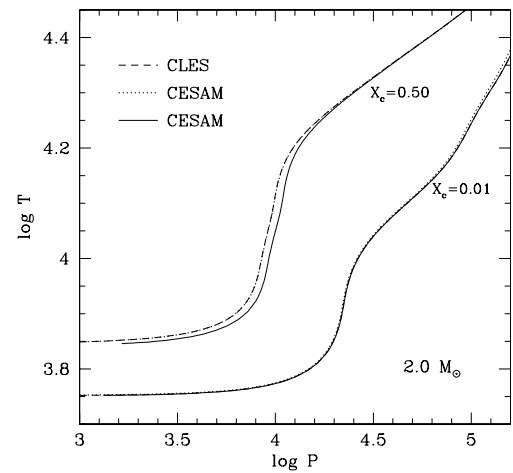}
\caption{Stellar structure of the external layers of 2~\msol\ models at two different evolutionary stages ($X_c=0.50$ and 0.01). Dashed lines: \cles\ models. Dotted lines: \cesam\ models with boundary conditions given by $P(\tau_{\rm min})$. Solid lines: \cesam\ models with boundary conditions given by $\rho(\tau_{\rm min})$ and default values (see the text).}

\label{fig:atmos}
\end{figure}

 Eddington's grey $T(\tau)$ law is used for the atmosphere calculation: $T=T_{\rm eff}[\frac{3}{4}(\tau+\frac{2}{3})]^\frac{1}{4}$ where $\tau$ is the optical depth.  \cesam\ integrates the hydrostatic equation in the atmosphere starting at the optical depth $\tau=\tau_{\rm min}$ ($\tau_{\rm min}=10^{-4}$ for solar like models) and makes the connection with the envelope at $\tau=10$ where the continuity of the variables and of their first derivatives are assured. The radius of the star is taken to be the bolometric radius, i.e. the radius at the level where the local temperature equals the effective temperature ($\tau=2/3$ for the Eddington's law). 

In the stellar structure integration \cles\ gets the  external boundary conditions   ( the values of density and temperature at a given optical depth $\tau$) by interpolating  a pre-computed table, and  the stellar radius is defined as the level where $T=T_{\rm eff}$.  The Eddington atmosphere table, which provides $\rho$ and $T$ at $\tau=2/3$ (therefore at $R=R_{\ast}$), was built  by integrating the hydrostatic equilibrium  equation in the atmosphere starting at an optical depth  that can vary between $10^{-4}$ and  $10^{-2}$. The atmosphere structure for a given model is computed afterwards, by integrating the same equations for the corresponding  values of $T_{\rm eff}$, $\log g$ and chemical composition.




{


\begin{figure*}[ht!]
\centering
\resizebox{\hsize}{!}{\includegraphics{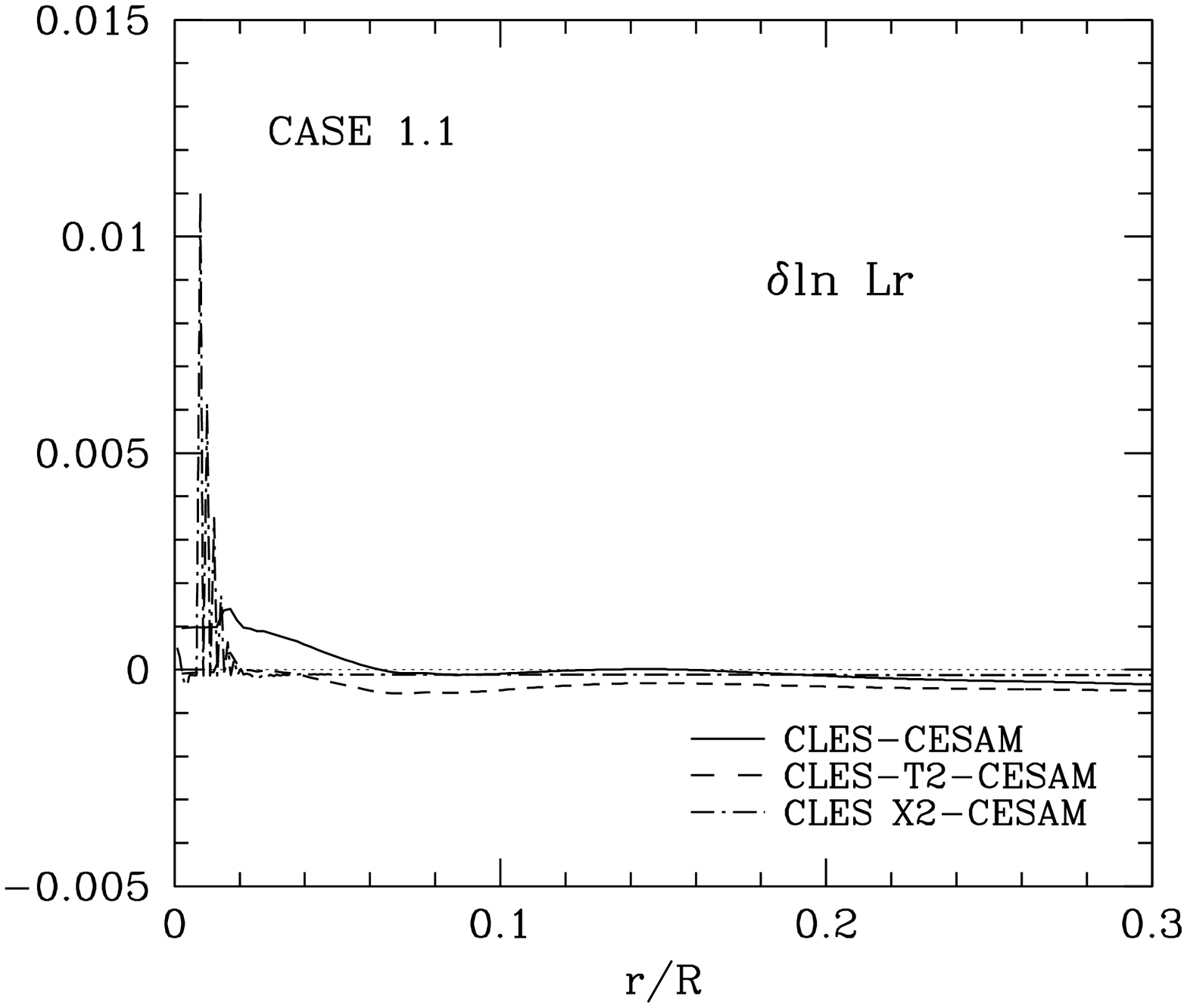}\includegraphics{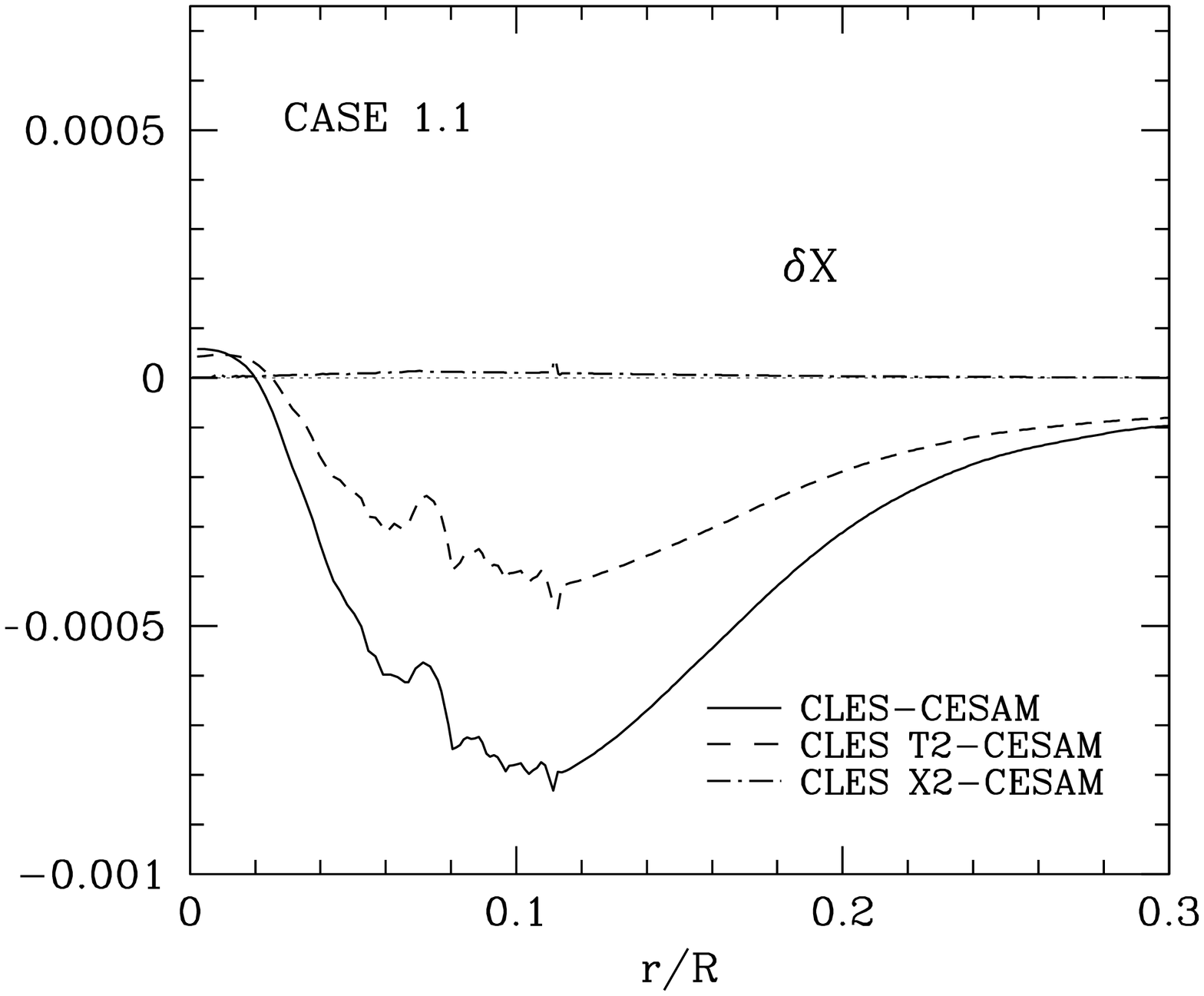}
\includegraphics{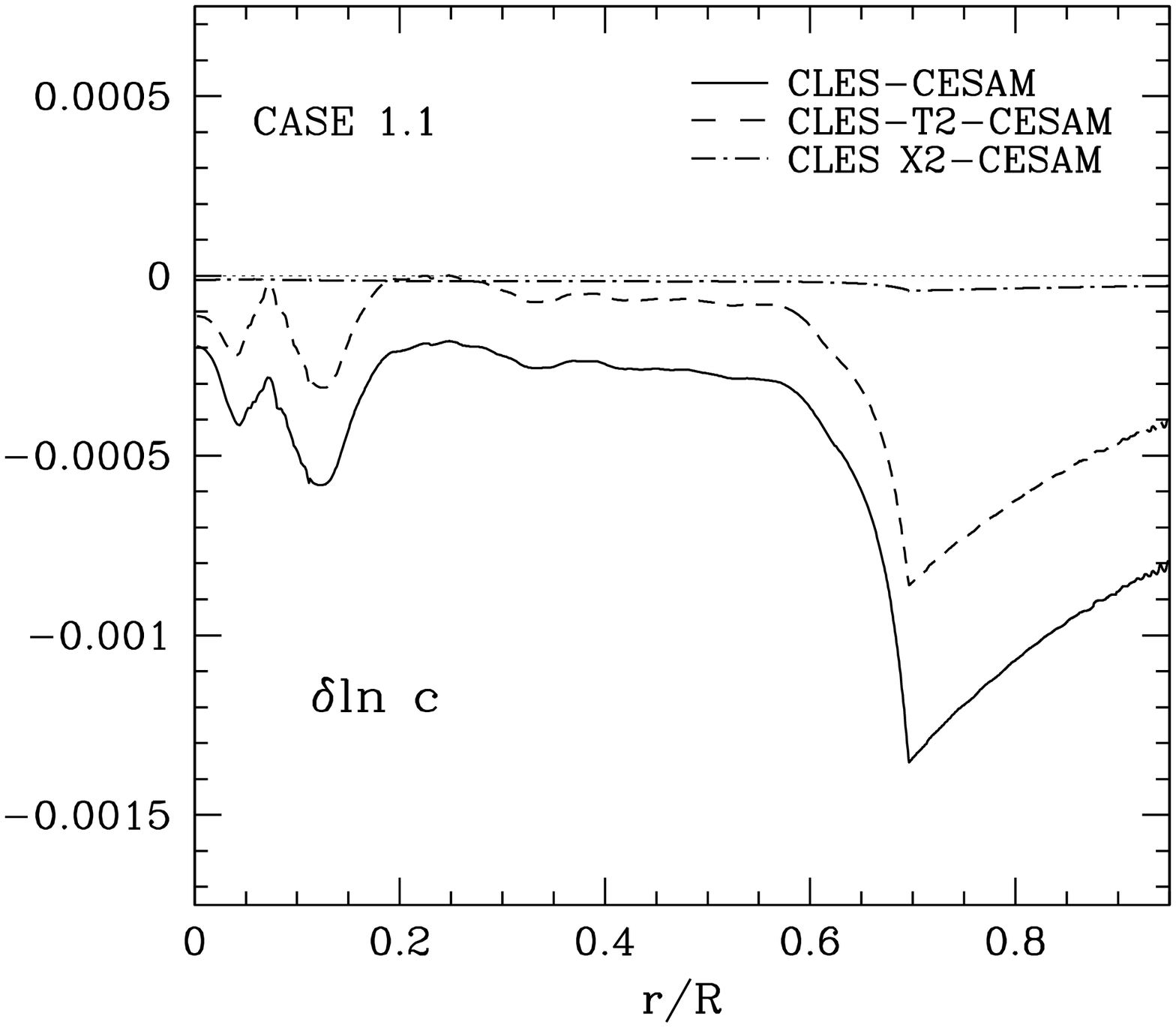} \includegraphics{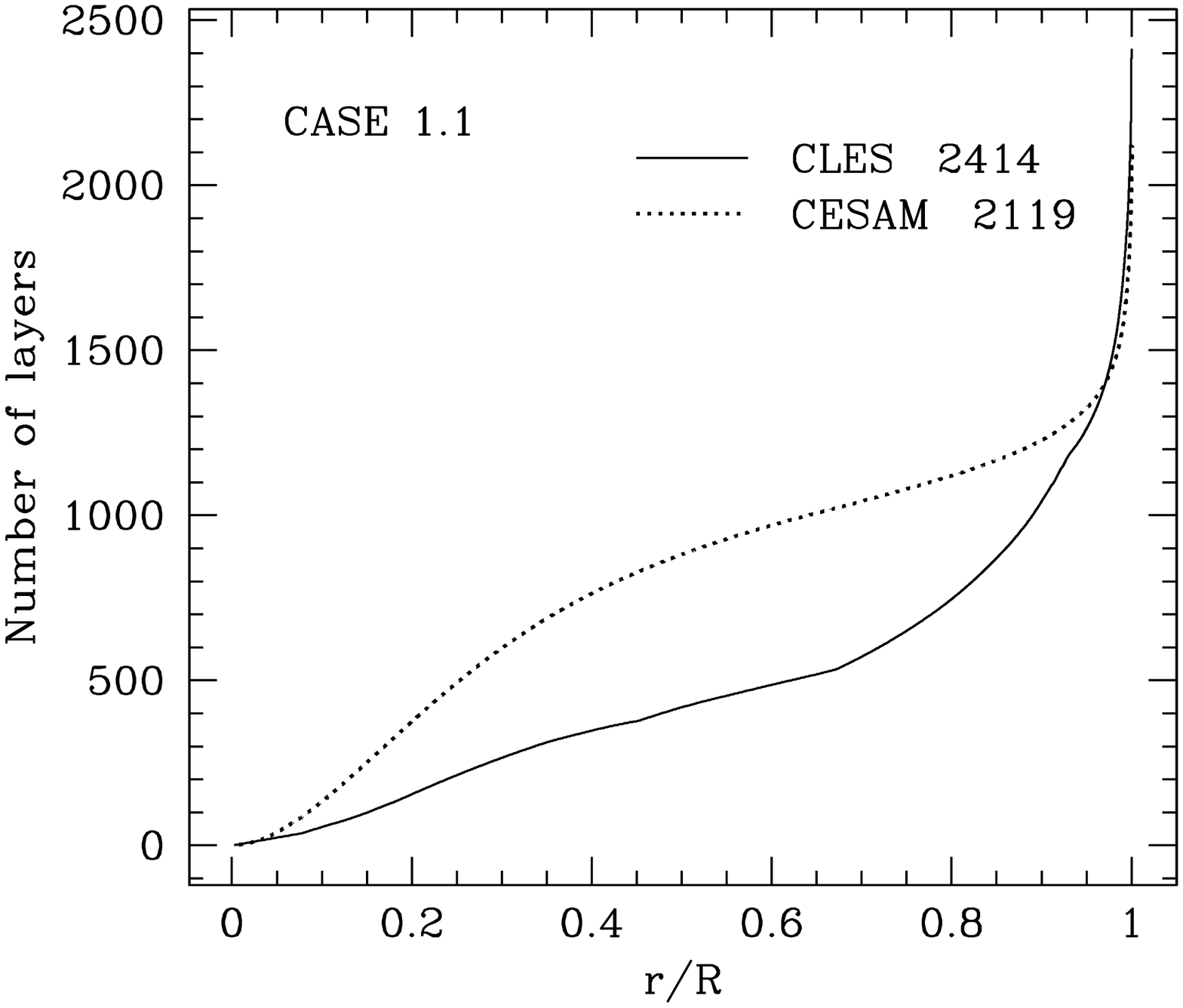}}
\resizebox{\hsize}{!}{\includegraphics{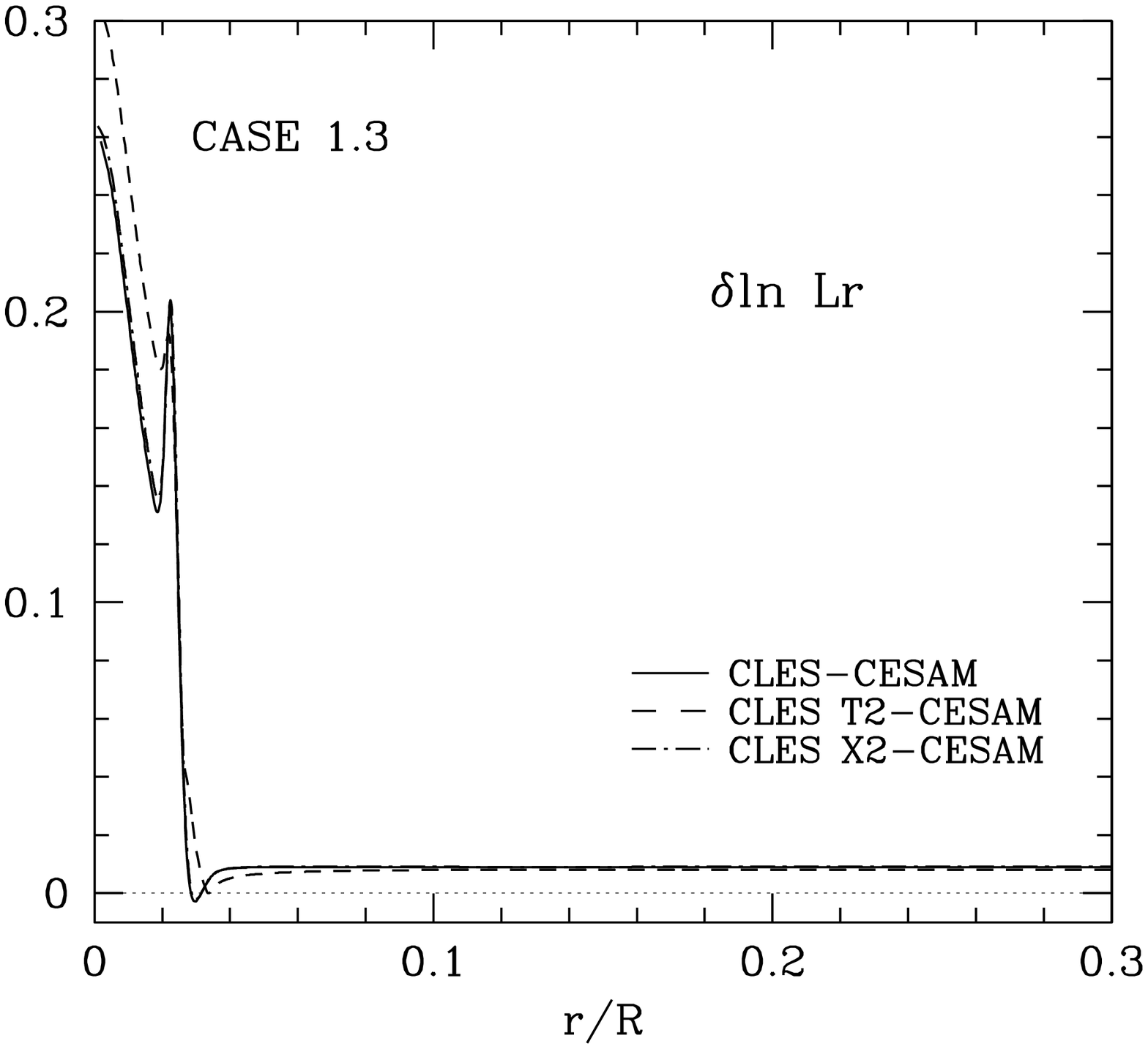}\includegraphics{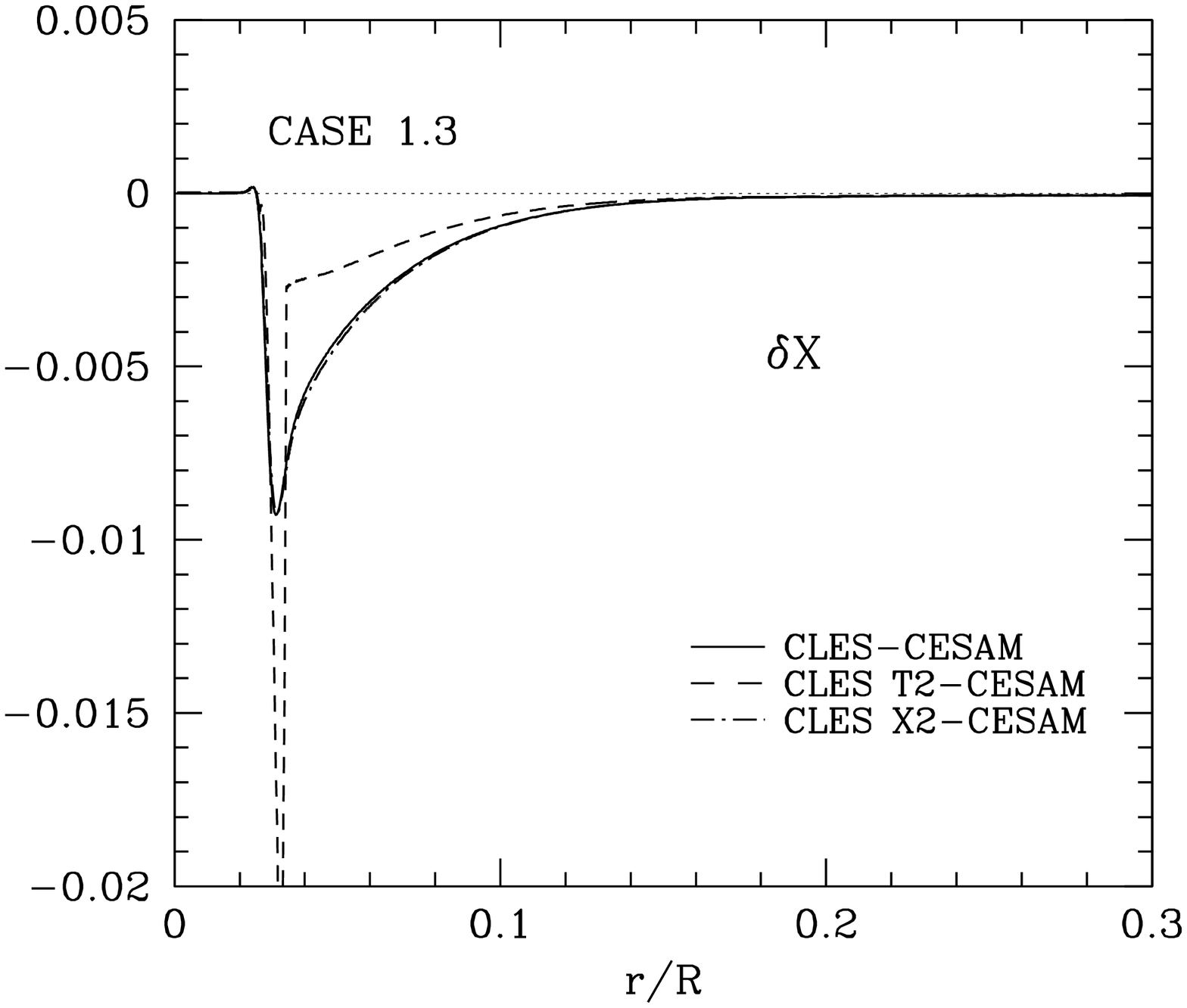}
\includegraphics{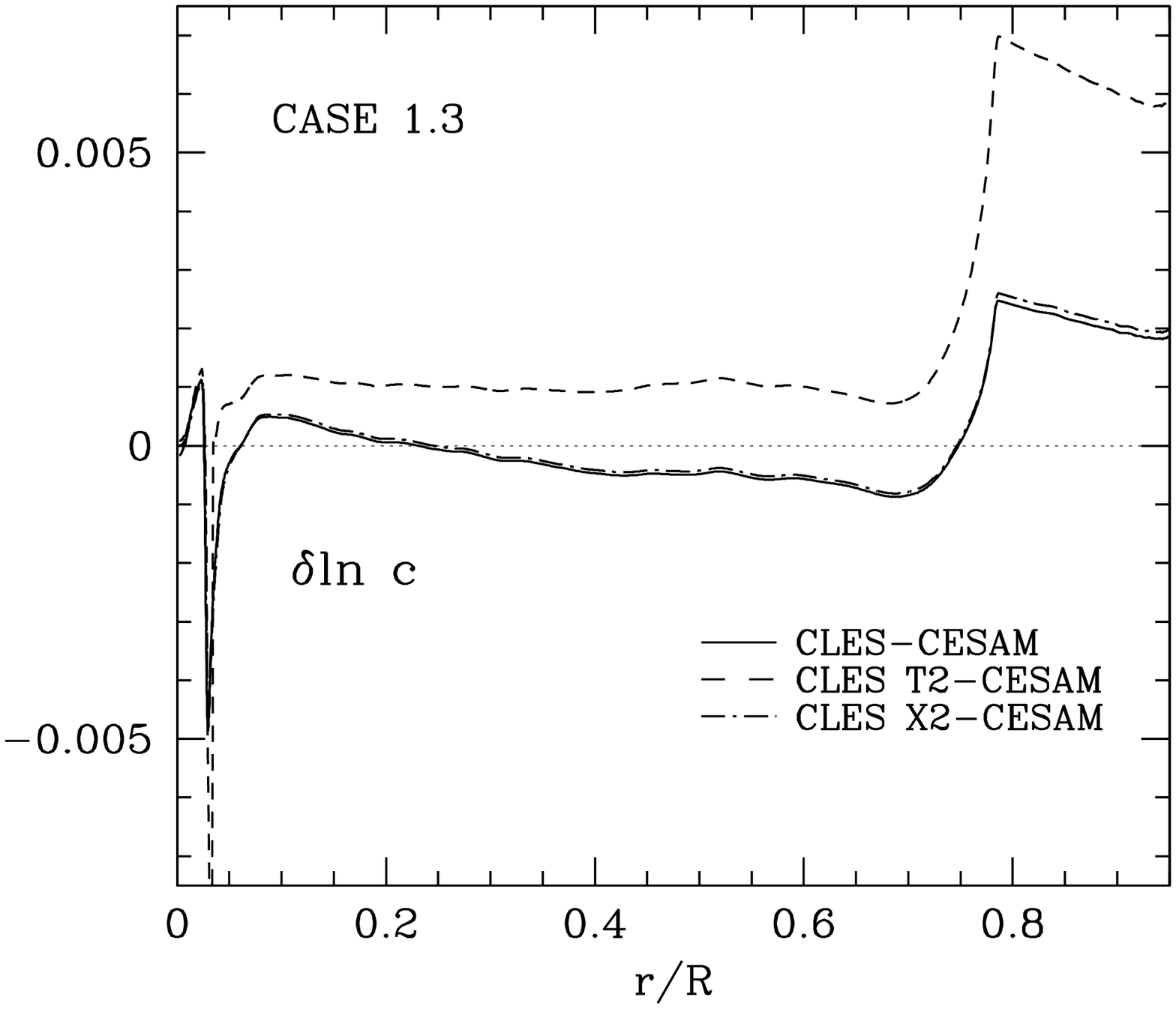} \includegraphics{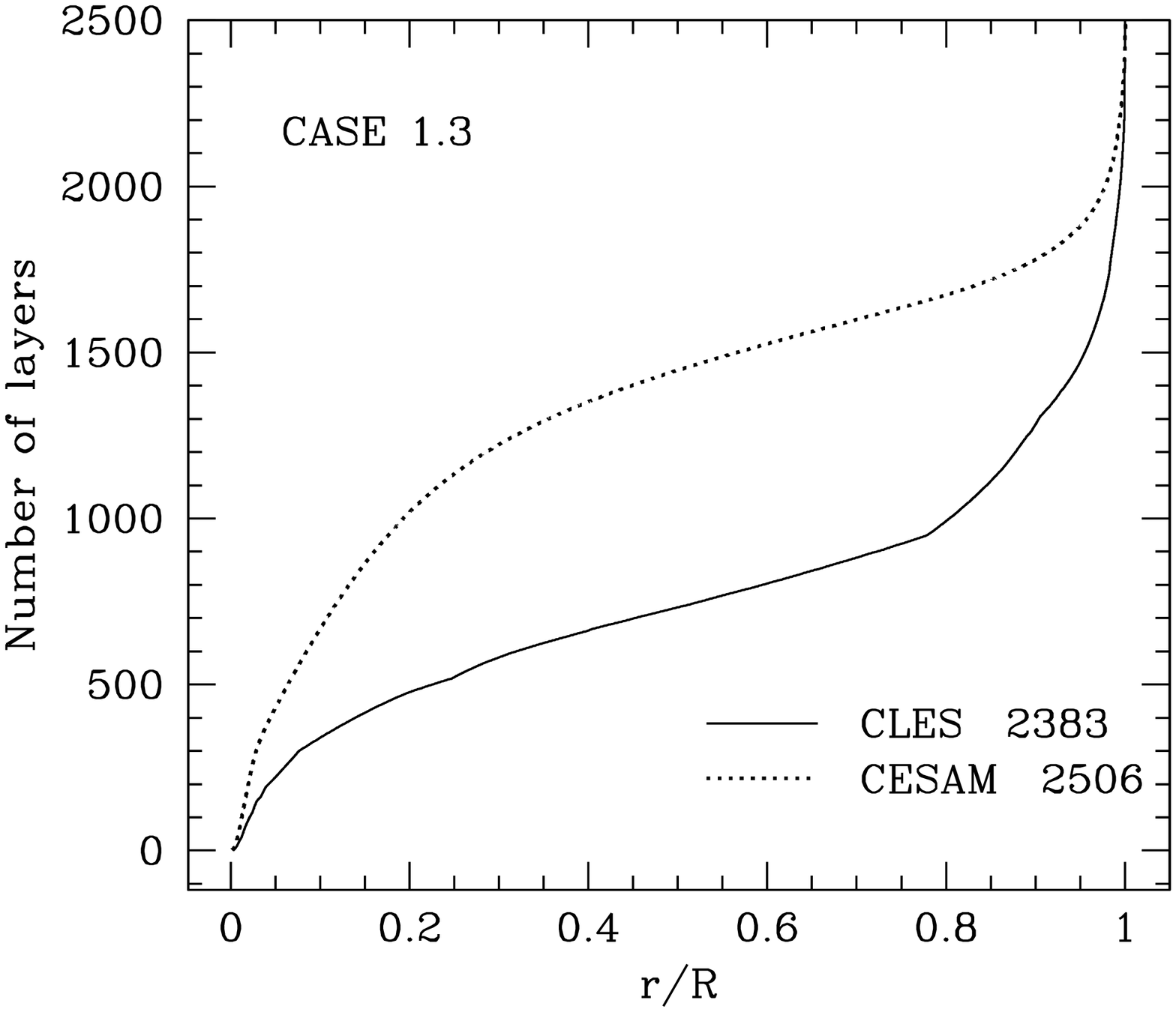}}
\resizebox{\hsize}{!}{\includegraphics{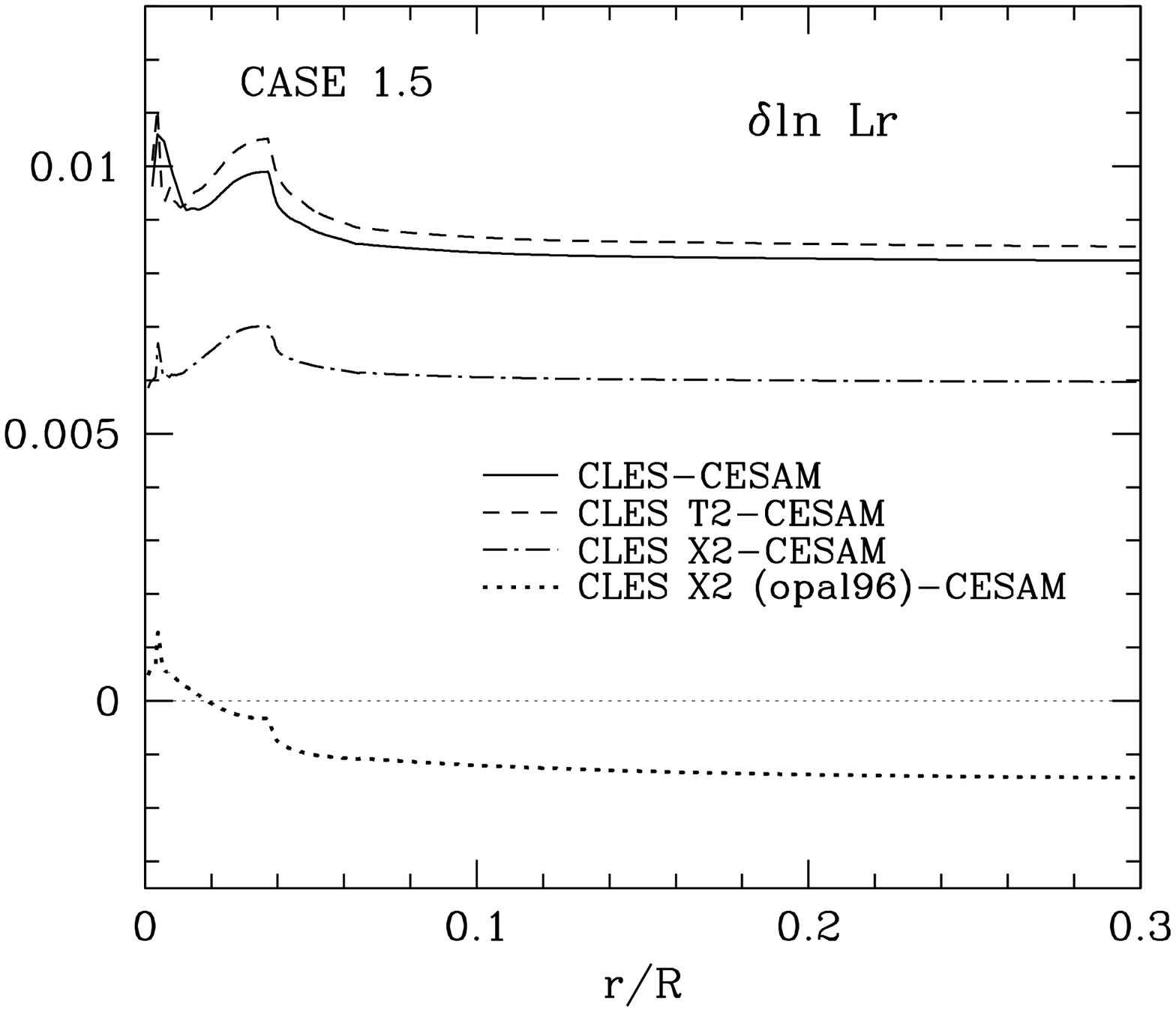}\includegraphics{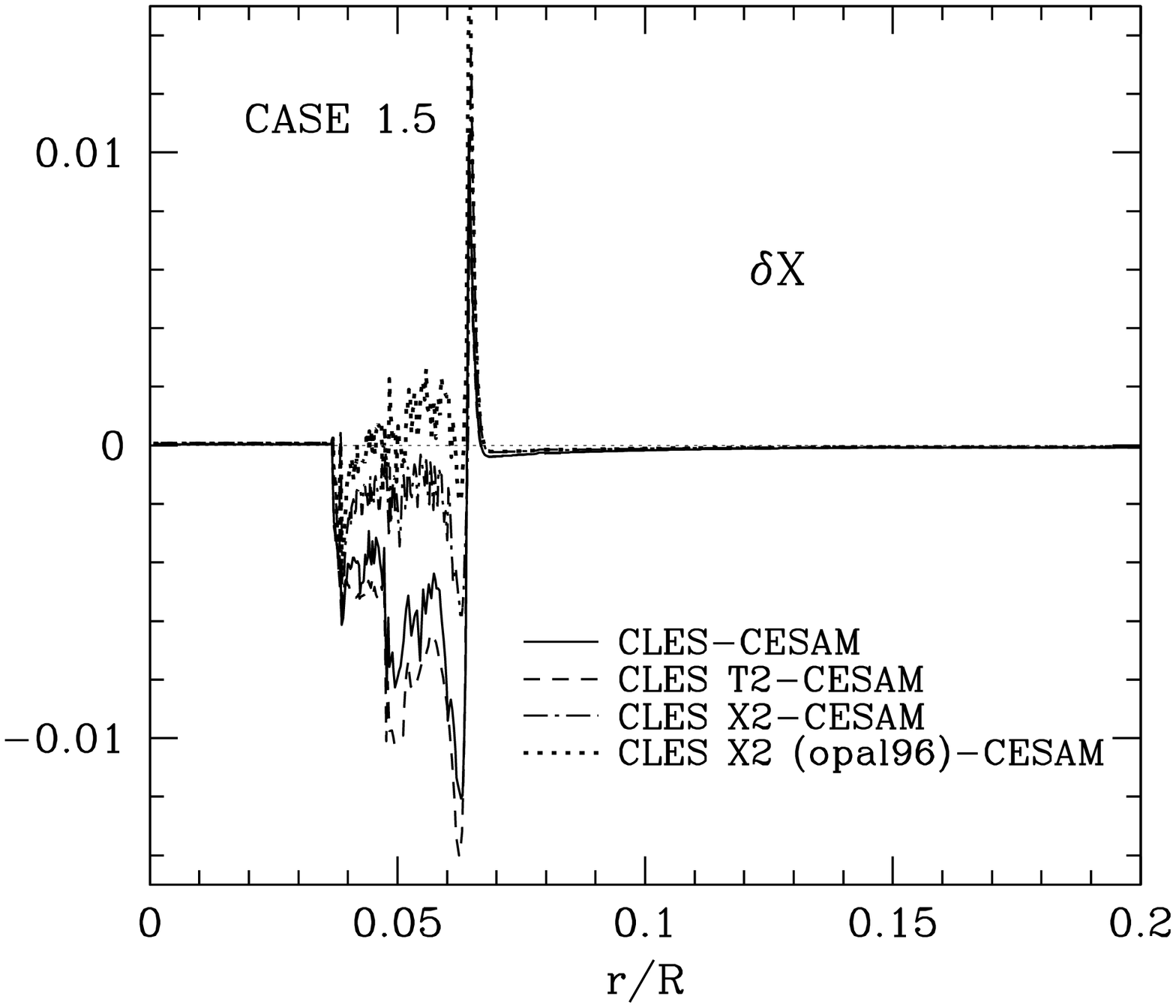}
\includegraphics{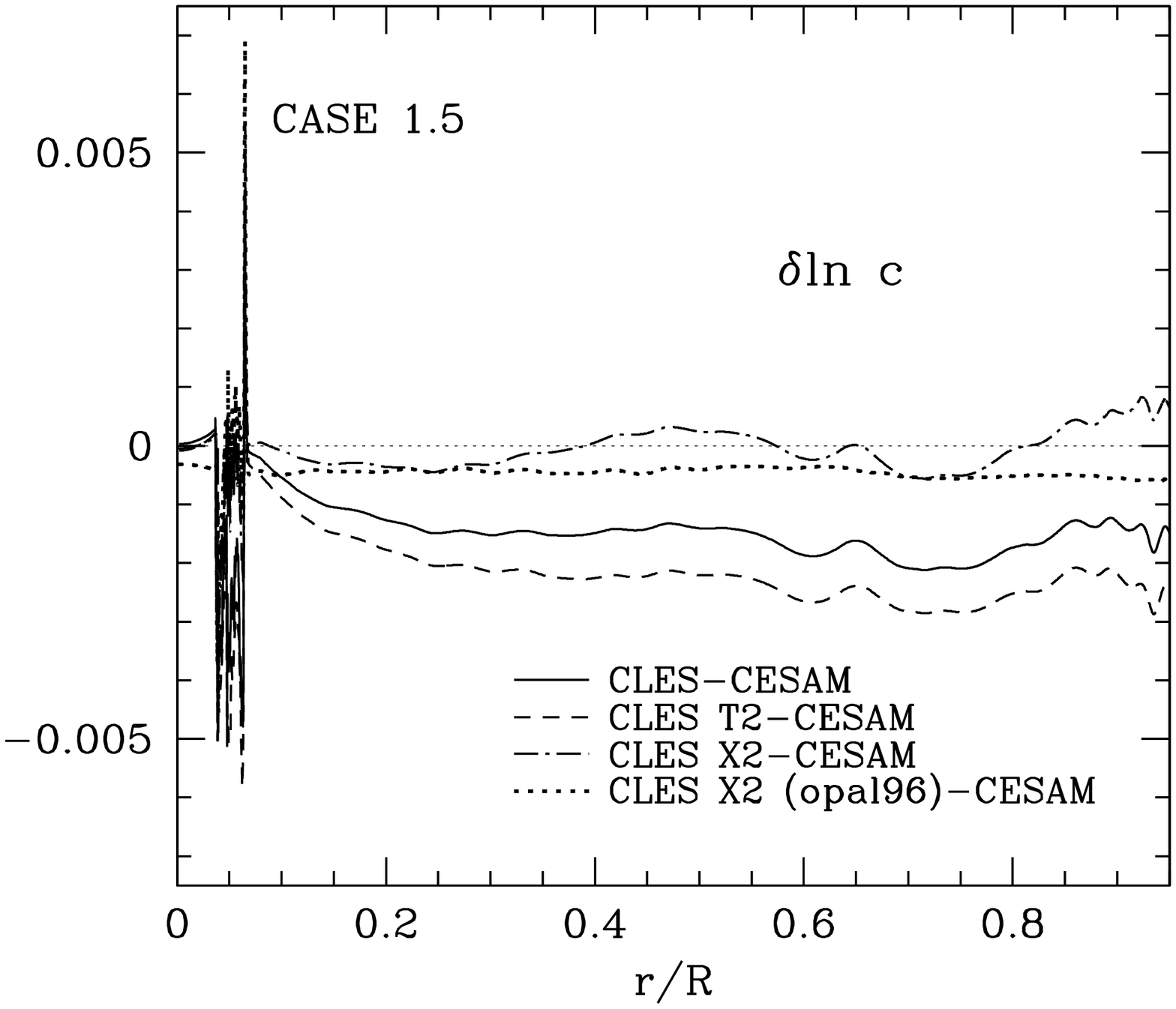} \includegraphics{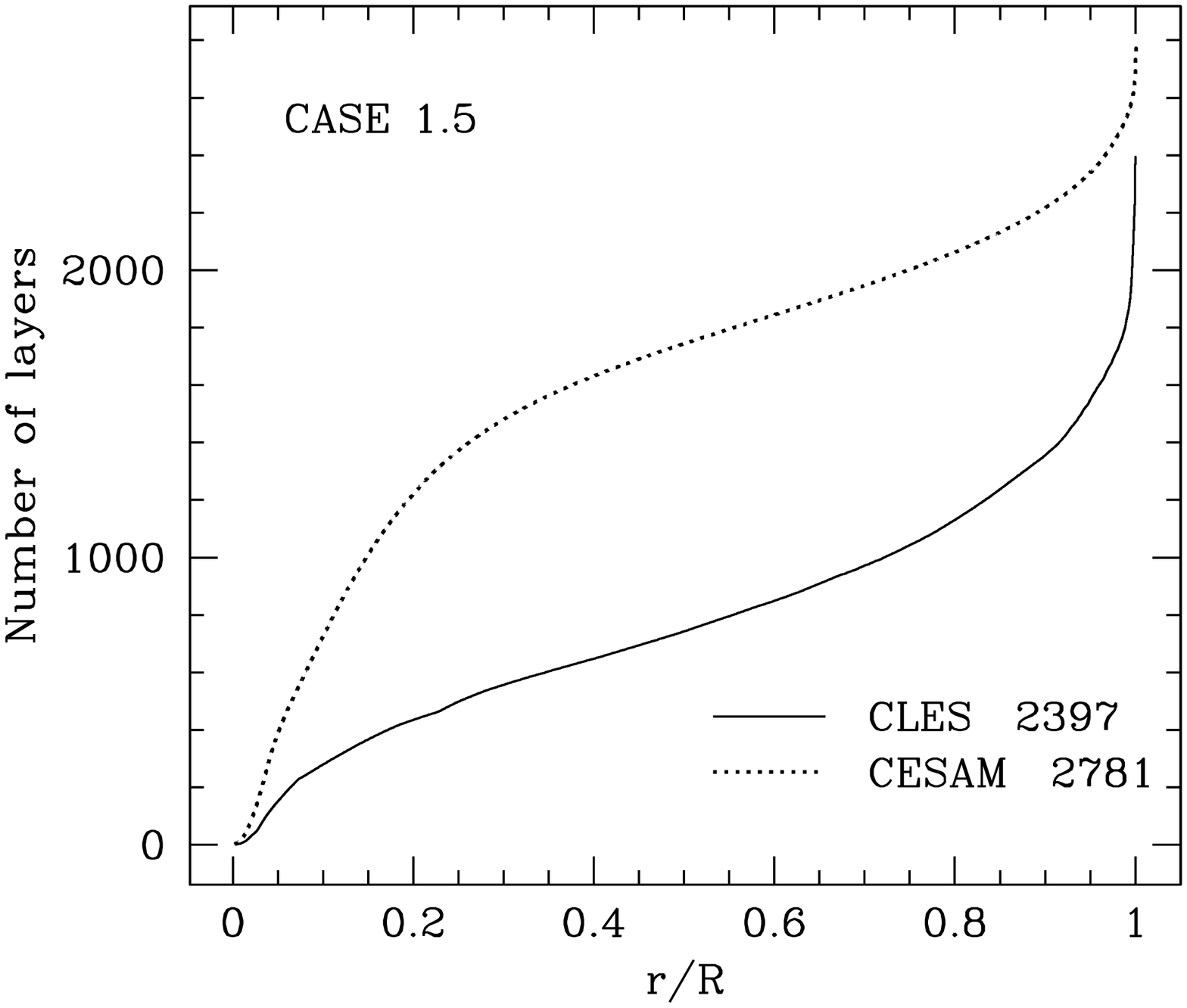}}
\vspace*{0.5cm}
\caption{Plots in terms of the relative radius of the differences at fixed relative mass (three left panels) between three different \cles\ computations  and \cesam\ models. Solid lines correspond to differences between standard \cles\ and \cesam. Dashed lines correspond to
differences between \cles\ models computed by doubling the number of time steps and \cesam\ ones. Dot-dashed lines correspond to differences between \cles\ models computed with a double number of mesh points.
In lower panels there the differences between \cesam\ and \cles\-X2 models computed using the opacity tables without smoothing (opal96) are plotted by using  dotted lines. The right column plots show the distribution of mesh points inside the \cesam\ models (dotted lines) and \cles\ ones (solid lines).} 
\label{fig:meshpoints}
\end{figure*}

While in \cles\ atmosphere computations the condition at the optically thin limit ($\rho(\tau_{\rm min})$) is determined for each ($T_{\rm eff}$, $\log g$, $X$, $Z$)  by a Newton-Raphson iteration algorithm, \cesam\ allows to integrate the atmosphere by fixing either $\rho(\tau_{\rm min})$ or $P(\tau_{\rm min})$. We think it is worth warning here about the relevance of an appropriate choice of $\rho(\tau_{\rm min})$ in \cesam\ calculations. As explicitly indicated in the corresponding  tutorial, the default values were determined for solar like models, and if much different physical conditions are considered, the boundary  conditions in the optically thin limit  should be coherently changed.

These comparisons have allowed to show the discrepancies in frequencies that a ``black-box'' use of an evolution code might lead to. For illustration, in Fig.~\ref{fig:atmos} we show the outer layers of 2~\msol\ models at two different evolutionary states ($X_{\rm c}=0.50$, and $X_{\rm c}=0.01$), with effective temperature  $T_{\rm eff}$=8337~K  and 6706~K respectively. While the atmosphere and sub-phostospheric structure  of \cles\ and \cesam\ models is quite close if  $P(\tau_{\rm min})$ option is used in \cesam, discrepancies that increase with the effective temperature of the model appear when  $\rho(\tau_{\rm min})$  with default values (derived for solar-like models) is adopted. The differences induced in the structure of these outer layers by the use of inappropriate limit values in the atmosphere integration  are much larger than  those due to opacity differences at $\log T \sim 4$. {\it These outer structure differences can lead in fact to frequency differences of the order of several $\mu$Hz}.

\section{Numerical aspects}
\label{sec:num}
The different  numerical techniques  in \cles\ and \cesam\ lead to different distribution of mesh points in the stellar structure and to different values of the time step between two consecutive models.
As pointed out in Sect.~\ref{sec:globalpar} \citep[see also][]{yl2-apss} the disagreement between \cles\ and \cesam\ models can be partially reduced in some cases by changing the  mesh. Even if both sets of models have a similar total number of mesh points, their distribution inside the star, as shown in Fig.~\ref{fig:meshpoints} (right column), is quite different. In this section we analyze  the effect of doubling the number of mesh points (\cles-X2) or  the number of time steps (\cles-T2) on the differences  between \cles\ and \cesam.
As shown in Fig.~\ref{fig:meshpoints}  the effects of these changes are not the same in all the considered stellar cases.
While for the CASE~1.1 the increase of mesh points makes almost disappear the  disagreement between  \cles\ and \cesam\ models ($\delta \ln {\rm Var} \sim 10^{-4} - 5\times10^{-5}$), the effect is almost negligible for the CASE~1.3.
Doubling the number of mesh points in CASE~1.5 leads to a significant effect in decreasing the luminosity of this TAMS model. As pointed out in   Sect.~\ref{sec:globalpar}, a larger number of mesh points near the boundary of the convective core  decreases the effect of the sort of ``numerical diffusion'' that changes the chemical composition gradient at the boundary of the convective core, and that works as a slightly larger overshooting. In fact, the differences of hydrogen abundance  in the region of chemical composition gradient ($r/R$ between 0.035 and 0.06, Fig.~\ref{fig:meshpoints} central-left panel for CASE~1.5) also decrease with respect to those obtained with the standard \cles\ models.
As already discussed in Sect.~\ref{sec:globalpar} a part of the disagreement between \cles\ and \cesam\ comes from the differences in the opacities used in both codes. In the lower panels of  Fig.~\ref{fig:meshpoints} we have also plotted (dotted lines)  the results of comparing the models computed with \cles\  doubling the number of mesh points and using the OPAL96 tables without smoothing. A  significant decrease of luminosity and  hydrogen-profiles differences is obtained when both the number of mesh  points and the  opacity tables are changed.
 
Decreasing the time step in the evolution models  does not  lead, in general,  to better agreement between \cles\ and \cesam.

\section{Conclusions}
In addition to the quantitative results of code comparison presented in \citet{mm2-apss} and \citet{yl2-apss}, the analysis  of stellar models computed with the codes \CESAM\ and \CLES\   has allowed us to reveal some interesting aspects about the two codes, as well as about the input physics, that only a thorough analysis might bring to light.  Some of these evidences have  led to changes or correction of bugs in the codes, other simply allowed us to understand the origin of some differences that were reported in the above mentioned papers.

\begin{itemize}
\item The inconsistencies among the thermodynamic quantities in OPAL2001 equation of state tables lead to differences in the stellar models and in the oscillation frequencies larger  than the uncertainties due to different interpolation tools.
Even if the quantity $C_V$ as tabulated in OPAL2001 tables is not used, the remaining inconsistencies among the three adiabatic indices lead to differences between the model computed with a code that takes the thermodynamic quantities directly from OPAL tables (such as \cesam) and a model computed with a code whose thermodynamic variables are derived by the  thermodynamic relations and a minimum of tabulated quantities (such as \cles). Furthermore, the discrepancies will depend on the choice of tabulated quantities.

\item The precision of the theoretical oscillation frequencies is seriously limited by the uncertainties in the opacity computations.

\item Different approaches used to estimate the electron density in \cesam\ and \cles\ lead to differences in the screening factors that have no relevant effects on the stellar models.

\item Even with a  simple physics, such as the Eddington's law for gray atmosphere, the details of numerical tools can have significant consequences on the seismic properties of the models.

\item The different distribution of mesh points in the models can explain  part of the  disagreement between \cesam\ and \cles\ models.
An increase of mesh points in the internal regions seems to be required in \cles\ to decrease the differences with \cesam.

\end{itemize}

 Apart from the discrepancies in the screening factors which does not significantly affect the oscillation  frequencies, the other factors analyzed here can affect the absolute oscillation  frequencies by up to several $\mu$Hz.

\begin{acknowledgements}
JM, AM, RS, and AN acknowledge financial support from the Belgium Science Policy Office (BELSPO) in the frame of the ESA PREODEX8 program (contract C90199) and from the Fonds National de la Recherche Scientifique (FNRS). P.M. thanks J.P. Marques (Coimbra University) and L. Piau  (Brussels University) for their contribution to the OPAL-EoS implementation in \cesam.
\end{acknowledgements}


\bibliographystyle{Spr-mp-nameyear_modif}
\bibliography{master,preprint}   

\end{document}